\documentclass{aa}

\usepackage[varg]{txfonts}

\usepackage{natbib}
\bibpunct{(}{)}{;}{a}{}{,} 

\defcitealias{2009A&AMordasinia}{M09a}

\defcitealias{2013A&AAlibert}{A13}

\defcitealias{NGPPS1}{Paper I}
\def\paperone{\citetalias{NGPPS1}}

\defcitealias{NGPPS11}{Paper XI}

\usepackage{amssymb, amsmath}
\usepackage{wasysym}

\usepackage{siunitx}
\DeclareSIUnit\erg{erg}
\DeclareSIUnit\year{yr}
\DeclareSIUnit\au{au}

\def\mearth{M_\oplus}
\def\rearth{R_\oplus}
\def\msun{M_\odot}

\def\mj{M_{\textrm{\tiny \jupiter}}}

\def\rj{R_{\textrm{\tiny \jupiter}}}

\def\mstar{M_\star}

\def\mcore{M_{\rm core}}

\def\menv{M_\mathrm{env}}

\def\rtot{R_\mathrm{tot}}

\def\aplanet{a_\mathrm{planet}}

\def\miso{M_\mathrm{iso}}

\def\mstart{M_\mathrm{emb,0}}

\def\betag{\beta_\mathrm{g}}

\def\rin{r_\mathrm{in}}
\def\rcutg{r_\mathrm{cut,g}}

\def\mwind{\dot{M}_\mathrm{wind}}
\def\mgas{M_\mathrm{g}}

\def\sigmasol{\Sigma_\mathrm{s}}
\def\sigmas0{\Sigma_\mathrm{s,0}}
\def\rcuts{r_\mathrm{cut,s}}
\def\betas{\beta_\mathrm{s}}

\def\fpg{f_{\rm D/G}}
\def\fpgsun{f_\mathrm{D/G,\odot}}
\def\feh{[\mathrm{Fe/H}]}

\def\nsystot{N_\mathrm{sys,tot}}

\def\fopa{f_{\rm opa}}

\def\nsys{N_\mathrm{sys}}
\def\npla{N_\mathrm{pla}}

\usepackage{hyperref}
\hypersetup{pdfauthor=A. Emsenhuber et al., pdftitle=The New Generation Planetary Population Synthesis (NGPPS). II., colorlinks=true, urlcolor=blue, linkcolor=blue, citecolor=blue}

\title{The New Generation Planetary Population Synthesis (NGPPS)}
\subtitle{II. Planetary population of solar-like stars and overview of statistical results\thanks{The data supporting these findings is available online at \url{http://dace.unige.ch} under section ``Formation \& evolution''.}}
\titlerunning{The New Generation Planetary Population Synthesis (NGPPS). II.}

\author{Alexandre Emsenhuber\inst{\ref{uofa},\ref{unibe},\ref{usm}} \and Christoph Mordasini\inst{\ref{unibe}} \and Remo Burn\inst{\ref{unibe},\ref{mpia}} \and Yann Alibert\inst{\ref{unibe}} \and Willy Benz\inst{\ref{unibe}} \and Erik Asphaug\inst{\ref{uofa}}}
\authorrunning{A.~Emsenhuber et al.}

\institute{
	Lunar and Planetary Laboratory, University of Arizona, 1629 E. University Blvd., Tucson, AZ 85721, USA\label{uofa}
	\and
	Physikalisches Institut, Universität Bern, Gesellschaftsstrasse 6, 3012 Bern, Switzerland\label{unibe}
	\and
	Universitäts-Sternwarte München, Ludwig-Maximilians-Universität München, Scheinerstraße 1, 81679 München, Germany\\
	\email{emsenhuber@usm.lmu.de}\label{usm}
	\and
	Max-Planck-Institut für Astronomie, Königstuhl 17, 69117 Heidelberg, Germany\label{mpia}
}

\date{Received 7 July 2020 / Accepted 5 October 2021}

\abstract
{Planetary formation and evolution is a combination of multiple interlinked processes. Constraining the mechanisms observationally requires statistical comparison to a large diversity of planetary systems.}
{We want to understand global observable consequences of different physical processes (accretion, migration, and interactions) and initial properties (like disc masses and metallicities) on the demographics of the planetary population. We also want to study the convergence of our scheme with respect to one initial condition, the initial number of planetary embryo in each disc.}
{We select distributions of initial conditions that are representative of known protoplanetary discs. Then, we use the Generation III Bern model to perform planetary population synthesis. We synthesise five populations with each a different initial number of Moon-mass embryos per disc: 1, 10, 20, 50, and 100. The last is our nominal population consisting of 1000 stars (systems) that we use for an extensive statistical analysis of  planetary systems around \SI{1}{M_\odot} stars.}
{The properties of giant planets do not change much as long as there are at least 10 embryos in each system. The study of giants can thus be done with simulations requiring less computational resources. For inner terrestrial planets, only the 100-embryos population is able to attain the giant-impact stage. In that population, each planetary system contains on average 8 planets more massive than \SI{1}{\mearth}. The fraction of systems with giants planets at all orbital distances is \SI{18}{\%}, but only \SI{1.6}{\%} are at \SI{>10}{au}. Systems with giants contain on average 1.6 such planets. The planetary mass function varies as $M^{-2}$ between 5 and \SI{50}{M_\oplus}. Both at lower and higher masses, it follows approximately $M^{-1}$. The frequency of terrestrial and super-Earth planets peaks at a stellar [Fe/H] of -0.2 and 0.0, respectively, being limited at lower [Fe/H] by a lack of building blocks, and by (for them) detrimental growth of more massive dynamically active planets at higher [Fe/H]. The frequency of more massive planets (Neptunian, giants) increases monotonically with [Fe/H]. The fast migration of planets in the \num{5}--\SI{50}{\mearth} range is reduced by the presence of multiple lower-mass inner planets in the multi-embryos populations. To assess the impact of parameters and model assumptions, we also study two non-nominal populations: in-situ formation without gas-driven migration, and a different initial planetesimal surface density.}
{We present one of the most comprehensive simulations of (exo)planetary system formation and evolution to date. For observations, the syntheses provides a large data set to search for comparison synthetic planetary systems that show how these systems have come into existence. The systems, including their full formation and evolution tracks are available online. For theory, they provide the framework to observationally test the global statistical consequences of theoretical models for specific physical processes. This is a important ingredient towards the development of a standard model of planetary  formation and evolution.}


\begin{document}
\maketitle

\section{Introduction}

Exoplanets are common. Results from the \textit{Kepler} survey show that, on average, there are more exoplanets than stars, at least in the galactic environment probed by the Kepler satellite  \citep[e.g.,][]{2018AJMulders,2021ARAAZhuDong}. The number of discovered exoplanets, principally through large surveys, either radial velocity (RV), such as HARPS \citep{2011MayorArxiv} and Keck \& Lick \citep{2021ApJSFulton}, or transit surveys, as \textit{CoRoT} \citep{2013IcarusMoutou} or \textit{Kepler} \citep{2010ScienceBorucki,2018ApJSThompson}, permits to constrain properties of exoplanetary systems, about their mass, radii, distances, eccentricities, spacing and mutual inclinations \citep[e.g.][]{2015ARA&AWinn}. In addition, various correlations with stellar properties have also been determined \citep{2003AASantos,2011MayorArxiv,2018AJPetigura}.

Yet, understanding how the formation and evolution of these planets work remains a challenge. Observations of the progenitors (circumstellar discs) are plentiful, but only few forming planets are known, such as PDS 70b \citep{2018AAKeppler,2018AAMueller}. Reliance on theoretical modelling for the formation stage is then necessary. A model that reproduces the final systems accounting for the initial state can provide valuable information about how planetary systems form and evolve.

For the constrains on planets, we can divide them in three main categories: 1) the characteristics of the planets themselves, for example their mass, radii, distances and eccentricities, 2) the properties of planetary systems and their diversity in terms of architecture, such as their multiplicity, mutual spacing and correlations between occurrences of different planet types, and 3) the correlations between the previous items and stellar properties, such as its metallicity.

Giant planets within 5--\SI{10}{\au} around FGK stars have a frequency of 10-\SI{20}{\percent} \citep{2008PASPCumming,2010PASPJohnson,2011MayorArxiv}.
Earlier works based on radial velocity surveys found that giants have increasing probability of occurrence in log(P) between 2 and 2000 days \citep{2008PASPCumming}, with an excess of hot-Jupiters, as they occur on 0.5-\SI{1}{\percent} of Sun-like star \citep{2010ScienceHoward,2011MayorArxiv,2012ApJWright}. More recent results based on the Kepler satellite survey also find an increase with distance \citep{2013ApJDongZhu,2016AASanterne}. There could be a peak at intermediate distances, possibly near the snow-line \citep{2019ApJFernandes,2021ApJSFulton}. Then there is a decrease in the occurrence rate with distance, where the onset of the reduction could be already at \num{3}-\SI{10}{\au} \citep{2016ApJBryan,2019AJNielsenA} and a \SI{\approx1}{\percent} occurrence rate for detectable distant (tens to hundreds of AUs), massive planets \citep{2016PASPBowler,2016AAGalicher,2021AAVigan}.

System-level statistics provide additional information about properties in a system versus the whole population level. Diversity within each system compared to the whole population is a good example. For instance planets in small-mass systems have similar masses \citep{2017ApJMillholland}, sizes and spacing \citep{2018AJWeiss}. Planet multiplicity tend to decrease for systems that host more massive planets \citep{2011ApJLatham}. For giant planets, hot Jupiters do not usually have nearby companions \citep{2012PNASSteffen}, but roughly half of them have more distant ones \citep{2014ApJKnutson}. Conversely, distant giants also have a multiplicity rate of roughly \SI{50}{\percent} \citep{2016ApJBryan,2019ApJWagner}. There are also correlations between Super Earths and Jupiter analogs \citep{2019AJBryan} and between Super Earths and cold giants \citep{2018AJZhuWu}. This will be the subject of a companion work \citep{NGPPS3}.
In addition, \citet{2016ApJBryan} observed that planets in multiple systems have on average a higher eccentricity than single giant planets; a different result from previous studies that found that planets in multiple systems had on average lower eccentricities \citep{2013ScienceHoward,2015PNASLimbach}.

Correlations between stellar and planetary properties provide important information on the formation mechanism. Protoplanetary discs properties, especially their heavy-elements content, is linked to the host star's metallicity \citep{2016ApJGaspar}, as they form from the same molecular cloud. Giant planets are preferentially found around metal-rich stars \citep[e.g.][]{1997MNRASGonzalez,2004A&ASantos,2005ApJFischer,2019GeoSciAdibekyan}. For low-mass planets, such a correlation still exists although it is weaker \citep{2008AASousa,2011A&ASousa,2012NatureBuchhave,2014NatureBuchhave,2015AJWang,2018AJPetigura}.

Further, we now have correlations between architecture and metallicity, with compact multi-planetary systems being more common on metal-poor stars \citep{2018ApJBrewer} while systems around metal-rich stars are more diverse \citep{2018AJPetigura}. Also, the eccentricities of giant planets around metal-rich stars tend to be higher than the one around metal-poor stars \citep{2018ApJBuchhave}.

From the survey of star forming regions, we can determine the distribution of some characteristics of protoplanetary disc. The percentage of stars with a disc decreases with age in an exponential fashion with a characteristic time of a few \si{\mega\year} \citep{2009AIPCMamajek,2010AAFedele}. Correlations were also found between disc masses and sizes \citep{2010ApJAndrews,2018ApJAndrewsA,2017ApJTripathi,2020ApJHendler}, stellar masses \citep{2013ApJAndrews,2016ApJAnsdell,2016ApJPascucci} and accretion rate onto the star \citep{2016AAManaraB,2019AAManara,2017ApJMulders}.

With these observations, it is possible to retrieve the characteristics at early stages of disc evolution \citep{2018ApJSTychoniec,2020ApJTobinA}, which are relevant for the initial conditions, and constraints on the transport mechanism in effect \citep{2017ApJMulders}.

To link protoplanetary discs to final systems, we need to use a formation model. Several approaches can be used: the study of individual Rosetta Stone systems, statistical studies on the population level as in the case here, or also studies of disc chemistry imprints for formation  \citep[e.g.][]{2011ApJObergB,2016ApJMordasini}. However, the constraints derived from observation for a single exoplanetary system compared to the model parameters does not permit to fully understand planetary formation at the individual system level. In addition, the diverse outcomes of \textit{N}-body simulations \citep[e.g.][]{2012ApJHansenMurray,2013ApJHansenMurray} renders the task even more difficult. Working at the population level, with planetary population synthesis \citep{2004ApJIda1,2009A&AMordasinia,2009A&AMordasinib,2012A&AMordasiniA} is a much more powerful tool to understand planetary formation in general. This allows to determine how the different mechanisms that occur during planetary systems formation of interact.

Modelling planetary formation is a complex task, as many physical effects occur concurrently: growth of micron-size dust to planetary-sized bodies, the accretion of gas, orbital migration and dynamical interactions for multi-planetary systems. In \citet[hereafter \paperone]{NGPPS1}, we present an update of the \textit{Bern} model of planetary formation and evolution. This is a global end-to-end model, i.e. it includes the relevant processes that occur from the initial accretion of the protoplanets starting at the planetesimal-embryo stage up to their long-term evolution, trying to address as many relevant physical processes as possible. By using an approach that is rich in physics, but low-dimensional numerically to keep the computational cost acceptable, this model can be used to compute synthetic planet populations. Our formation model is based on the core accretion paradigm with planetesimals. The early phases of the evolution of the solids from dust to pebbles to planetesimals to embryos and pebble accretion \citep[e.g.][]{2010AAOrmelKlahr} are currently not included, but will be taken into account in future work based on \citet{2020AAVoelkel}.

Theoretical models that are able to reproduce the characteristic of the observed exoplanets can be used to make predictions about the real population, which is helpful when designing future observations and instruments. For discovered planets, they can be used to propose a pathway for their formation \citep{2020NatureArmstrong}, or point to other formation mechanisms if they cannot be reproduced at all \citep{2019ScienceMorales}.

In this work, we apply the Generation III \textit{Bern} model of planetary formation and evolution described in \paperone{} to obtain synthetic populations of planetary systems. We provide the methods that we use to perform population synthesis, which are an update from \citet[hereafter \citetalias{2009A&AMordasinia}]{2009A&AMordasinia}.

We then present five synthetic planet populations for solar-like stars where we vary the initial number of embryos per system, which represent the oligarchs at the end of the planetesimals runaway growth. They act similarly to the large bodies in \textit{N}-body studies, such as \citet{2006IcarusOBrien} or \citet{2009RaymondIcarus}. As we do not model their formation in our work, we treat their number as a free parameter. The goal is to test the convergence of our model with respect to this parameter. The populations with a larger number of embryos are capable to follow the formation of terrestrial planets (\paperone) but they are expensive to compute. On the other hand, the populations with a lower number of embryos are much cheaper to compute (with the extreme case of a single embryo per system), but fails to follow properly terrestrial planets. This test will be useful for future works in this series about the effects of the parameters of the model or physical processes, which requires the computation of multiple populations.

\section{Formation and evolution model}
\label{sec:model}

The model is described in \paperone; so we will give here only a brief summary. In our coupled formation and evolution model, we first model the planets' main formation phase for a fixed time interval (set to \SI{20}{\mega\year}) during which planets accrete solids and gas, migrate, and interact via the N-body.  Afterwards, in the evolutionary phase, we follow the thermodynamical evolution of each planet individually to \SI{10}{\giga\year}.

The formation model derives from the work of \citet{2004A&AAlibert,2005A&AAlibert}. It follows the evolution of a viscous accretion disc \citep{1952ZNatALust,1974NMRASLyndenBellPringle}. The turbulent viscosity is provided by the standard $\alpha$ parameter \citep{1973A&AShakuraSunyaev}. Solids are represented by planetesimals, whose dynamical state is given by the drag from the gas and the stirring from the other planetesimals and the growing protoplanets \citep{2004AJRafikov,2006IcarusChambers,2013A&AFortier}. This disc provides gas and solids from which the protoplanets can accrete while also affecting the bodies that are inside it, by gas-driven planetary migrations.

The formation of the protoplanets is based on the core accretion paradigm \citep{1974IcarusPerriCameron,1980PThPhMizuno}, assuming planetesimal accretion in the oligarchic regime \citep{1993IcarusIdaMakino,2002IcarusOhtsuki,2003A&AInaba,2006IcarusChambers,2013A&AFortier}.
Gas accretion is initially governed by the ability of the planet to radiate away the potential energy \citep{1996IcarusPollack,2015ApJLeeChiang}, and so the envelope mass is determined by solving the internal structure equations \citep{1986IcarusBodenheimerPollack}. Once the planet is massive enough (on the order of \SI{10}{\mearth}), cooling becomes efficient, and runaway gas accretion can occur. In that situation, the envelope is no longer in equilibrium with the surrounding gas disc and contracts \citep{2000IcarusBodenheimer} while gas accretion is limited by the supply of the gas disc.

Multiple embryos can form concurrently in each system, and the gravitational interactions are modelled using the \texttt{mercury} \textit{N}-body package \citep{1999MNRASChambers}.

Once the formation stage is finished, the model transitions to the evolutionary phase, where planets are followed individually to \SI{10}{\giga\year}. The planetary evolution model is based on \citet{2012A&AMordasiniB} and includes atmospheric escape \citep{2014ApJJin} and migration due to tides raised on the star \citep{2011AABenitezLlambay}.

\section{Population synthesis}
\label{sect:popsynt}

\begin{table}
    \centering
	\caption{Fixed parameters for the formation and evolution model.}
	\label{tab:params}
	\begin{tabular}{ll}
			\hline
			Quantity & Value \\
			\hline
			Stellar mass $\mstar$ & \SI{1}{M_\odot} \\
			Disc viscosity parameter $\alpha$ & \num{2e-3} \\
			Power law index of the gas disc $\betag$ & 0.9 \\
			Power law index of the solids disc $\betas$ & 1.5 \\
			Characteristic radius of the solids disc $\rcuts$ & $\rcutg/2$ \\
			Planetesimal radius & \SI{300}{\meter} \\
			Planetesimal density (rocky) & \SI{3.2}{\gram\per\cubic\centi\meter} \\
			Planetesimal density (icy) & \SI{1}{\gram\per\cubic\centi\meter} \\
			Embryo mass $\mstart$ & \SI{1e-2}{\mearth} \\
			Opacity reduction factor $\fopa$ & \num{3e-3} \\
			\hline
	\end{tabular}
\end{table}

To perform a population synthesis of planetary systems, we use a Monte Carlo approach for the initial conditions of the discs, in a similar fashion that has been performed in \citetalias{2009A&AMordasinia} and \citet{2012A&AMordasiniC}. The Monte Carlo variables are selected as:
\begin{itemize}
    \item The initial mass of the gas disc $\mgas$
    \item The external photo-evaporation rate $\mwind$
    \item The dust-to-gas ratio $\fpg=M_\mathrm{s}/\mgas$
    \item The inner edge of the gas disc $\rin$
    \item The initial location of the embryos
\end{itemize}
Here, $M_\mathrm{s}$ is the initial mass of solids in a disc. The other fixed parameters used in this study are provided in Tab.~\ref{tab:params}. These are taken to remain the same in all systems.

In the rest of this section, we discuss each Monte Carlo variable and their distributions, as well as the related fixed parameters. The significant number of parameters in global end-to-end models like the one used here is a notoriously difficult aspect of this approach. The issue naturally results from global models combining many sub-models, each coming with its own model parameters. Some of these parameters are at least to some extent constraint by observations, while others are based on theoretical considerations only, and some are merely educated guesses. When interpreting the results presented in this work, like for example the key demographic predictions of planet occurrence rates or the general shape of the planet mass-distance diagrams, it is important to keep in mind that these results are clearly functions of the chosen parameters and base assumption underlying the formation model. Thus, these results always have to be seen as the predictions made in the context of the current model and for the chosen (nominal) parameter values, and that large systematic uncertainties exist.

Ideally, one would quantitatively assess the impact of all these parameters by running numerous syntheses were the values of the parameters, as well as important underlying model assumptions, are varied systematically. This would give an understanding of the systematic uncertainties in the model predictions. In practice, this is not easily feasible, because the computational cost of the multi-embryo syntheses is very significant ($\sim$1 M CPU h), especially for a high number of initial embryos per disc. To still elucidate the impact of parameters at least for two of them (besides the initial number of embryos per disc), we present in Appendix \ref{sec:modparams} two non-nominal populations: one, where the initial solid surface density of planetesimals has a different slope, and one where gas-driven orbital migration is neglected. In the appendix, we study how this changes the mass-distance diagram, and key  demographic properties.

\subsection{Gas disc mass}

\begin{table}
    \centering
	\caption{Mean and standard deviation of the normal distribution of the disc mass for different observational sample.}
	\label{tab:mdisc}
	\begin{tabular}{lcc}
			\hline
			Source & $\mu$ & $\sigma$ \\
			\hline
			Fit to Taurus\textsuperscript{\it a} & -1.66 & 0.74 \\
			Fit to Ophiuchus\textsuperscript{\it a} & -1.38 & 0.49 \\
			\citet{2010ApJAndrews} & -1.66 & 0.56 \\
			Fit to class I from \citet{2018ApJSTychoniec} & -1.49 & 0.35 \\
			Class I from \citet{2019ApJWilliams} & -2.94 & 0.86 \\
			\hline
	\end{tabular} \\
	\raggedright\textsuperscript{\it a} Fit to the values obtained by \citet{1996NatureBeckwithSargent} performed by \citetalias{2009A&AMordasinia}.
\end{table}

\begin{figure}
	\centering
	\includegraphics{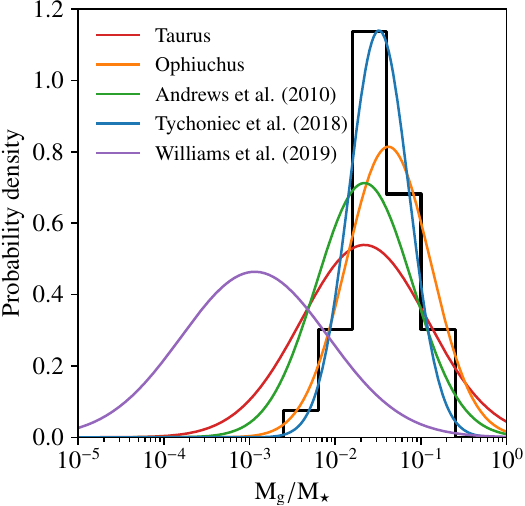}
	\caption{Probability density functions for the different distributions given in Table~\ref{tab:mdisc}. In addition, we show the histogram of Class~I discs from Fig.~12 of \citet{2018ApJSTychoniec} in black. All the curves are normalised so that the surface below them is unity.}
	\label{fig:mdsic}
\end{figure}

It is very difficult to observe directly H\textsubscript{2} in protoplanetary discs, and so the most reliable method to determine disc masses remains the measurement of the continuum emission of the dust. To recover the gas mass, a dust-to-gas ratio similar to the interstellar medium is applied \citep{1996NatureBeckwithSargent,2005ApJAndrewsWilliams,2010ApJAndrews}.

Several observational data for protoplanetary disc masses are reported in Table~\ref{tab:mdisc} and plotted in Fig.~\ref{fig:mdsic}. The first two values, for the fits on the distributions of Taurus and Ophiuchus star-forming regions were obtained by \citetalias{2009A&AMordasinia} by fitting log-normal distributions on the results of \citet{1996NatureBeckwithSargent}. The third value was directly given in \citet{2010ApJAndrews}, while for the fourth one, we applied the same procedure as for the first two, but using the histogram of Class~I disc masses reported in Fig.~12 of \citet{2018ApJSTychoniec}. Finally, we provide ALMA data of Class~I discs in the Ophiuchus star-forming region from \citet{2019ApJWilliams}. The latter was converted to gas masses using a gas-to-dust mass ratio of 100:1 (as in \citealp{2018ApJSTychoniec}).

There is more than one order of magnitude difference between the results from ALMA for the Ophiuchus star-forming region \citep{2019ApJWilliams} and others, such as those obtained with the VLA for Perseus \citep{2018ApJSTychoniec}. These differences are discussed in \citet{2020AATychoniec}, where the authors argue that 1) their median masses from VLA are more complete and 2) Class~0/I objects are more likely to be representative of the discs at early stage of planetary formation. The second point is related to our modelling, as the model used in this work begins once the protoplanetary disc is formed and dust has grown into planetesimals. Class I discs are hence the most relevant for our study. Thus, the work of \citet{2018ApJSTychoniec} is then the best suitable for our initial conditions, and this is the one we select. To avoid extreme values, we only allow disc masses between \num{4e-3} and \SI{0.16}{\mstar}. With this upper mass limit, the discs are always self-gravitationally stable.

Compared to the populations obtained with earlier versions of the model, our disc masses are smaller than the ones from \citetalias{2009A&AMordasinia}, which used the parameters derived from fitting the values in the Ophiuchus star-forming region from \citet{1996NatureBeckwithSargent}. It should noted that unlike \citetalias{2009A&AMordasinia}, we model the entire disc and not only the innermost \SI{30}{\au}, so we do not need to scale the disc masses to obtain only the innermost region. However, the distribution we adopted has a higher mean than what was obtained by \citet{2010ApJAndrews}; so we have overall larger disc masses than in the works of \citet{2012A&AMordasiniB,2012A&AMordasiniC}. \citetalias{2013A&AAlibert}, \citet{2013A&AFortier}, and \citet{2014AAThiabaud,2015A&AThiabaud} also used the results from \citet{2010ApJAndrews}, albeit in a different fashion, where initial masses were bootstrapped from the specific values of the observed discs.

\subsection{Initial gas surface density: Spatial distribution}
\label{sec:pop-gas-spatial-dist}

With spatially resolved discs it is possible to estimate the distribution of the material with respect to the distance from the star. The surface density typically goes with $r^{-1}$ until a characteristic radius where it relates more to an exponential decrease \citep{2008ApJHughes,2009ApJAndrews,2010ApJAndrews}. While in principle both the index of the power law and the characteristic radius would require their own distributions, we decided against adding more parameters for the initial conditions of our populations.

The power law index is fixed to $\betag=0.9$, which is consistent with the results from \citet{2010ApJAndrews}. For the characteristics radius $\rcutg$ as a function of disc mass, we use the following relationship, which is taken from Fig.~10 of \citet{2010ApJAndrews},
\begin{equation}\label{eq:mdiskrdisk}
    \frac{\mgas}{\SI{2e-3}{\msun}}=\left(\frac{\rcutg}{\SI{10}{\au}}\right)^{1.6}.
\end{equation}
The relationship is somewhat different than the $\mgas\propto\rcutg^2$ found in more recent work \citep{2017ApJTripathi,2018ApJAndrewsA}. Further, the latter is however not universal across different stellar forming regions with various ages \citep{2020ApJHendler}. The results of that work also suggest that the relationship becomes shallower with age, and the power-law index we use is similar to the youngest stellar forming regions, and thus more appropriate as an initial condition.

A complication arises from the fact that the observational relation of Eq.~\ref{eq:mdiskrdisk} was derived for dust disc radii of Class II discs, and not for gas discs at early times (e.g., after the end of infall and potential gravitational instabilities). On one hand, this could mean that our approach leads to too  small initial gas disc radii in our synthetic disc population given the effect of inward drift of dust \citep{2018ApJAnsdell}. On the other hand, the discs observed by \citet{2010ApJAndrews} were specifically selected for good observability with SMA and span the upper half of the millimeter continuum luminosity distribution only. This could mean that the disc radii are on the large side compared to a more representative sample. These effects could partially cancel each other. To elucidate this, we present in Sect.~\ref{sect:comptobin} a comparison of our theoretical disc gas radii with the dust radii of younger discs, as found with the more recent VANDAM survey \citep{2020ApJTobinA}.

\subsection{External photo-evaporation rate}
\label{sec:mwind}

\begin{figure}
	\centering
	\includegraphics{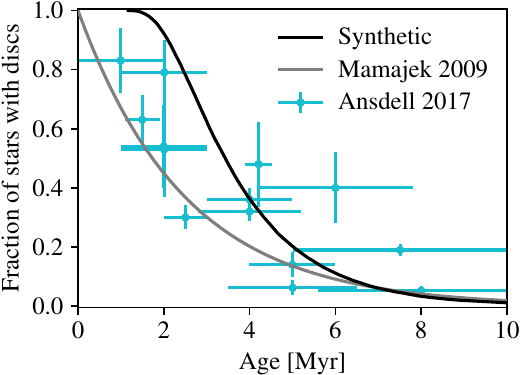}
	\caption{Fractions of stars with a protoplanetary disc as function of their age. The black line shows our results, while the blue line follow the exponential decay with a time scale of \SI{2.5}{\mega\year} from \citet{2009AIPCMamajek}. The purple points are from \citet{2017PhDAnsdell}.}
	\label{fig:lifetimes}
\end{figure}

The photo-evaporation rate $\mwind$ and the viscosity parameter $\alpha$ are the main parameters that determine the life time of the gas discs. This is a degenerate problem, as increasing either $\alpha$ or $\mwind$ leads to shorter disc life times. However, $\alpha$ also constrains the mass that is accreted onto the star, which we can use to lift the degeneracy. Our aim is then to find combinations of $\alpha$ and $\mwind$ that provide accretion rates onto the star and disc life times that are in agreement with observations.

\citet{2017ApJMulders} combined the \textit{ALMA} observations of the disc mass $M_\mathrm{disc}$ from \citet{2016ApJPascucci} and the \textit{X-Shooter} accretion rate onto the star $\dot{M}_\mathrm{acc}$ from \citet{2016AAManaraA,2017AAManara} for the Chamaeleon~I star-forming region and \textit{ALMA} from \citet{2016ApJAnsdell} and \textit{X-Shooter} from \citet{2014AAAlcala,2017AAAlcala} for the Lupus region. The $M_\mathrm{disc}$--$\dot{M}_\mathrm{acc}$ relation obtained by the combination of the two region is shallower than linear, indicating that another effect than viscous dissipation is potentially at play. Nevertheless, they obtained that for $\alpha$ values between \num{e-3} and \num{e-2}, it is possible to find relations that are comparable with observation.

\citet{2019AAManara} compared the $M_\mathrm{disc}$--$\dot{M}_\mathrm{acc}$ relation predicted in a population synthesis obtained with an earlier version of the formation used in this work to an extended sample relative to \citet{2017ApJMulders}. The synthetic disc population for a constant $\alpha$ fails to reproduce the whole scatter observed in the actual $M_\mathrm{disc}$--$\dot{M}_\mathrm{acc}$ relationship. Nevertheless, the synthetic population of discs is able to retrieve the observed correlation of $M_\mathrm{disc}$ and $\dot{M}_\mathrm{acc}$. Thus, to avoid introducing one more Monte Carlo variable in our population synthesis scheme, we will stick to a single $\alpha$ value for all discs. We selected a value of $\alpha=\num{2e-3}$, which is the same as the comparison shown in \citet{2019AAManara}. This leaves only the value of the external photo-evaporate to determine the life times of the discs.

Proptoplanetary discs have a lifetime in the 3--\SI{7}{\mega\year} range \citep{2001ApJHaisch,2010AAFedele,2018MNRASRichert}. Fitting the results with an exponential law gives time constants of \SI{2.5}{\mega\year} \citep{2009AIPCMamajek} or \SI{2.7}{\mega\year} \citep{2017PhDAnsdell}.

Given the fixed $\alpha=\num{2e-3}$ and the fixed distribution of initial disc masses described above, we determine an empirical distribution of external photoevaporaiton rates that leads to a distribution of the lifetimes of the synthetic discs that is in agreement with the observed distribution of disc lifetimes.

In this way, we find a log-normal distribution with parameters $\log_{10}(\mu/(\si{\msun\per\year}))=-6$ and $\sigma=\SI{0.5}{dex}$. Note that these rates would give the actual photoevaporation rates if the modelled discs would have a size of \SI{1000}{\au} (\paperone). In reality, their outer radius are smaller ($\sim\SI{100}{\au}$) and given dynamically by the equilibrium of viscous spreading that acts to increase the outer radii and external photoevaporation which reduces the radii.

The selection of those values was made so that we have a cluster of disc life times at about \SI{3}{\mega\year}. We show in Fig.~\ref{fig:lifetimes} the corresponding life times obtained using our model for the disc masses, $\alpha$ and $\mwind$ that we selected. While we miss the short-lived discs (less than \SI{1}{\mega\year}), our distribution is more able to reproduce some longer-lived clusters in the range of \SIrange{4}{6}{\mega\year}.

\subsection{Dust-to-gas-ratio}

\begin{table}
    \centering
	\caption{Mean and standard deviation of the normal distribution of $\feh$ for different observational sample.}
	\label{tab:feh}
	\begin{tabular}{lcc}
			\hline
			Source & $\mu$ & $\sigma$ \\
			\hline
			\citet{2005AASantos} & -0.02 & 0.22 \\
			\citet{2017AJPetigura} & 0.03 & 0.18 \\
			\hline
	\end{tabular}
\end{table}

The initial mass of the solids disc is linked to that of the gas disc by a factor $\fpg$. To determine the distribution of this parameter, we assume that stellar and disc metallicites are identical. Hence we have the relation \citep{2001ApJMurray}
\begin{equation}
    \frac{\fpg}{\fpgsun}=10^{\feh}.
\end{equation}

Furthermore, we now assume that the dust-to-gas of the Sun, $\fpgsun=0.0149$ \citep{2003ApJLodders}. It should be noted that this value is quite lower than in the first generation of our planetary population syntheses \citep{2009A&AMordasinia,2012A&AMordasiniC}, where it was taken to be a factor roughly three times greater.

There are multiple possibilities for the distribution of the parameter; as stellar metallicities vary among different regions in the galaxy. The choice depends on the kind of observational survey we aim to compare to. RV surveys will favour stars in the neighbourhood of the Sun, while transit and in particular microlensing surveys can reach greater distances. For instance, the \textit{Kepler} survey targets stars only in one specific direction towards Cygnus and Lyra. We provide the parameters of a normal distribution from different sources in Table~\ref{tab:feh}. The two distribution are similar, with the difference in their mean corresponding to \SI{12}{\percent}. The selection of either distribution should therefore not affect significantly our results. For the population syntheses presented below, we use the distribution from \citet{2005AASantos} for the Coralie RV search sample.

One thing to mention is that the normal distribution is unbound on both sides. Hence to avoid modelling system that have metallicities not occurring in the solar neighbourhood given galactic chemical evolution, we restrict the selection of the parameter to the $-0.6<\feh<0.5$ range.

\subsection{Inner edge}

\begin{figure}
	\centering
	\includegraphics{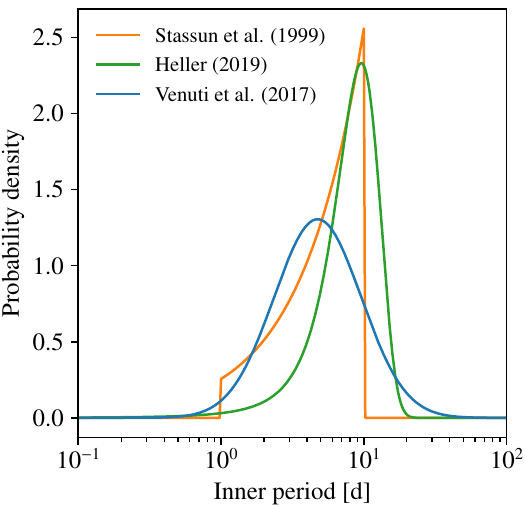}
	\caption{Probability density functions for the different distributions of inner radius as given in the text. All the curves are normalised so that the surface below them is unity.}
	\label{fig:inner-rad}
\end{figure}

The position of the inner edge of the gas disc plays an important role for the final location of the close-in planets. For planets that form and then migrate inward, migration will stall when the planet reaches a location where gas is no longer present. If planets rather form in-situ, then the inner edge is also linked to where planets are able to accrete.

There are various possible ways to determine the inner edge of protoplanetary discs, for example 1) determining stellar rotation rate and assuming the disc is truncated at the corotation radius, 2) from the continuum near infra-red (NIR) emissions, and 3) from emission lines.

We chose to use the corotation radius to determine the inner edges of protoplanetary discs. Apart from the good agreement to observations, the main reason of this choice over a prescription for the magnetospheric truncation radius is that the magnetic field strengths of young stars are not very well constrained. \citet{2019AAHeller} recently investigated planet formation scenarios using either inner edges at the corotation or at a prescribed magnetospheric truncation radius. In any case, the two radii lie very close to each other: \citet{2019AAHeller} used a magnetic truncation period of 4 days motivated by works of \citet{2006ApJRomanovaLovelace} and \citet{2002ApJKuchnerLecar}.

The continuum NIR mainly originates from the hot dust and not from dust-depleted gas. However, as indicated by observations from \citet{2005ApJEisnerA,2009ApJEisnerA} and \citet{2008AAIsella} and in detail modeled by \citet{2017ApJFlockA}, the temperatures at the inner edge of the disc are larger than the evaporation temperature of silicate dust grains. Therefore, the gas extends closer to the star than the silicate evaporation line. Thus, NIR might not be able to trace the inner edge of the gas discs. This is the likely reason put forward by \citet{2005ApJEisnerA} who found consistently larger radii by NIR interferometry than the corotation and the magnetospheric truncation radii.

As for emission lines, they are able to trace the gas disc. \citet{2007IAUSCarr} found a factor of 0.7 smaller disc radii (using the CO $v=1-0$ transition near \SI{4.7}{\micro\meter}) than the corotation radius. This is a reasonable agreement given the scatter of the distribution. The largest dataset of this sort by \citet{2009ApJEisnerA,2010ApJEisnerA} consists of 15 discs around stars of various masses (including 7 T tauri stars). This is a low number of observation, from which it is difficult to extract a full distribution. It can nevertheless be noted that the values are in good agreement with magnetic corotation truncation discussed below.

By using the corotation radius, the location of the inner edge can be derived from rotation rates of young stellar objects (YSOs). We show several distributions of those values in Fig.~\ref{fig:inner-rad}: a uniform distribution in the period between \num{1} and \SI{10}{\day} that is compatible with the results of \citet{1999AJStassun}, a normal distribution with parameters $\mu=\SI{8.3}{\day}$ and $\sigma=\SI{5}{\day}$ derived by \citet{2019AAHeller} based on the work of \citet{2008MNRASIrwin}, and a log-normal distribution with a mean $\log_{10}(\mu/\mathrm{d})=0.67617866$ and deviation $\sigma=\SI{0.3056733}{dex}$ that is derived from the work of \citet{2017AAVenuti}.

In the present work, we adopt the last one, based on \citet{2017AAVenuti}. Here, the mean corresponds to a rotation period of 4.7 days or a distance of \SI{0.055}{\au}. To avoid that some discs have inner edges that are smaller than the initial stellar radius predicted by the stellar evolution model (\paperone), we truncate the distribution so that no inner edge can be within \SI{1.65e-2}{\au}, which corresponds to a period of 0.77 days. We use the period as the main variable to obtain the inner radius as it is largely independent of the stellar mass at young ages \cite[e.g.][]{2012ApJHendersonStassun}.

It should be noted that the means of all the distributions presented here are lower than what is obtained in other works, such as 10 days in \citet{2017ApJLeeChiang}. The value of 10~days also correspond the peak of the location of the innermost planet as found by \textit{Kepler} \citep{2018AJMulders}.

Concerning the solids disc, we do not place planetesimals inside the iron evaporation line as given by our condensation model (\paperone). Therefore, if the iron evaporation line is further out than the inner edge of the gas disc, a disc has two edges: one for the gas and one for the planetesimals. The region inside the planetesimal inner edge will not contribute to solid accretion, but it can be an important region for orbital migration. However, it is found that the inner part is hot enough only for the most massive gas discs with small inner edges. In most cases, the temperature at the inner edge of the gas discs is less than about 1700 K, meaning that the inner edge of gas and solid disc coincide.

\subsection{Planetesimal disc masses}

\begin{figure}
	\centering
	\includegraphics{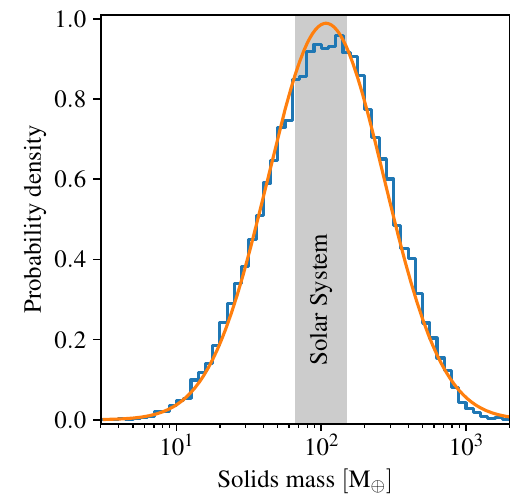}
	\caption{Distribution of initial planetesimals disc masses. The blue curve is an histogram of the actual values while the yellow curve show a log-normal fit to the data, whose mean (in log-space) is \SI{108}{\mearth} and a standard deviation of \SI{0.40}{dex}. The gray area denotes the possible range of values for the minimum-mass solar nebula (MMSN).}
	\label{fig:pla-disc-masses}
\end{figure}

\begin{table}
	\centering
	\caption{Adopted values for our calculation of the minimum-mass solar nebula (MMSN).}
	\begin{tabular}{l|c|c|l}
	    Items & Min. & Max. & Ref. \\
	     & [\si{\mearth}] & [\si{\mearth}] \\
	    \hline
	    \hline
	    Terrestrial & 2 & 2 & \\
	    Jupiter & 24 & 46 & \citet{2017GRLWahl} \\
	    Saturn & 16 & 23 & \citet{2019ApJMilitzer} \\
	    Uranus & 10 & 14 & \citet{2000PSSPodolak} \\
	    Neptune & 14 & 16 & \citet{2000PSSPodolak} \\
	    Migration & -- & 50 & \citet{1999AJHahnMalhotra} \\
	    \hline
	    Total & 66 & 151
	\end{tabular}
	\label{table:mmsn}
\end{table}

The total mass in solids is not itself a Monte Carlo variable, but the product of the gas disc mass $\mgas$ with the dust-to-gas ratio $\fpg$. However, it is one of the most important quantities that determines the types of planets that will be formed. Thus, it is still worth discussing. The distribution of the total mass in solids is shown in Fig.~\ref{fig:pla-disc-masses}. The disc masses were computed using the disc model, in a similar fashion than for the disc life times (Sect.~\ref{sec:mwind}). As the distribution of solids mass is the product of two log-normal distributions (the gas disc mass and the dust-to-gas ratio), it also close to a log-normal distribution (because the two underlying distributions are truncated plus the reduction of solids mass by volatiles being in the gas inside the corresponding ice lines). We therefore fitted a log-normal distribution, whose parameters are a mean (in log-space) of \SI{108}{\mearth} and a standard deviation of \SI{0.40}{dex}.

To compare the obtained masses with the solar system, we overlay the distributions with a possible range of values for the minimum-mass solar nebula. The lower boundary was chosen according to the lowest estimates for the core masses of the giant planets, at \SI{66}{\mearth} while the upper boundary was calculated as \SI{101}{\mearth}, from the higher estimates, plus \SI{50}{\mearth} needed for the outward planetesimals-driven migration of Neptune \citep{1984IcarusFernandezIp,1993NatureMalhotra,1999AJHahnMalhotra}. The calculation of these values is provided in Table~\ref{table:mmsn}.

\subsection{Initial solids surface density: spatial distribution}
\label{sec:pop-plan-disc-sizes}

\begin{figure}
	\centering
	\includegraphics{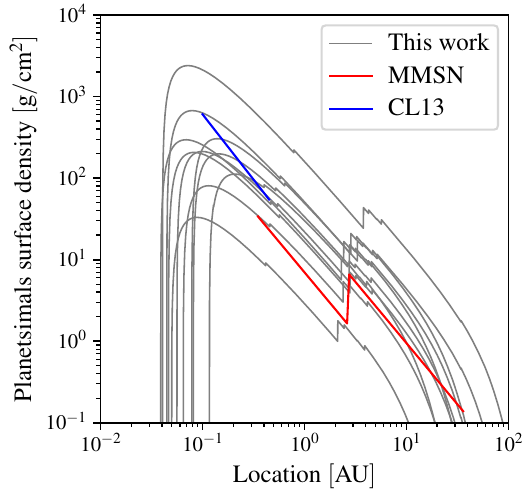}
	\caption{Initial planetesimals surface density profiles for 10 discs, which were selected using the quantiles of the disc mass distribution, to be representative of the entire population. The top and bottom grey lines thus show the most and least massive disc. The red line is the minimum-mass solar nebula (MMSN, \citealp{1977ApSSWeidenschilling,1981PThPSHayashi}), while the blue line is the minimum-mass extrasolar nebula \citep[CL13]{2013MNRASChiangLaughlin}.}
	\label{fig:pla-disc-profiles}
\end{figure}

To account for the inward drift of solids and the effect of planetesimal formation, we select an initial profile of the planetesimal disc that is different to that of the gas disc. The first difference is that the characteristic radius of the planetesimals disc is set to half that of the dust disc that was observed by \citet{2010ApJAndrews}. This follows the planetesimal formation simulations of \citet{2020AAVoelkel}, who found that the planetesimals disc is smaller than that of the dust. In this work, we chose to still keep the relationship of \citet{2010ApJAndrews} to provide the characteristic radius of the gas disc and use a smaller radius for the planetesimals disc. In future work, thanks to the addition of the dust-pebble-planetesimal growth phase in \citet{2020AAVoelkel}, such approximations will no longer be necessary. The second difference is that the power law index is steeper for the solids disc ($\betas=1.5$) than that of the gas disc ($\betag=0.9$) following \citet{2019ApJLenz} and \citet{2020AAVoelkel}.

To show how the difference in the characteristic radius affects the planetesimal surface density distribution in the discs, we provide in Fig.~\ref{fig:pla-disc-profiles} a comparison between the synthetic discs of our populations, the minimum-mass solar nebula \citep[MMSN;][]{1977ApSSWeidenschilling,1981PThPSHayashi} and the minimum-mass extrasolar nebula of \citep{2013MNRASChiangLaughlin}. The ten discs were selected using the quantiles of the planetesimals disc mass distribution so that they are representative of the overall distribution of the discs in our populations. Outside the ice line, the median surface density is larger by a factor of roughly two compared with the MMSN. Due to the larger jump in surface density at the ice line in the MMSN compared to our populations, we find a larger difference inside the ice line. Nevertheless, our profiles are compatible or even smaller than the minimum-mass extrasolar nebula obtained by \citet{2013MNRASChiangLaughlin}. It is derived from close-in planets discovered by the Kepler satellite, assuming as for the MMSN in-situ formation.

Thus, despite the relatively small characteristic radii we selected for our planetesimal discs, which increase the surface density at given total mass, the planetesimals surface density are, on average, only larger than the MMSN by a factor of roughly 2 outside the ice line, and there are also discs with lower surface densities. The region outside of the ice line is a location of great importance for the formation of giant planets, as most of the cores of these planets are formed there. Regarding the regions close to the star, most synthetic discs have surface densities lower than the minimum-mass extrasolar nebula.

\subsection{Comparison with \citet{2020ApJTobinA}}
\label{sect:comptobin}
\begin{figure}
	\centering
	\includegraphics{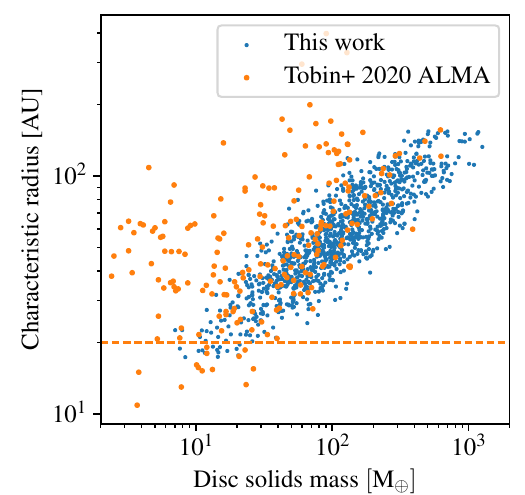}
	\includegraphics{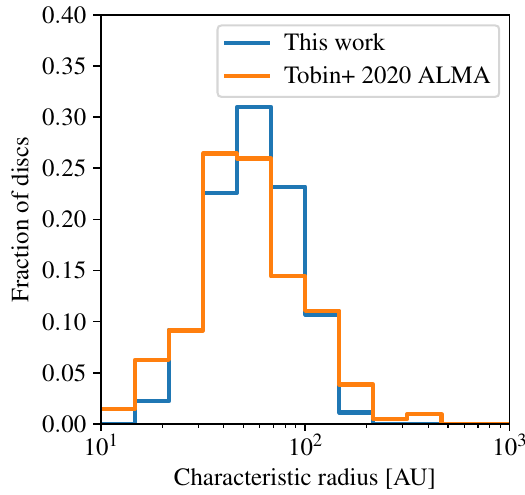}
	\caption{Comparison of characteristic initial gas disc radii versus disc masses (top) and disc radii alone (bottom) between this work (in blue) and the observational results for Class 0/I/flat spectrum dust discs of non-multiple protostars using ALMA  \citep[in orange]{2020ApJTobinA}. The dashed orange line represents half the typical spatial resolution of the survey.}
	\label{fig:comptobin}
\end{figure}

As mentioned already in Sect.~\ref{sec:pop-gas-spatial-dist}, the definition of initial conditions from disc observations is not trivial because of the differences of dust, planetesimal, and gas discs and different ages. The picture is further complicated because dust radii found in simulations versus those observed may substantially differ because observations (also) depend on sufficient opacity to detect matter \citep{2019MNRASRosottiB}.

With the properties of both the synthetic gas and planetesimal disks introduced, we here compare our approach with a more recent observational paper, \citet{2020ApJTobinA}. In their multiwavelength VANDAM survey (ALMA and VLA), they observed several hundred protostellar discs. These younger discs should be more representative of initial conditions in which we are interested here than older Class II discs. The ages of the Class 0 and I/flat spectrum discs should roughly be 100 and \SI{200}{\kilo\year}, respectively \citep{2020ApJTobinA}. At these early times, the chances are higher that evolution has not yet led to significant
differences in the dust and gas radii. To which extent this holds is a function of several parameters like the turbulence level or the strength of external photoevaporation, as indicated by simulation of  dust evolution \citep{2012A&ABirnstiel,2020AAVoelkel}. For the time being, we follow \citet{2020ApJTobinA} and compare our initial gas disk radii with their observed dust radii of Class 0/I/flat spectrum disks of non-multiple protostars. We furthermore make the rough assumption that our initial planetesimal masses are representative of their observed dust masses. This obviously only holds if the  planetesimal formation process is efficient. Some planetesimal formation models do produce such a high efficiency of dust-to-planetesimal conversion if the turbulence level in the discs is low \citep{2019ApJLenz,2020AAVoelkel}, whereas others rather find a \SI{\sim10}{\percent} efficiency \citep{2021MNRASColeman}. In the absence of a description for the early phases of the growth from dust over pebbles to planetesimals in our current model, this is the comparison that can currently be made that at least does not involve additionally also converting dust masses into gas masses, which would add even more uncertainties.

The result is shown in Fig.~\ref{fig:comptobin}. One sees that our disc radii overlap well with that of \citep{2020ApJTobinA}, even though we are using the Class II relation of \citet{2010ApJAndrews}. We also note that our solid mass distribution does not extend to the lowest masses seen in VANDAM. At these low masses, many observed discs have still significant radii of about 40-\SI{50}{\au}, while our relation would predict sizes of 10-20 AU. Here, it might be relevant that the spatial resolution of the survey was about 40 AU, meaning that it could be incomplete at the small sizes. However, we note that there is a discrepancy between our disc masses (which are based on the VLA measurements of Perseus by \citealp{2018ApJSTychoniec}) and those of \citet{2020ApJTobinA} for Orion. \citet{2020ApJTobinA} discuss this difference and suggest that the opacity law used in previous studies needs to be revised. However, they can not rule out an underlying discrepancy between the two regions.

\subsection{Embryos}

The embryos are initialised in the following way: we place a predetermined number of bodies of initial mass $\mstart=\SI{e-2}{\mearth}$ with a uniform probability in the logarithm of the distance between $\rin$ and $\SI{40}{\au}$. This spacing was selected to reproduce the outcomes of \textit{N}-body studies of runaway and oligarchic growth where embryos have a constant spacing in terms of Hill radius \citep{1998IcarusKokubo}. We further enforce that no pair of embryos can lie within 10 Hill radii of each other, which is the usual spacing at the end of runaway growth \citep{1998IcarusKokubo,2006IcarusChambers,2010IcarusKobayashi,2019IcarusWalshLevison}. We thus begin with planetesimals plus embryos, as other studies by, for instance, \citet{2006IcarusOBrien} or \citet{2009RaymondIcarus}, although the planetesimals in our case are treated in a fluid-like description (surface density with a dynamic state).

The starting mass was selected such that 1) it is somewhat larger to where embryos start to repulse each other giving the 10 Hill radii separation \citep{2000IcarusKokuboA}, 2) the mass is below the threshold where gravitational interactions start to play a role, as we found in \paperone, and 3) the mass is also below the onset of envelope effects (such as the increase of the planetesimals capture radius, \paperone). The selected starting mass is, however, larger than the transition from planetesimal runaway to oligarchic growth from \citet{2010ApJOrmel}, which for our planetesimals size is usually between \num{e-4} and \SI{e-3}{\mearth}.

The embryos start right at the beginning of the simulation. This means we assume that they form in a negligible time compared to evolution of the gas disc. This is obviously a strong assumption and will be revised in future generations of the model by addressing the evolution of the solids at early times (drift, planetesimal formation, embryo formation, see \citealp{2020AAVoelkel}). In our populations, we place a maximum of 100 lunar-mass embryos per system. With this number of embryos, the mean separation is roughly 28 Hill radii. This also means that a maximum of 280 lunar-mass embryos per systems could be placed while enforcing a minimum separation of 10 Hill radii.

\subsection{Other parameters}

The formation  model has several other parameters that are kept constant throughout this work (Table \ref{tab:params}). The grain opacity reduction factor in protoplanetary atmospheres, which is important for the efficiency of gas accretion in the attached phase, was set to $\fopa=0.003$. This was selected according to the numerical simulations in \citet{2014AAMordasiniA}, which showed that this reduction factor produces the best agreement with detailed numerical simulations of grain dynamics and resulting opacities \citep{2010IcarusMovshovitz}. This value also leads to the best match of planetary metal content between numerical models and observations \citep{2014AAMordasiniA}. The choice of the planetesimal radius follows \citet{2013A&AFortier}, who found that small planetesimals are required to reproduce the occurrence rate of exoplanets. This is a strong assumption of the model and deviates from studies that found that planetesimals are formed big, for instance \citet{2009IcarusMorbidelli}. For expanded discussions of the comparison of this values with constraints of the Solar system, the reader is referred to \paperone{} and \citet{NGPPS3}. The density of the planetesimals inside the ice line is set to be \SI{3.2}{\gram\per\cubic\centi\meter}, similar to values used in \citet{1993AJHillsGoda} or \citet{1988IcarusPodolak}, while the density outside the ice line is taken to be \SI{1}{\gram\per\cubic\centi\meter} following \citet{1988IcarusPodolak}.

Overall, the most uncertain parameters of our population synthesis are the planetesimals radius and the opacity reduction factor $\fopa$. These are the  least constrained by observations, and were selected according to previous theoretical studies and population syntheses. Underlying theoretical model concepts that likely also come with large uncertainties are the general description of the gas disc as a constant-$\alpha$ viscous accretion disc, neglecting recent result on wind-driven accretion \citep{2014PPVITurner}, or the description of gas-driven orbital migration, a process that is still not fully understood \citep[e.g.][]{2016SSRvBaruteau}. The treatment of the early phases of the evolution of the solids is, as mentioned, also a model aspect that will be improved in future work.

\subsection{Results}\label{subsect:resultspops}

In this work, we perform five population syntheses, that differ only by the initial number of planetary embryos per system: 100 (NG76), 50 (NG75), 20 (NG74), 10 (NG84), and 1 (NG73). Here, per system also means per star and per disc. We will use the terms interchangeably in the following discussion. The names in parentheses refer to populations identifiers on the online archive DACE\footnote{\url{https://dace.unige.ch/populationSearch/}}.

For the populations with multiple embryos per system, we model $\nsystot=\num{1000}$ systems, whereas the single embryo population includes $\nsystot=\num{30000}$ systems to compensate the overall lower number of embryos. In the remainder of this section, we will discuss results at the population level without taking into account how planets are distributed in the systems. System-level statistics will be discussed in Sect.~\ref{sec:types}.

In addition, we compute two non-nominal populations with initially 100 embryos per disc but varied parameters (one with a different power-law index of the planetesimal disc and one without gas-driven migration) that we discuss in Appendix~\ref{sec:modparams}. We use them to assess to what extent the relative results obtained with the nominal populations (like, for instance, the relative occurrence rates of super Earths versus giant planets) and general emerging trends (like correlations with stellar metallicity), are robust when changing model parameters and assumptions. The main results of this analysis are that a steeper power law index (which result in an increased concentration of mass near the star) results in more super Earths being ice-poor. We think that such relative trends should be less affected by the specific chosen model parameters than absolute results like the (absolute) occurrence rates and multiplicities of certain planet types. However, it seems still important to report them also, as first, this allows to calculate the relative frequencies, and second, when keeping the caveat in mind that these are results for given parameters only, they can still be directly confronted with observations. We also find that gas-driven migration affects the mass distribution and location of the planets. For instance, migration is necessary to bring giant planets close to the observed peak near \num{2}--\SI{3}{\au} \citep{2019ApJFernandes,2021ApJSFulton}, however the giant planets in our nominal model are too close-in comparatively. Further, the inclusion of migration reduces the number of planets whose masses are between that of Neptune and Jupiter, which is in contradiction with certain analysis of radial velocity results \citep[e.g.][]{2021AJBennett}. Both cases contain elements that are consistent with observations, which suggests that migration is weaker than previously thought, as \citet{2018ApJIda} already pointed out.

\section{Mass-distance diagrams and formation tracks}
\label{sec:res-am}

\begin{figure*}
	\centering
	\includegraphics{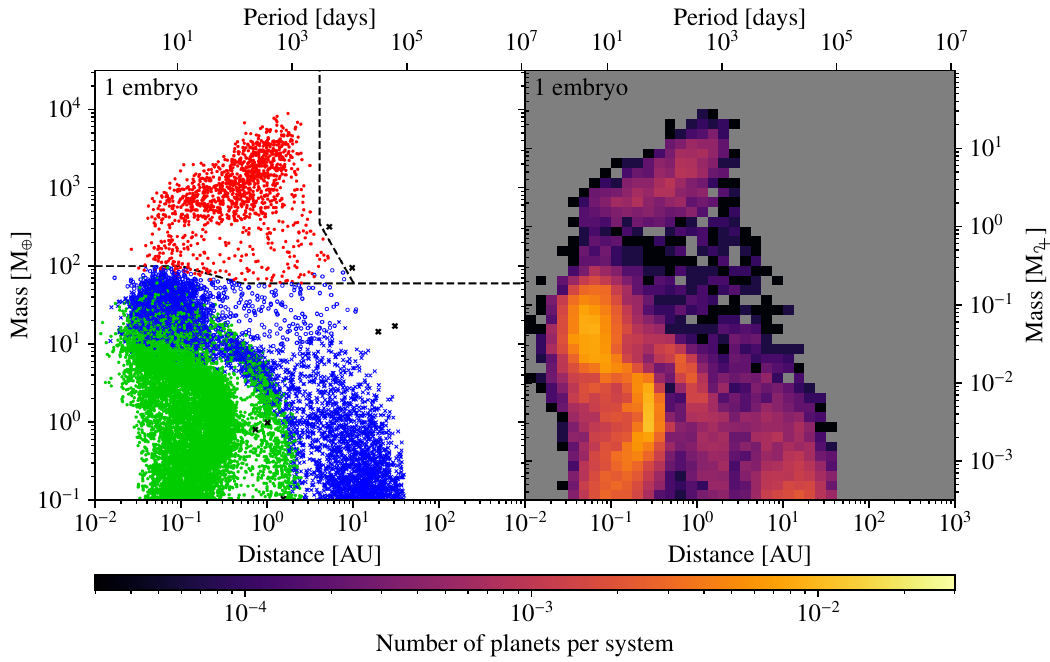}
	\caption{Mass-distance diagram (\textit{left}) and the corresponding histogram (\textit{right}) for the population with a single embryo per system. The colours and shapes of the symbols show the bulk composition: Red points are giant planets with $\menv/\mcore>1$. Blue symbols are planets that have accreted more than \SI{1}{\percent} by mass of volatile material (ices) from beyond the ice line(s). The remainder of the planets are shown by green circles. Open green and blue circles have 0.1$\leq\menv/\mcore\leq1$ while filled green points and blue crosses have $\menv/\mcore\leq0.1$. Black crosses show the Solar system planets. The dashed black line highlights the change of planet regime (from core-dominated, blue, to envelope-dominated, red) at \SI{100}{\mearth} inside \SI{0.1}{\au} to \SI{60}{\mearth} beyond \SI{0.5}{\au}. The vertical dashed line shows the outer limit for giant planets (\SI{4}{\au} above \SI{350}{\mearth} to \SI{10}{\au} at \SI{60}{\au}).}
	\label{fig:1emb}
\end{figure*}

\begin{figure*}
	\centering
	\includegraphics{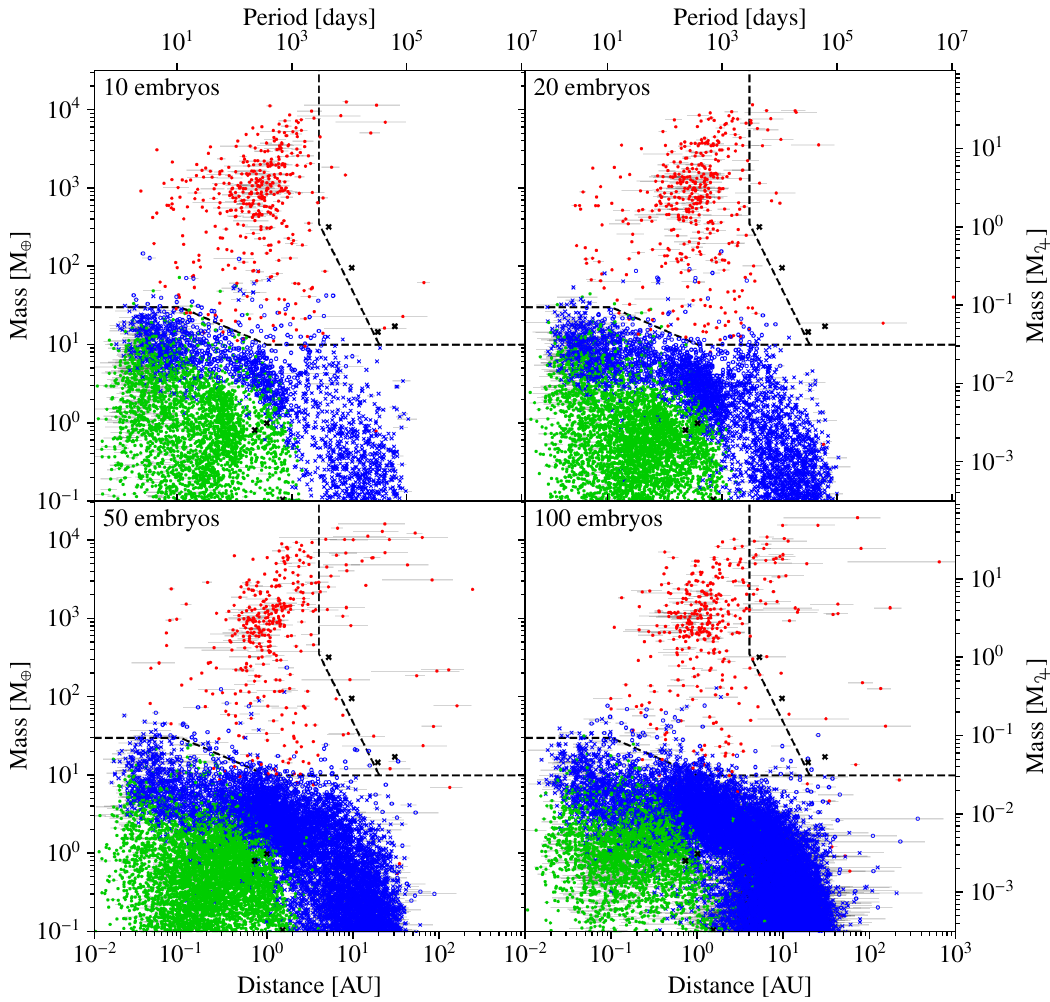}
	\caption{Mass-distance diagrams of the populations with initially 10 (top left), 20 (top right), 50 (bottom left) and 100 (bottom right) \SI{e-2}{\mearth} embryos per disc. The symbols are identical as in the left panel of Fig.~\ref{fig:1emb}. In addition, the grey horizontal bars go from $a-e$ to $a+e$. Dashed black lines show distinct regions in the diagrams, with the change from core-dominated (blue or green) to envelope dominated (red) at \SI{30}{\mearth} inside \SI{0.1}{\au} to \SI{10}{\mearth} outside \SI{10}{\mearth}. The vertical dashed line show the same division for giant planets in Fig.~\ref{fig:1emb}.}
	\label{fig:ame}
\end{figure*}

\begin{figure*}
	\centering
	\includegraphics{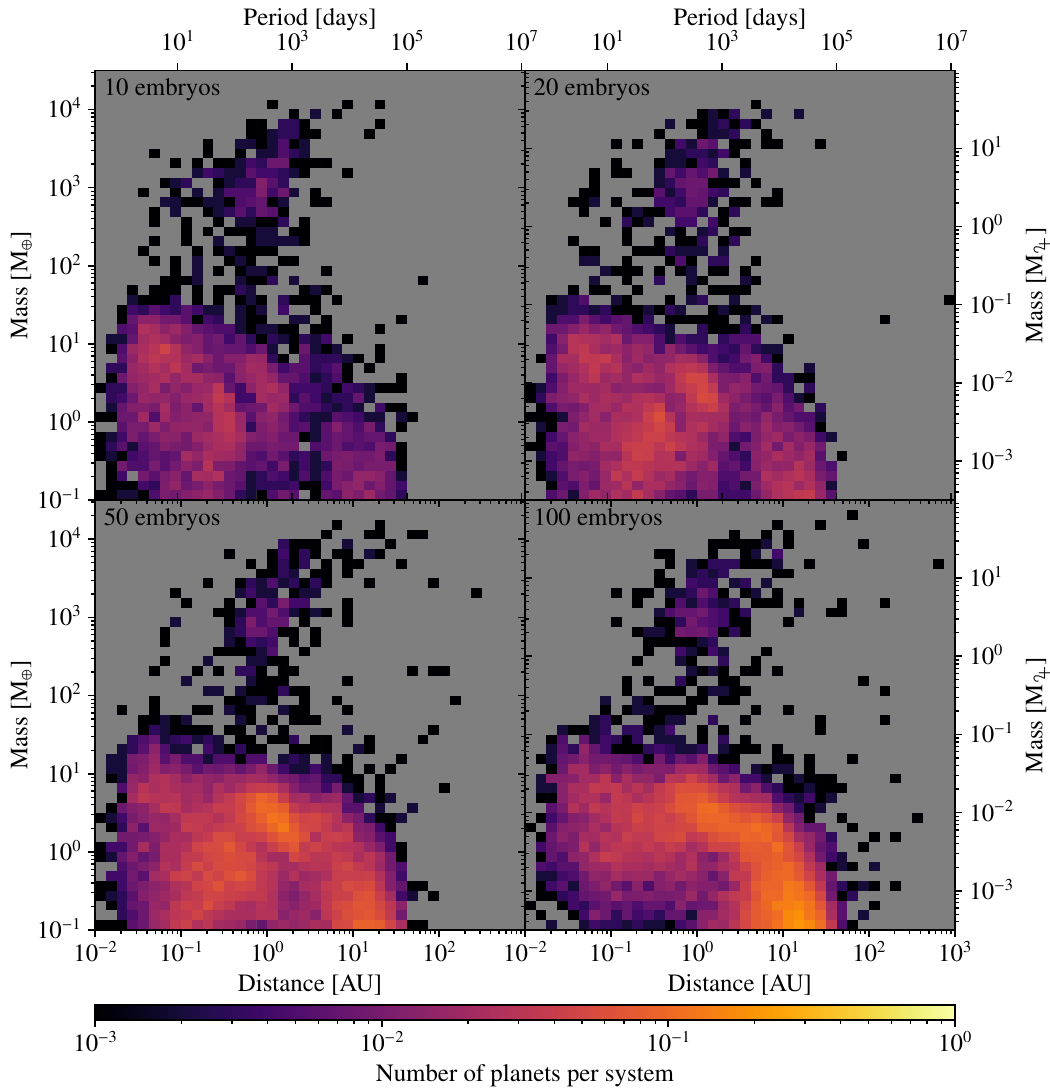}
	\caption{Two dimensional histogram of mass-distance relationship of the populations with initially 10 (top left), 20 (top right), 50 (bottom left) and 100 (bottom right) \SI{e-2}{\mearth} embryos per disc (as in Fig.~\ref{fig:ame}). The number of planets has been normalised by the number of systems. The colour scale is the same in all populations, but different than in Fig.~\ref{fig:1emb}. Gray regions have no planets.}
	\label{fig:amh}
\end{figure*}

\begin{figure*}
	\centering
	\includegraphics{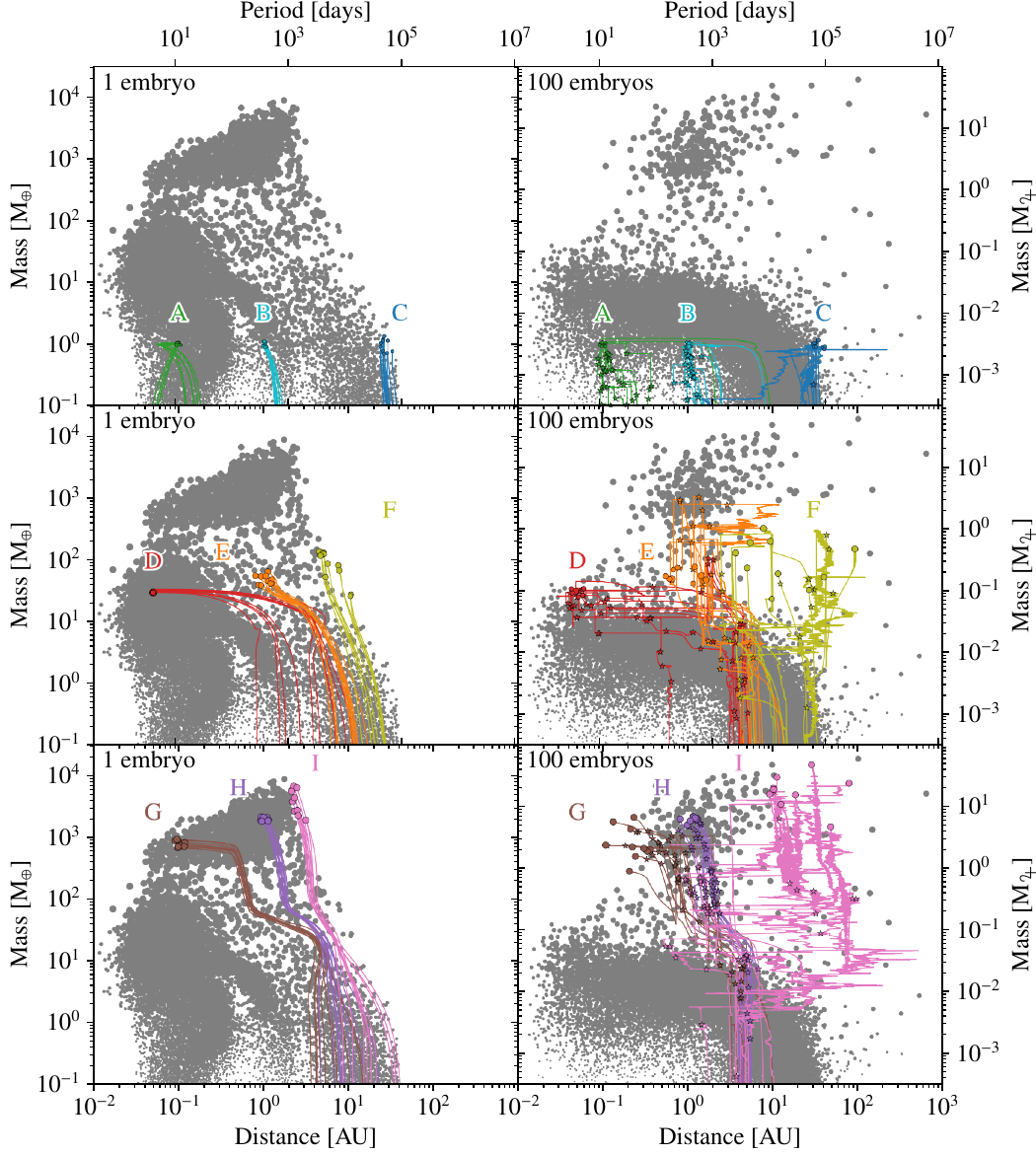}
	\caption{Comparison of the formation tracks between the population with initially 1 (\textit{left}) and 100 (\textit{right}) embryos per system and 9 different group of planets labelled A through I, each shown with a different colour. The positions of the groups in the mass-distance diagram are explained in the text. The stars in the 100-embryos population denote the instant at which the planets were hit by other protoplanets (giant impacts).}
	\label{fig:am_tracks}
\end{figure*}

A key result of synthetic populations is the mass-against-distance diagram of the final planets. It shows what kind and where the formed planets are. This and the corresponding 2D histogram for the single embryo population are provided in Fig.~\ref{fig:1emb}. For the four populations with multiple embryos per system, the diagrams are shown in Fig.~\ref{fig:ame}, and the corresponding histograms in Fig.~\ref{fig:amh}. To generate these snapshots, we used the state at \SI{5}{\giga\year}. For the mass-distance diagrams, the time at which the results are plotted has a limited effect, as long as it is during the evolution stage (after \SI{20}{\mega\year}). Only the close-in planets may be affected, either by tidal migration or photo-evaporation.

To better understand the differences between the populations and how the interactions between embryos affect planetary formation, we also analyse how different types of planets form in two populations. We therefore show in Fig.~\ref{fig:am_tracks} formation tracks in the mass-distance diagram of selected groups of planets, for two populations: the one with a single embryo per system and the one with 100. In that figure, there are nine groups, divided in 3 series. The first series in the top panels shows Earth-mass planets close to the inner edge of the gas disc (group A in green, about \SI{0.1}{\au} and \SI{1}{\mearth}), Earth-like planets (group B in light blue, about \SI{1}{\au} and \SI{1}{\mearth}), and at the end of the region where such planets are found (group C in dark blue, about \SI{40}{\au} and \SI{1}{\mearth}). The middle panels show intermediate-mass planets in the ``planetary desert'' (see below), in the pile-up at the inner edge of the disc (group D in red, about \SI{0.05}{\au} and \SI{30}{\mearth}), at the position of the Earth (group E in orange, about \SI{1}{\au} and \SI{50}{\mearth}), and at large separation (group F in yellow, about \SI{20}{\au} and \SI{100}{\mearth}, with a minimum mass of \SI{50}{\mearth}). The bottom panels show giant planets, with hot-Jupiters (group G in maroon, about \SI{0.1}{\au} and \SI{800}{\mearth}), at the location of the Earth (group H in purple, about \SI{1}{\au} and \SI{2e3}{\mearth}), and distant giants (group I in pink, about \SI{40}{\au} and \SI{5e3}{\mearth}, with a minimum of \SI{2e3}{\mearth}). For selecting the planets that belong in each group, we use the following procedure: we search for the ten closest planets to the given point, the metric being the difference in the logarithm of both quantities (possibly with a second criterion on the minimum mass). This ten planets are highlighted and their formation tracks are superimposed on the overall mass-distance diagram.

In all populations, planets whose masses are between that of Neptune and Jupiter are less common than smaller or larger planets. This results is contrary to results from radial velocity \citep{2021AJBennett} and microlensing \citep[e.g.][]{2016ApJSuzuki} surveys and is an area where the model could be improved in the future. As this range is where planets reach the critical mass to undergo runaway gas accretion. Planet accrete mass rather quickly here, and it is therefore unlikely that the gas disc vanishes during the short period of time planets spend in this mass range. \citet{2004ApJIda1} called this deficit of planets the ``planetary desert''. Another common feature is the gradual inward migration of icy planets (shown in blue symbols on the diagrams) for intermediate masses causing planets with masses higher than \num{3} to \SI{10}{\mearth} to reach the inner edge of the disc. This formation of this morphological feature is similar to the ``horizontal branch'' of planets found first in \citet{2009A&AMordasinia}, as we will see in Sect.~\ref{sec:am-mig-acc}. As the Type~I migration rate is proportional to the planet's mass \citep[e.g.][]{2002ApJTanaka,2014PPVIBaruteau}, more massive planets will tend to end up at locations that are further inward from their original position than lower-mass planets, as long as the planets are not so massive that they migrate in slower Type~II migration regime. An important consequence of this is that the ice content of planets when starting in the left bottom corner increases not only when going outwards to larger orbital separations as it is trivially expected, but also when moving upwards to higher masses.

Coming to the differences between the populations, we see that the single-embryo population stands out compared to the others. Among the major differences we can cite: 1) the presence of a pile up of planets between \num{4} and \SI{100}{\mearth} at the inner edge of the disc (about \num{0.02} to \SI{0.2}{\au}), 2) a different mass for the transition to envelope-dominated planets as visible in the transition from the blue to the red points (\num{60} to \SI{100}{\mearth} in single-embryo population compared to \num{10} to \SI{30}{\mearth} in the multi-embryos case, as shown by the horizontal dashed lines on Figs.~\ref{fig:1emb} and~\ref{fig:ame}), 3) the effect of the convergence zone for Type~I migration (see for instance \citealp{2010ApJLyra,2014A&ADittkrist}; \paperone) which are most visible in the single-embryo population and less as the number of embryos per system increases, and 4) a total lack of distant giant planets in the single embryo population (the upper right region on the left panel of Fig.~\ref{fig:1emb}).

The first two effects are due to the intricate link between accretion and migration that we discuss in Sect.~\ref{sec:am-mig-acc}. The following two effects are extension of the changes we see in the multi-embryos populations. If we look at all of them, we see gradual changes in the imprint of migration, the masses and locations of the giant planets. These will be discussed in Sect.~\ref{sec:am-nemb}. The last effect is due to close encounters resulting in planet-planet scatterings that cannot happen in the presence of only a single protoplanet. In addition, only the 100-embryos population shows one important feature about the inner low-mass planets, namely that inside of \SI{1}{\au}, there are less planets of very low mass (\SI{\sim0.1}{\mearth}) than planets of \SI{\sim1}{\mearth}. In the populations with less than 100 embryos, there are in contrast many embryos inside \SI{1}{\au} that have not grown significantly. We will discuss this in Sect.~\ref{sec:res-form-low}.

\subsection{The interplay between migration and accretion: single versus multiple embryos per disc}
\label{sec:am-mig-acc}

\begin{figure}
	\centering
	\includegraphics{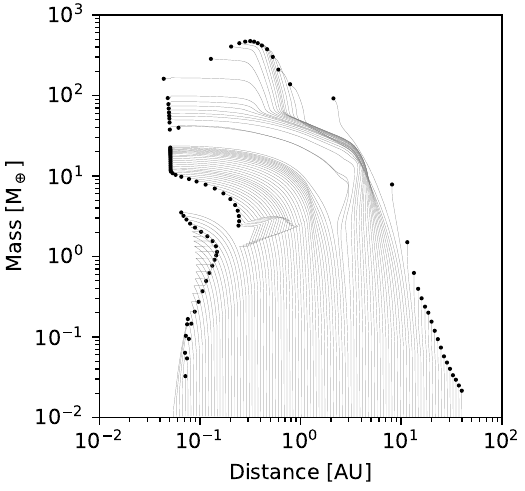}
	\caption{Possible formation tracks for the case on a single embryo per system for one given disc. This figure shows 100 systems with the same initial conditions except for the initial location of the embryo. The initial location was varied from the inner edge of the disc to \SI{40}{\au} with an even spacing in the logarithm of the distance to reflect our choice of initial embryos location in the overall population.}
	\label{fig:track_single_438}
\end{figure}

\begin{figure}
	\centering
	\includegraphics{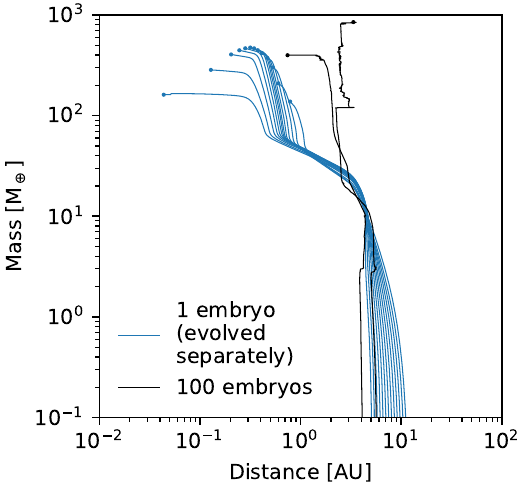}
	\caption{Comparison of the formation tracks between single embryos (in blue) and the corresponding 100-embryos system (in black). Only planets whose masses are larger than \SI{100}{\mearth} are shown.}
	\label{fig:track_438_am}
\end{figure}

The single-embryo population stands out with respect to the multi-embryos in several ways. A major difference is that the single-embryo population completely lacks dynamical interactions. This means that the only possibility for a planet to change its location is through orbital migration, and migration is not hampered by the presence of other embryos.

To compare the formation tracks of planets in the single-embryo population with that of the muti-embryos populations, we chose only particular to study. We came down to system 438, which has a solids disc of \SI{287}{\mearth} and can form planets up to about \SI{2}{\mj} in the single-embryo population, depending on the initial location of the embryo. The possible formations tracks of a single embryo in this system are provided Fig.~\ref{fig:track_single_438}. There, we show a grid of 100 distinct systems whose initial conditions are identical, except for the initial location of the embryo. The initial location was set using a uniform spacing in the logarithm of the distance, from the inner edge to \SI{40}{\au}. This means that each point of the grid represents the same probability in the population.

It can be seen in Fig.~\ref{fig:track_single_438} that many of the intermediate-mass planets, between \num{10} and \SI{100}{\mearth}, end up at the inner edge of the protoplanetary disc. This is due to migration being most efficient in this mass range. The low- and middle-mass planets (up to a few tens of Earth masses, see Fig.~10 of \paperone) will undergo type~I orbital migration, whose rate is proportional the planet's mass \citep[e.g.,][]{2002ApJTanaka,2014PPVIBaruteau}. Thus, the least massive planets (below \SI{1}{\mearth}) will remain close to their original location.

Now, in the single-embryo population, this fast migration will only stop under two conditions: 1) when the planet reaches the inner edge of the disc, or 2) when the planet grows sufficiently to open a gap in the disc and switches to type~II migration, which is significantly slower.

Thus, to avoid being taken to the inner edge, planets must grow rapidly while they are in the \num{10} to \SI{100}{\mearth} range. The planets are still in the planet-limited gas accretion regime at this epoch (that is, the attached phase): gas accretion is limited by the ability of the planet to radiate away the gravitational energy gained by accretion of both solids and gas. Thus, if the planet is still accreting solids, its ability to bind a large amount of gas is severely limited. To be able to undergo runaway gas accretion, the planet must either strongly decrease its solids accretion rate or attain a mass large enough so that cooling (and therefore contraction which allows gas accretion) become efficient, hence being able to accrete gas despite the solids accretion rate remaining large.

In contrast, multi-embryos populations have additional mechanisms to prevent rapid inward migration:
\begin{itemize}
    \item mean-motion resonances with closer-in protoplanets that will slow migration down (because the torque is spread over multiple bodies when planets migrate in resonances, see e.g. \citealp{2002ApJLeePeale,2012ARAAKleyNelson}), and
    \item a decrease of the accretion rate of solids that will enable the protoplanet to undergo runaway gas accretion.
\end{itemize}

The combined effect of these two processes can be been in Fig.~\ref{fig:track_438_am}. The protoplanets migrate more slowly in the multi-embryos populations, which prevents the rapid inward migration and at the same time reduces the planetesimals accretion rate, enabling runaway gas accretion at lower core masses. The resulting giant planets are located further out and have smaller core masses that the planets formed in the single-embryo case.

\subsubsection{Mean-motion resonances as a way to reduce inward migration}

\begin{figure}
	\centering
	\includegraphics{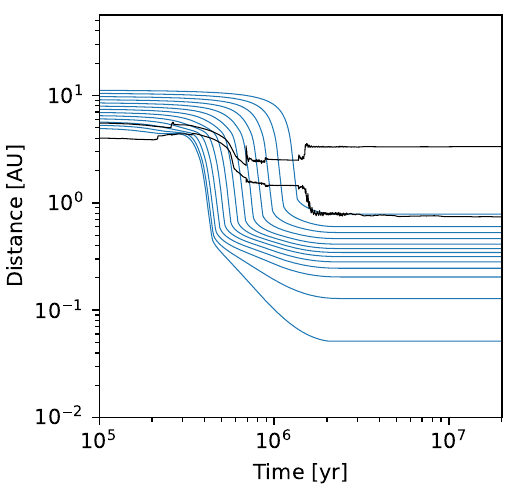}
	\caption{Comparison of the distance as function of time between single embryos (in blue) and the corresponding 100-embryos system (in black). Only planets whose masses are larger than \SI{100}{\mearth} are shown.}
	\label{fig:track_438_ta}
\end{figure}

In the single-embryo population, planets follow precisely the migration prescription, as shown in Fig.~\ref{fig:track_single_438}. For the multi-embryos populations however, dynamical interactions have to be taken into account too. One possible dynamical interactions is the trapping in mean-motion resonances (MMRs). What we found is that the trapping in MMRs can significantly reduce the inward migration of intermediate-mass planets; we show an example of this in Fig.~\ref{fig:track_438_ta}. That figure shows the same planets as in Fig.~\ref{fig:track_438_am}, but instead provides the time evolution of the planet's distances.

Figure~\ref{fig:track_single_438} shows regions of outward migration for planet masses between \num{1} and \SI{20}{\mearth}. These are caused either by opacity transition near the iceline \citep{2010ApJLyra} or structure in the gas disc \citep{2012ApJKretke}. A migration map depicting these feature is shown in Fig.~10 of \paperone.

In Fig.~\ref{fig:track_438_ta}, it can be seen that the two giant planets formed in the 100-embryos system (in black) migrate much more slowly that planets that are alone in their system (in blue). A sketch of how the effect happens is the following:
\begin{itemize}
    \item embryo growth happens inside out, as accretion time scale is shorther in the inner region of the disc,
    \item more distant embryos become larger than closer-in ones as the isolation mass increases with distance, and
    \item once embryos start to feel the torque from the disc that leads to migration, they will be trapped into MMRs with closer-in, lower-mass planets, which will result in a reduced migration speed for the largest embryos.
\end{itemize}

As closer-in, lower-mass planets will have a lower intrinsic torque, they would migrate more slowly that outer, more massive planets. Thus, to allow the outer planets to migrate, the latter have to transfer torque to the inner, low-mass planets through MMRs. This will lead to a reduced migration speed because the torque has to be spread across multiple planets.

A consequence is that planets in the \num{20}--\SI{50}{\mearth} will have more time available before ending in the inner region of the disc. Another consequence is that planets can be pushed out of the convergence zones of orbital migration. For instance, we observe in the single-embryo population, three different zones with a lack of planets in between. The first two are for rocky planets, while the latter contains mostly icy planets. In the 10, 20, and 50-embryos we still see some imprint of the convergence zones, each time with a decreased intensity. In the 100-embryos population, the effects of the convergence zones have nearly vanished.

\subsubsection{A reduction of the solids accretion rate}

\begin{figure}
	\centering
	\includegraphics{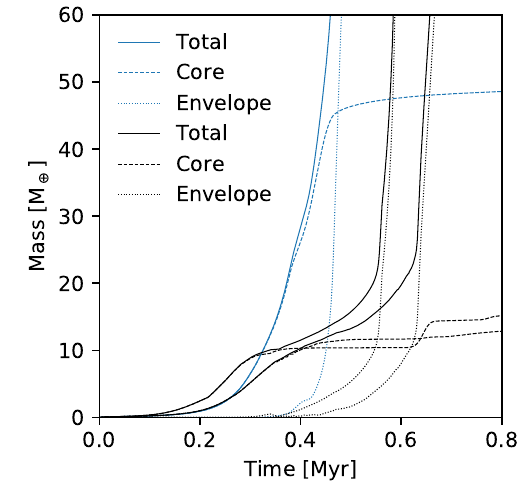}
	\caption{Comparison of the mass (core, envelope, and total) as function of time between one single embryos (in blue) and the corresponding 100-embryos system (in black, with only planets whose masses are larger than \SI{100}{\mearth} are shown). The scales are linear to compare to \citet{1996IcarusPollack}.}
	\label{fig:track_438_tm}
\end{figure}

With a single embryo per system, there can be no reduction of the accretion rate of solids once the planet starts to migrate. This is because the planet will always find new material to accrete from as it enters regions full of pristine planetesimals. It will most likely end only when the planet comes to the inner edge of the disc. The thermal support of the envelope because of strong continuous planetesimal accretion is sufficient to prevent runaway gas accretion except for the most massive cores. Hence, giant planets in that population always have a massive core, because it is the only way for them to undergo runaway gas accretion quickly. This effect also requires that a large amount of solids is present where the planet forms, so that it can accrete a very massive core without migrating too much.

On the other hand, when multiple embryos are present, the competition for solids provides a different pathway for giant planets to form. In this scenario, the initial part of the accretion of the core, until planets start to migrate, remains similar as in the single-embryo case. However, once the core experiences inward type~I migration, it will enter at some point a region where another embryo has grown and depleted the planetesimals. This will deprive the first core of material to accrete from and cause a sudden decrease in the accretion rate of solids. As consequence, there will be a drop in the luminosity released by the accretion of solids, which opens the pathway to trigger runaway gas accretion at lower masses.

This difference is able to explain the first two items we mentioned above, namely the pile-up of massive close-in planets at the inner edge in the single-embryo population and the difference for the transition to envelope-dominated planets (\SI{\sim100}{\mearth} versus \num{10} to \SI{30}{\mearth}). Also, the more embryos there are, the less migration is needed to enter the region where another embryo has already accreted, as the embryos are more tightly packed. This results in a lower extent of Type~I migration in the many-embryos populations, as the planets will undergo gas runaway more rapidly and switch to the slower Type~II migration.

This effect also means that the multi-embryos populations have a way to limit the accretion of planetesimals as it would occur if the embryos ``shepherd'' the planetesimals while they migrate \citep[e.g.,][]{1999IcarusTanakaIda}. Thus, the single-embryo populations does not represent the true situation with the efficient accretion of planetesimals during planetary migration.

To see such effects, it is necessary to calculate the interior structure to get the accretion rate dependent on the core accretion rate and the corresponding luminosity. With a model where the envelope mass depends only on the core mass, such an effect cannot be reproduced. It should also be noted that collisions with other embryos are included in our model. An additional contribution to the luminosity by collisions is included in the internal structure calculation (\paperone), and it does not provide a continuous luminosity source. They do not hinder gas accretion on the long term as does a relatively continuous accretion of planetesimals \citep{2012A&ABroeg}.

\subsection{Other effects of the number of embryos}
\label{sec:am-nemb}

There other, gradual changes that arise as the number of embryos increase. These include the distant giants and planets with masses below \SI{2}{\mearth} and distance below \SI{0.02}{\au} (i.e., inside the inner edge of the disc). These effects are mainly due to gravitational interactions between the protoplanets.

MMRs can push planets inside the inner edge of the disc by the inward migration of another planet which is still located within the disc, hence we find planets closer to the star than the inner edge of the disc in the corresponding populations.

We set the limit for the transition between ice-poor (rocky) and ice-rich planets at \SI{1}{\percent} of volatiles by mass in the core. This is to avoid having planets with extremely low amount of volatile appearing as icy in the diagram. The limit was set according to the amount of water (the main component of volatiles) to obtain high-pressure ice at the bottom of oceans of a \SI{1}{\mearth} planet \citep{2014AAAlibert}. There is nothing particular happening in the model at this limit, it is only set for visualisation. Comparing the different population, we find that in the 50 and 100-embryos populations, ice-rich planets are found in regions populated only with ice-free planets in the 10-embryos population. This can be seen at the position of the Earth on Fig.~\ref{fig:ame}. In the 10-embryos population, the Earth lies in a region harbouring only ice-free planets. In the 20 and 50-embryos populations, the Earth lies at the transition between the two, while in the 100-embryos population, it is in the ice-rich region. Further, dynamical interactions are able to send icy low-mass planets in the inner region of the disc (inside \SI{1}{\au})

As the number of embryos increases, we observe a greater mixing of the rocky and icy planets. In the single embryo population, the two are well separated, while as the number of embryos grows, we note more and more icy planets of a few Earth masses in the inner part of the disc. This affects only planets of more than a few Earth masses, or regions directly inside of the ice line; for instance, we do not obtain icy planets less than an Earth mass within fractions of an AU. As the single embryo population shows, bringing icy low-mass bodies to the inner part of systems is not possible by migration only, so there must be multi-body effects, such as close encounter and capture in resonance, that send part of the icy planets forming outside of the ice line in the inner part of the disc.

At large orbital distances, the populations with multiple embryos per systems contain planets located outside the outer limit of embryo starting locations (\SI{40}{\au}) while the population with a single embryo does not. The only possibility for planets to end at those position are scattering events due to close encounters with other planets, as outward migration does not happen at these locations. The black horizontal bars of these planets on Fig.~\ref{fig:ame} show their eccentricities. We see all of these planets have a periapsis inside \SI{40}{\au}, indicating that these planets come in a location where other planets are present at some point during their orbit. We then might find planets formed by core accretion at large separation, but with our model these planet remain on eccentric orbit, as circularisation does not happen on a sufficiently short time scale before the dispersal of the gas disc. We could thus explain directly-imaged planets at large separation, such as HIP 65426b \citep{2017A&AChauvin} only if they have a significant eccentricity to have a periapsis at a distance where core accretion is efficient, i.e., inside of \SI{\sim10}{\au}. This formation scenario was studied extensively in \citet{2019AAMarleau}.

\subsection{Formation of low-mass planets}
\label{sec:res-form-low}

The formation tracks of the low-mass planets in the single-embryo case (top left panel of Fig.~\ref{fig:am_tracks}) is straightforward. As we already mentioned in \paperone{} and Sect.~\ref{sec:res-am}, gas-driven migration is weak for these planets, so that they end close to where they started, with minimal inward migration. We can still note that the close-in group (A, in green), there is either outward migration all the way through, or inward migration followed by an episode of outward migration without accretion. This effect is caused by the presence of the innermost outward migration zone for low-mass planets (see Fig.~7 of \paperone). We are in the presence of two scenarios that depend of the disc characteristics: either planets are inside the outward migration zone from the beginning and they will move out while they accrete, or they are in the inward migration zone at the beginning and pass in the outward zone later on during the disc's evolution. In the latter case, there is no accretion during the second pass in a region because all planetesimals were previously accreted.

In the 100-embryos population, the formation of the same resulting planets are more varied. For the two innermost groups (A and B), we divide the planets in two groups. First, for 16 planets, there is growth by giant impacts that we had anticipated and discussed in \paperone{}. These planets have starting distances of \num{0.1} to \SI{0.3}{\au}. The second (4 planets) is growth by accretion of planetesimals at a much larger distance (starting distances of \num{2} to \SI{10}{\au}) followed by a strong inward migration combined with limited accretion. This pathway is unseen in the single-embryos population because the inward migration is caused by the trapping in resonance chains with other more massive planets (around \SI{10}{\mearth}) that experience stronger migration. Clearly, these different formation pathways could result in diversity in terms of the composition of the planets, as we will discuss below. For the outermost group, we see that there also two formation pathways, but they are not the same as for the inner groups. The first pathway is the same as in the single-embryo population, where the only effect is limited inward migration. The second is growth stirred by more massive planets, which causes jitter in their location and occasional scattering events. However, we see that these planets undergo much less giant impacts that their close-in counterparts. This implies that these planets growth mostly from the accretion of planetesimals, in a similar way than planets in the single-embryo case.

There is also a specific feature absent in the other population, for planets within \SI{1}{\au}: a decrease of the occurrence rate with decreasing mass at masses less than \SI{3}{\mearth}, with a near total absence of bodies of the mass of Mars (\SI{\sim0.1}{\mearth}). This feature relates to the formation of the terrestrial planets we discussed in \paperone. It applies to systems with a low metallicity, where migration is unimportant because growth is slow. Only in the 100-embryos population, the inner region of the disc is fully depleted in planetesimals and the embryos end their growth with a ``giant impact'' stage, similar to the terrestrial planets in the solar system \citep{1985ScienceWetherill,2002ApJKokubo}. In the other populations, the spacing between the embryos is too large and they end up growing as if they were isolated. This means that they grow only to masses that are much less compared to the case that all solids in the inner disc end up in planets as in the 100-embryos population, instead of remaining in planetesimals.

The 100-embryos populations should hence be representative of the formation of planets spanning the entire mass range, at least with \SI{1}{\au}. This will enable use to compare architectures of low-mass (i.e. terrestrial) systems with observations to determine if planet pairs have similar masses \citep{2017ApJMillholland}, radii and spacing \citep{2018AJWeiss}; see \citet{NGPPS5}. Accretion through giant impacts is stochastic in nature, and planets may well have collided with bodies originate from beyond the ice lines, or that have themselves had giant impacts with such distant embryos. This explains why we obtain close-in planets that have some content of icy material. For the second pathway (forced migration), we see that those planets have accreted most of their mass before experience the strong migration. We can thus expect that they harbour a large amount of icy material.

There are two caveats with our model for the formation of terrestrial planets. Firstly, we set the limit for the formation stage to \SI{20}{\mega\year}, where planets can accrete and dynamically interact via \textit{N}-body integration. As the accretion time scale increases with distance, this limitation affects more the outer part of the system. In the case of a system with an initial mass comparable to MMSN, we found in \paperone{} that by about \SI{20}{\mega\year} the instability phase had mostly finished inside of \SI{1}{\au}. This is comparable to what \citet{2001IcarusChambers} found is required for Earth-like planets to accrete half of their mass for a similar scenario. Growth is more rapid in systems with a higher solid content though. It follows that for a disc with MMSN content of solids, the low-mass planets obtained in our population are mostly at the end of their formation in the inner region, roughly a factor two too small around \SI{1}{\au}, and that much longer times are required for more distant planets. The limited time for the \textit{N}-body interactions (which in our model occur only during the formation) can also mean that we miss dynamical instabilities at late times. For instance, \citet{2021AAIzidoro} found that it can take up to \SI{100}{\mega\year} for systems to go unstable.

The second caveat is that terrestrial planets accrete predominantly from other embryos, rather than planetesimals as it the case for the giant planets. However, giant impacts, due to similar size of the involved bodies have a variety of possible outcomes \citep{2010ChEGAsphaug}. Accretion (or merging) is not the most common result of such a collision \citep{2012ApJStewart,2013IcarusChambers,2016ApJQuintana}. Despite this, our collision model is unconditional merging when it comes to terrestrial planets (\paperone). For instance, we see that we are unable to form equivalents of the smaller terrestrial planets of the Solar System, Mercury and Mars as a majority of terrestrial-planet forming systems give planet masses similar to $\SI{1}{\mearth}$. Mercury might be the result of a series of erosive collisions \citep{2007SSRvBenz,2014NatGeoAsphaug,2018ApJChau} that are not part of our model.
Using a more realistic collision model, \citet{2013IcarusChambers} found that the resulting planets have slightly lower mass and eccentricities, and that the overall time required to form these planets increases, as collisional accretion is not as efficient.

\subsection{Formation of intermediate-mass planets}
\label{sec:res-form-mid}

The formation of the intermediate-mass bodies in the single-embryo population (left middle panel of Fig.~\ref{fig:am_tracks}) has already been largely discussed in Sect.~\ref{sec:am-mig-acc} for the innermost group (D) and in Sect.~8 of \paperone{} for the more distant groups (E and F). They are in the range where migration is most efficient, massive enough to undergo strong migration, but not massive enough to open a gap and migrate in the slow Type~II regime. The inner group (D, in red) is similar to the ``horizontal branch'' of \citet{2009A&AMordasinia}.

In the 100-embryos population, some of the planets in the innermost group (D) form in a similar fashion as in the single-embryos case, with the exception of some giants impact near the end of their migration. However, a few of these planets had their envelope removed in the aftermath of giant impacts that occurred at about \SI{2}{\au}. Were it not for the impacts, these planets would have most certainly ended up being giants at larger separation.

The more distant groups (E and F) have, however, a different formation history. Half of the planets in the mid-distance group had at some point a mass larger than \SI{100}{\mearth} while their final mass is close to \SI{50}{\mearth}. The mass loss is due to the removal of the envelope following a giant impact, due to the burst of luminosity caused by the sudden accretion of the impactor. The mass loss is delayed by about \SI{3e4}{\year} from the time of impact due to the timescale of release of the supplementary luminosity following \citet{2012A&ABroeg}. It is then possible for the giant impact to not be marked exactly at the location of the mass loss in the formation tracks. We also see that some of these planets spent time around \SI{10}{\au} after beginning the runaway gas accretion closer to \SI{1}{\au}. The formation tracks of these planets also show sudden change in their position, both outward and inward. Thus, not only migration is responsible for their final locations, but also close encounters. Actually, migration plays a lesser role in the 100-embryos population, as most of these planets began inside \SI{8}{\au}, while in the single-embryo population most of the embryos come from outside \SI{10}{\au}.

It is obvious that the effects induced by the concurrent formation of several planets introduces strong additional diversity in the formation pathways of planets. In the single-embryo case, planets may undergo gas runaway only once, and this must be late in the evolution of the gas disc to not accrete too much gas. With multiple embryos, the possibility of giant impacts means that it is possible to undergo gas runaway multiple times, provided the envelope is removed in between.

\subsection{Formation of giant planets}
\label{sec:res-form-giant}

The formation of giant planets in the single-embryos population (left bottom panel of Fig.~\ref{fig:am_tracks}) has also been discussed in Sect.~8 of \paperone{}. They follow a similar pattern than for the intermediate masses at the beginning, but accretion dominates over migration, as indicated by the different slopes of the tracks in the mass-distance plane. Starting with roughly \SI{10}{\mearth}, the formation tracks of the different group begin to be well separated from each other. Then, the planets that finish closer-in will undergo larger migration up to about \SI{50}{\mearth} before undergoing runaway gas accretion. Migration takes over again later and, during later times, the planet migrate with only little accretion, as indicated by the different slopes of the tracks in the mass-distance plane. For the more distant planets (groups H and I), a similar structure of the formation tracks is observed, but migration remains overall less efficient. This is because accretion is very fast, as they must undergo runaway gas accretion in the early times so that they can accrete gas during most of the life time of the protoplanetary disc. As accretion is so fast, it leaves only little time for migration to act.

For the 100-embryos population, we first note that gas-driven migration is overall less efficient than in the single-embryo population. For instance, there are no giants inside \SI{0.1}{\au}, so planets in the innermost group (G) are on average further away than in the single-embryo case. These planet undergo gas runaway at little bit further out than in the single-embryo case: the former are slightly outside \SI{1}{\au}, while the latter are inside of that mark. The difference comes later, with the planets in the 100-embryo population not migrating as much afterwards. The initial location of the seeds forming these planets is also slightly different: in the 100-embryos population they are concentrated between 3 and \SI{6}{\au} (and one case at about \SI{10}{\au}) while in the single-embryo close-in giant form from seeds between 3 and \SI{15}{\au}, only half of them being inside of \SI{6}{\au}. For the intermediate-distance planets (group H), in the case of the 100-embryos population, we see many giant impacts occurs. However, since these are giant planets, most of them did not loose their envelope. The few that did loose it have accreted a secondary envelope, and since they had already migrated inward, ended closer-in that we would have if they had not lost the envelope. The final part of their formation track is otherwise very similar to the single-embryo population. The initial part is different, with the same observation as the for the close-in giants: the seeds come from closer-in. Actually, the seeds for the close-in and intermediate-distance giants are from the same region of the disc in the 100-embryos population (between 3 and \SI{6}{\au}). In the single-embryo population however, these are centred at about \SI{8}{\au}.

For the giant planets ending up at \SI{0.1}{\au} and \SI{1}{\au}, we see by the numerous stars along the brown and violet tracks that once the giant planets have triggered runaway gas accretion and masses \SI{>100}{\mearth}, they are hit by numerous planets. The mass of the impactors are mostly \SI{5}{\mearth} and lower. These impactors also mostly come from nearby the giant planet they collide with. The collisions are due to the destabilisation of other less massive planets by the forming giant because of its strong mass growth coupled with migration. The effect of the impacts on the luminosity both during the formation and evolution phases will be investigated in a future publication of this series.

Distant giant planets in the 100-embryos population (group I) grow even less smoothly than the other groups. Interestingly enough, they come again from the same region of the disc as the two other groups in the 100-embryos population. They are then scattered outward by other massive planets in the same system. The fact that there are other massive planets in the same system is recognisable by the jitter in their distance. The large diversity of formation tracks of planets with very similar final mass and orbital distance in the 100-embryos population illustrates how difficult it is to infer the formation tracks of a specific planet just from the final position of an individual planet in the mass-distance diagram.

\subsection{Formation time}
\label{sec:res-time}

\begin{figure*}
	\centering
	\includegraphics{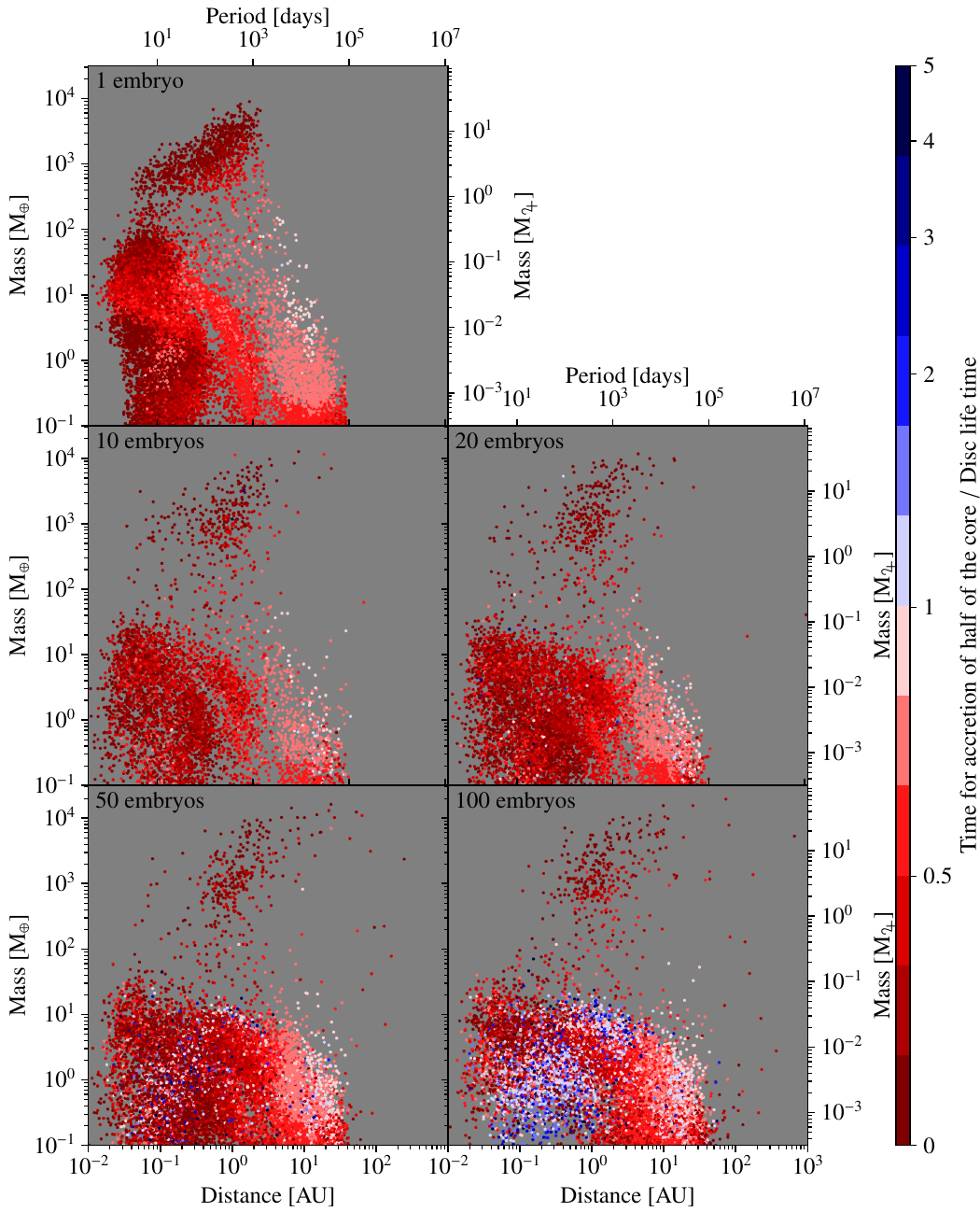}
	\caption{Two dimensional histogram of mass-distance relationship of the five populations shown in this study, with the initial number of embryos per system given in each panel}. The dot colours denote the time needed for each planet to accrete half of their final core mass, given in terms of the disc life times. The scale is linear from 0 to 1, and then logarithmic. The ratio can be larger than unity because we model formation to \SI{20}{\mega\year} while the median disc life time is about \SI{4}{\mega\year}.
	\label{fig:am_hcm}
\end{figure*}

To further characterise the differences between the populations, we seek to determine the time it takes for the planets to form. For this, we plot in Fig.~\ref{fig:am_hcm} the time at which the core's mass is half of its final value, itself given in terms of the life time of the protoplanetary disc. We show only the two end members of our populations: the single-embryo one of the left, and the 100-embryos population on the right.

In all populations, we note that the core of the giant planets formed early. In the case of the single-embryo population it is not particularly different from other types of planets (principally the close-in ones), but in the 100-embryos population it does stand out compared to the rest of the population, where most of the other planets form much later. Further, the single-embryo planet shows a consistent trend for the formation time, with the most massive giant planets forming their core earliest, while the less massive ones, and the planets in the desert forming late. This trend is much less perceptible in the 100-embryos population. Some of the planets in the desert formed quite late and show a formation history similar to what is obtained in the single-embryo population, that is, a smooth growth and slight inward migration. Others however had their core formed early, in the same way as for the most massive giant planets. However the planet-planet interactions enable other pathways for the formation of these planets. For instance, some have had lost partly or entirely their envelope following a giant impact, leaving the planets with a limited time to accrete gas again. Others have been trapped in mean-motion resonances with massive planets still in the disc. These resonances will prevent the fast inward migration, thus enabling a formation process that is somewhat similar to one of giant planets without migration \citep[e.g.,][]{1996IcarusPollack}, that is, with a significant delay between the accretion of the bulk of the core and the onset of the runaway gas accretion.

For the intermediate-mass planets, we observe that the ones that are close-in had their core formed quite early. In the single-embryo population, these are the planets that are close to the inner boundary of the disc, while for the 100-embryos population, this concerns some close-in planets in the \num{1} to \SI{30}{\mearth} range. This is related to the discussion in Sect.~\ref{sec:am-mig-acc}, about the planets that formed close to or beyond the ice line and rapidly migrated inward without undergoing runaway gas accretion, similar to the ``horizontal branch'' of \citetalias{2009A&AMordasinia}.

For the inner low-mass planets, (up to \SI{1}{\au} and \SI{30}{\mearth}), we obtain that the time taken for the formation increases along with the number of embryos. This can be seen in the right panel of Fig.~\ref{fig:am_hcm}, where many of these planet only attain half of their final mass after the dispersal of the gas disc. Thus a large part of the accretion process occurs in a gas-free environment. This helps explain for instance the disappearing of the features related to the gas disc, such as the migration traps.

We hence come back to the discussion of Sect.~\ref{sec:res-form-low} about the integration time required to model the formation of these systems. Our limit of $\SI{20}{\mega\year}$ is still too short for planets beyond $\sim\SI{2}{\au}$, as here we have a increasing fraction of planets that still accreted most of their mass while the gas disc was still present. This indicates that there have not been interactions after the dispersal of the latter, hence that a longer integration time could lead to fewer, more massive planets. This process however takes a long time to happen, on the order of \SI{100}{\mega\year} or more, as we have seen with the terrestrial planets. Thus the direct modelling would be a very expensive task computationally in the context of a population synthesis, i.e. for around \num{1000} planetary systems.

One consequence of the late formation of the low-mass planets, which turns out to be more similar to the formation of the solar system's terrestrial planets is the (in)ability of these planets to retains an envelope. Giant impacts that occur after the dispersal of the gas disc may lead to the ejection of the planet's envelope, but it cannot be reaccreted any more. Atmospheric escape is then not the only mean to loose the envelope, and the evaporation valley \citep{2009AALammer,2012ApJLopez,2012MNRASOwenJackson,2014ApJJin,2018ApJJin} is not as clear as in the populations with a lower initial number of embryos per disc.

In future work, we will improve the way how impact stripping of gaseous envelopes is dealt with \citep{2018SSRvSchlichting,2020MNRASDenman}. As described in \paperone, at the moment the impact energy is added into the internal structure calculation. This however neglects the mechanical removal of some gas during the impact via momentum exchange, and also assumes that the entire impact energy is deposited evenly deep in the planet's envelope, at the core-envelope boundary. In reality, only a part of the envelope close to the impact location might be strongly heated.
Both these effects  affect the efficiency of impact stripping. Interestingly enough, in a population synthesis, the emptiness of the valley can be used to observationally constrain the efficiency of impact stripping. The fact that the valley seems to be too populated compared to observations in the 100-embryos model is an indication that the current model for impact stripping in the Bern model is too efficient.

\section{Planet radii and luminosity}
\label{sec:rad}

\subsection{Mass-radius relationship}
\label{sec:mass-rad}

\begin{figure*}
	\centering
	\includegraphics{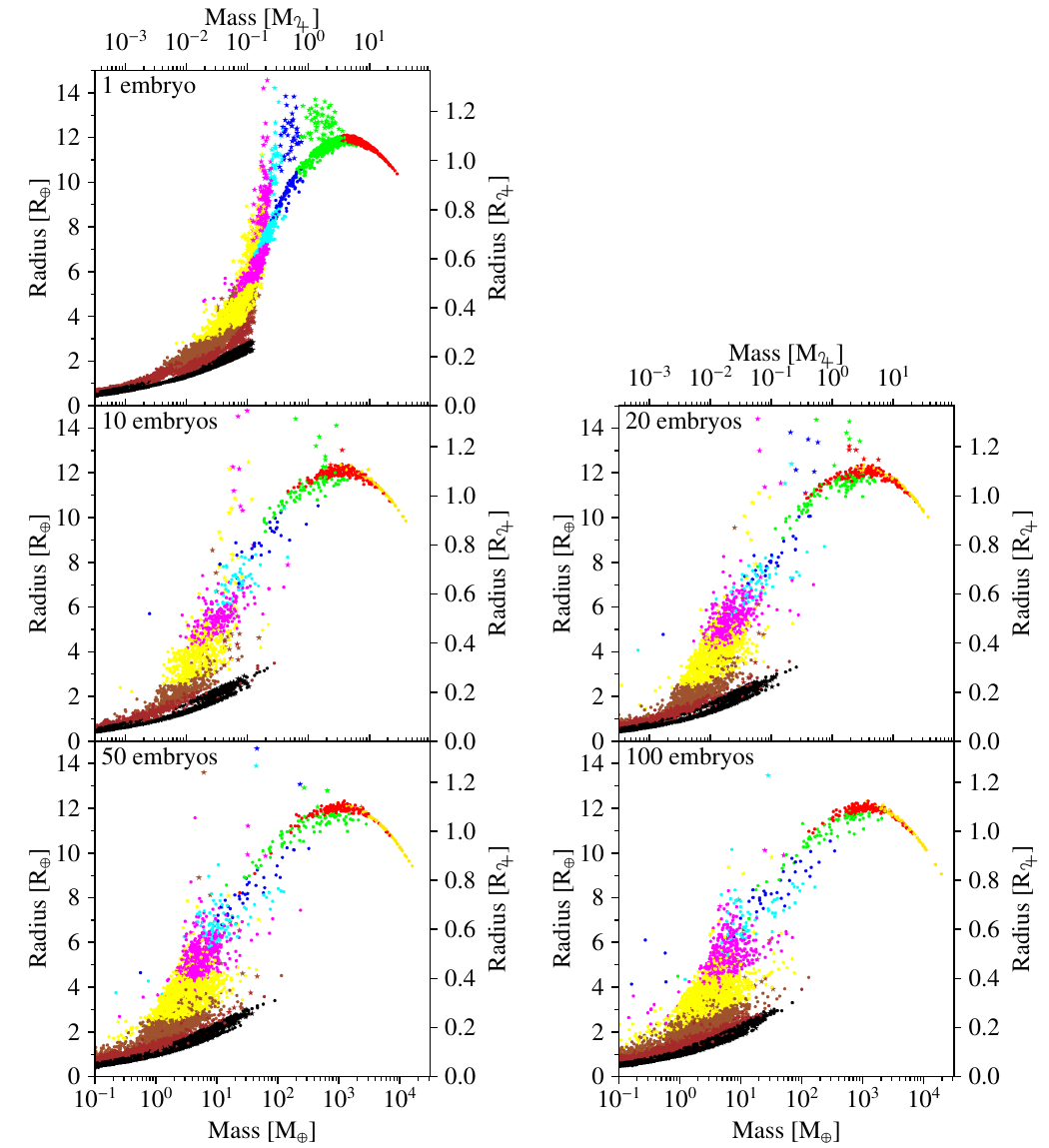}
	\caption{Mass-radius diagram at \SI{5}{\giga\year} obtained from following the long-term thermodynamic evolution of the five populations, as marked in each panel. Atmospheric loss by impacts, atmospheric escape driven by XUV photoevaporation, and bloating are included. The iron and ice mass fraction in the core is considered when calculating the radius. The colours show the bulk solid mass fraction $\mcore/M$, the rest being the H/He envelope: orange: <0.01, red: 0.01-0.05, green: 0.05-0.2, blue:  0.2-0.4, cyan: 0.4-0.6, magenta: 0.6-0.8, yellow: 0.8-0.95, sienna: 0.95-0.99, brown: 0.99-1. Black: no H/He envelope. Star symbols indicate planets who have additional luminosity from bloating, while circles indicate planets that do not.}
	\label{fig:mrad}
\end{figure*}

The mass-radius relation in the context of formation and evolution of planetary systems with 1 embryo per disc was extensively discussed in \citet{2012A&AMordasiniC} and \citet{2014AAMordasiniA}. To compare that scenario with the case of many embryos per disc, we show in Fig.~\ref{fig:mrad} the five population that are part of this study.

As discussed in \citet{2012A&AMordasiniC}, the global structure of the mass-radius relationship is caused by the combined effects of the properties of the equation of state of the main planetary forming materials (iron, silicates, ices, H/He), and the increase of the H/He mass fraction with the planet mass. The overall lower core masses in the population with multiple embryos per disc result in comparatively increased radius and lower metallicity for a given planet mass as the number of embryos increases. The spread of radii for a given mass is due to both different planet metallicities, $\mcore/M$, and distance to the central star. The last effect is more important in our work that in \citet{2012A&AMordasiniC} due to the prescription for the bloating of the close-in planets following \citet{2018AJThorngrenFortney} with a criterion for a minimum stellar flux of \SI{2e8}{\erg\per\second\per\square\centi\meter} from \citet{2011ApJSDemorySeager}. Planets that satisfy this criterion and have their symbol set to a star in Fig.~\ref{fig:mrad}. We note in the single-embryo population, there are two branches in the mass-distance diagram, with the bloated planets having a radius few \SI{0.1}{\rj} larger than their more distant counterpart. This branch does not continue for masses above \SI{e3}{\mearth} because we do not have that massive planets at close-in locations. The 100-embryos population, however, does not show the second branch in the mass-distance diagram for the bloated planets because there are no giant planets at close-in locations, and only few for less massive bodies.

The most bloated planets have a mass of about \SI{60}{\mearth}. Observationally, the most bloated planets have in contrast masses larger than \SI{100}{\mearth}. This reflects that using the empirical bloating model of \citet{2018AJThorngrenFortney} for planets of any mass leads in our model to a M-R relation that differs from the observed one. There could be several reasons for this: the actual physical bloating mechanism has a mass dependency (or dependency on a parameter linked to the mass like the magnetic field strength or the metallicity) not accounted for in the empirical model which was derived mostly on giant planets. The discrepancy could also be due rather to the evaporation model in the sense that atmospheric loss for bloated $\sim\SI{60}{\mearth}$ planets is more efficient than predicted by our evaporation model. This would reduce the radii of these planets. The morphology of the close-in population will be studied further in a dedicated NGPPS publication.

The presence of multiple embryos in a disc lead to more diverse formation tracks, as we have seen in Fig.~\ref{fig:am_tracks}. This is reflected in the bigger spread of radii and envelope mass fraction for a given mass. The spread works in both directions. Planets in the multi-embryos population can have higher envelope contents, for instance, at $M=\SI{10}{\mearth}$, the largest radius is around $\SI{5}{\rearth}$ for the single embryo population, whereas for 100-embryos case, planets can have radii up to $\SI{8}{\rearth}$. Planets in the single-embryo population have a smooth formation and similar tracks for similar final positions and masses. It follows that these planets have similar core masses, which limits the core-mass effect. The 100-embryos, however, has two more effects that can change the core mass fraction in opposite ways, that we will now discuss in more details.

The first effect to alter the core mass fraction in the 100-embryos population is the competition for solids. We have discussed in Sect.~\ref{sec:am-mig-acc} that giant planets in the multiple-embryos populations have lower core masses. This is reflected in the mass-radius diagram by the difference in the core mass fractions. For instance, there are no planets with a core mass fraction of less than \SI{1}{\percent} in the single-embryo population while we frequently obtain this value in the 100-embryo population
for planets above \SI{10}{\mj}. Even if the radius is only weakly dependent on the metallicity for these masses, we see that the maximum radius for the non-bloated planets is slightly larger in the 100-embryos population.

The second effect is giant impacts. Due to their random nature, they add some spread to the core mass fractions at a given mass. Some planets suffered from collisions with other bodies relatively late during their formation. These collisions can lead to the loss of a significant part of the planet's envelope. As consequence, these planets have a lower envelope mass fraction for a given mass, because there is no loss of solids during such an event. Thus a collision does not simply reset the planet back to an earlier time, rather it can induce an increase of the bulk metallicity. We can see examples of such planets at intermediate masses. In the single-embryo population, the minimum envelope mass fraction increases significantly starting with about \SI{40}{\mearth} and no planet more massive than that values remains without envelope. In the 100-embryos however, we have several examples of planets with higher masses that exhibit small radii, including one roughly \SI{70}{\mearth} core without any surrounding envelope. One can also note a few giant planets in the 100-embryos population that have smaller radii than in the single embryo case. They are also caused by giant impacts.

The timing of the collision is important. Early-on events when the gas disc is still important, may even lead to a more massive envelope that there could have been if no collision occurred, because the collision enables the core to cool more efficiently, thus increasing the gas accretion rate \citep{2012A&ABroeg}. Otherwise, the lack of a gas reservoir prevents the re-accretion of an envelope, namely when the collision occurs during the late stage of the gas disc presence in which case the envelope will not grow back to its previous mass.

Collisions are also the reason why there is a more extended range of planet masses without any envelope in the populations with multiple embryos per system. In the single embryo population, where only atmospheric escape acts, there are no planets without an envelope past $\SI{40}{\mearth}$, whereas we do have such cases in the other populations.

\subsection{Distance-radius plot}

\begin{figure*}
	\centering
	\includegraphics[width=\textwidth]{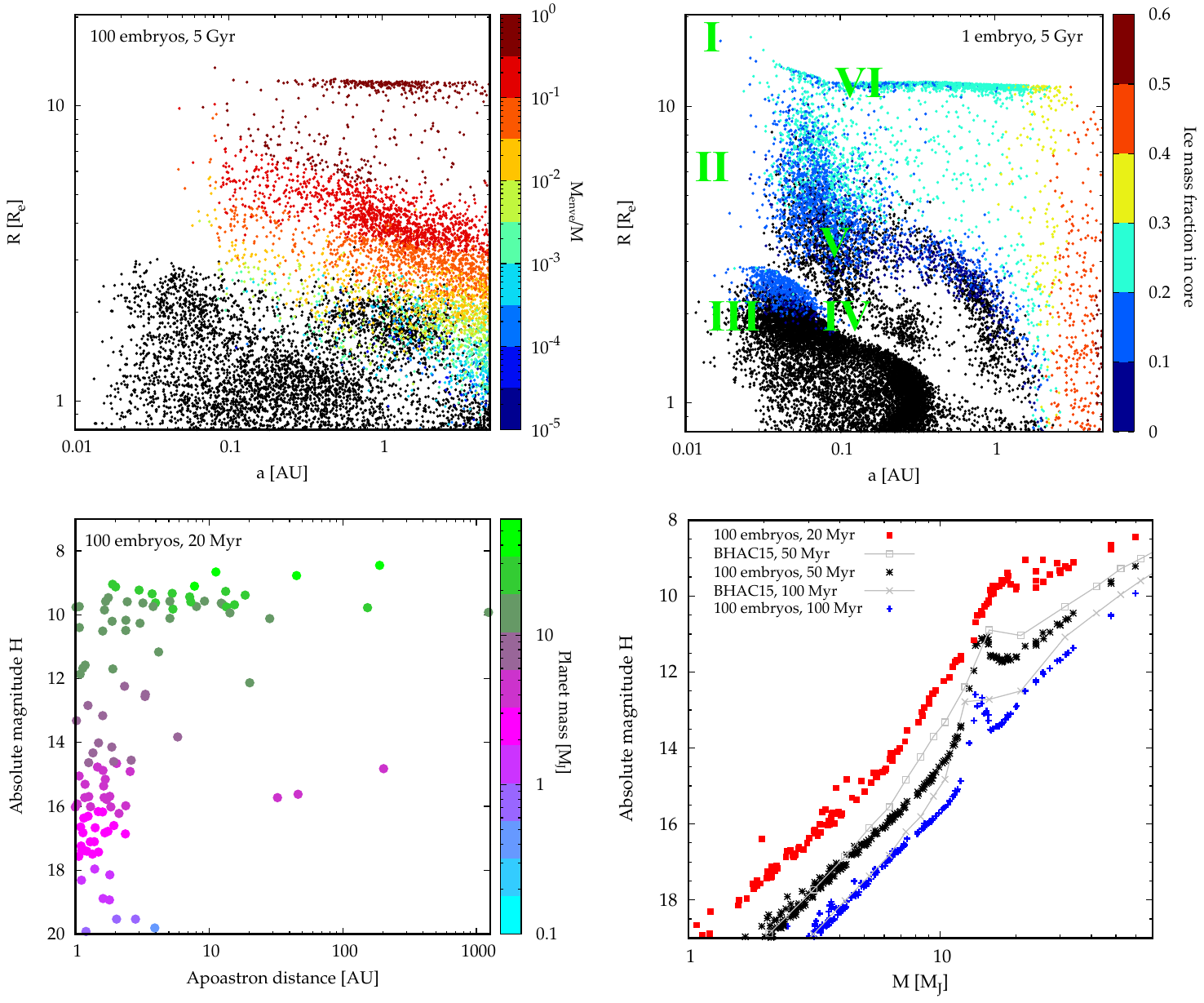}
	\caption{The synthetic populations in the eye of transit and direct imaging surveys. \textit{Top left:} distance-radius plot of the 100-embryos population at \SI{5}{\giga\year}. The color code gives the H/He envelope mass fraction. Planets without H/He are shown in black. \textit{Top right:} distance-radius plot of the single-embryo population, also at \SI{5}{\giga\year}. Roman numerals show important morphological features. The color code shows here the ice mass fraction in a planet's core. Black points are planets that do not contain any ice. \textit{Bottom left:} Apoastron distance versus absolute magnitude in the H band for the 100-embryos population at an age of \SI{20}{\mega\year}. The colors show the planets' mass. \textit{Bottom right:} H band absolute magnitude as a function of planet mass and age. The isochrones of \citet{2015AABaraffe} are also shown for comparison with grey lines.}
	\label{fig:araH}
\end{figure*}

In Fig.~\ref{fig:araH} we show the population NG73 (single embryo) and NG76 (100 embryos initially) as they would appear to transit and direct imaging surveys, that is, by showing the planes of orbital distance versus radius and apoastron distance versus absolute H magnitude.

A first major goal of the New Generation Planetary Population Synthesis was to predict directly and self-consistently all important observable characteristics of planets in multi-planetary systems, and not only masses and orbital elements as in previous generations of the Bern model. To achieve this, we have included (see \paperone) in the Generation 3 Bern model the calculation of the internal structure of all planets in all phases, in particular also in the detached phase, which was not done in Generation 2.  We have also coupled the formation phase to the long-term thermodynamic evolution phase (cooling and contraction) over Gigayear timescales. With this we can predict also the radius, the luminosity, and the magnitudes for each planet, from its origins as a \SI{0.01}{\mearth} seed to potentially a massive deuterium burning super-Jupiter. In this way it becomes possible to compare one population to all major observational techniques (radial velocity, transits, direct imaging, but also microlensing). These techniques probe all distinct parameters spaces of the planetary populations, and thus constrain different aspects of the theory of planet formation and evolution. Taken together, they lead to compelling combined constraints, and help to eventually derive a standard model of planet formation that is able to explain all major observational findings for the entire population, as opposed to a theory that is tailored to explain a certain sub-type of planets, but fails at other planets.

A second major goal of the new generation population synthesis was to be able to simulate planets ranging from Mars mass to super-Jupiters, and from star-grazing to very distant, or even rogue planets. For close-in planets for which the stellar proximity strongly influences the evolution, this meant that we had to include the effects of atmospheric escape, bloating, and stellar tides. Only then it becomes possible to meaningfully link formation and observations at an age of typically several Gigayears.

The top panels of the figure show the $a-R$ diagram of the two populations, at an age of \SI{5}{\giga\year}. The quantitative description of the radius distribution, the formation tracks leading to the radii, and the statistical comparison to transit surveys will be the subject of a dedicated NGPPS paper (\citealp{NGPPS5}, see also \citealp{2019ApJMulders}), therefore we here only give a short qualitative overview.

In the right plot, the roman numerals shown important morphological features of the close-in population.

(I) are the bloated hot Jupiters. The bloating model is the empirical model of \citet{2018AJThorngrenFortney}, leading to an increase of the radii with decreasing orbital distance inside of about \SI{0.1}{\au} \citep{2011ApJSDemorySeager,2020AASestovicDemory}.

(II) is the (sub)Neptunian desert, which is an absence of very close intermediate mass planets that was observationally characterized for example by \citet{2016NatCommLundkvist}, \citet{2016AAMazeh}, or \citet{2018AABourrier}. It is likely a consequence of atmospheric escape \citep[e.g.,][]{2007AALecavelierDesEtangs,2014ApJKurokawaNakamoto,2019ApJMcDonald}. In the plot, it is not very well visible,  but the hot Jupiters ``above'' are indeed found to down to smaller orbital distances than the intermediate mass planets.

(III) corresponds to the the hot and ultra hot solid planets like Corot-7b \citep{2009AALeger} or Kepler-10b \citep{2011ApJBatalha}.

(IV) is the evaporation valley \citep{2017AJFulton} which was predicted theoretically by several planet evolution models including atmospheric escape \citep{2013ApJOwenWu,2013ApJLopezFortney,2014ApJJin}. Super-Earth planets below the valley have lost their H/He as the temporal integral over the stellar XUV irradiation absorbed by these planets exceeded the gravitational binding energy of their envelope in the potential of the core \citep{2020AAMordasini}.

(V) are the Neptunian and sub-Neptunian planets above the valley. They appear numerous in the single embryo population, but less so in the 100-embryos population. In the analysis of \citet{2019ApJMulders} of an earlier generation of the Bern model, it was found that this class is the only one occurring with a significantly different rate (a lower one) than inferred observationaly from the Kepler survey.

(VI) are the giant planets. In the synthetic population, outside of about \SI{0.1}{\au} (that is, where no bloating is acting), the giant planets lead to an almost horizontal, thin pile-up of radii (but note the logarithmic y-axis). This concentration is the consequence of the following \citet{2012A&AMordasiniC}: the mass-radius relationship in the giant planet mass range has a maximum at around 3 Jovian masses, and is relatively flat. This causes many planets from a quite wide mass range to fall in a similar radius range, close to \SI{1}{\rj}. In the synthetic population, this concentration effect is artificially accentuated: during both the formation and evolutionary phase, the molecular and atomic opacities \citep[from][]{2014ApJSFreedman} correspond to a solar-composition gas, identically for all planets. In reality, the atmospheric compositions and thus opacities differ, inducing via different contraction timescales \citep{2007ApJBurrows} a certain spread in the mass-radius relation that cannot occur in the synthesis. Similarly, in reality planets do not have all exactly the same age.

Several of these features are also visible in the top left panel showing the 100-embryos population, albeit often in a less clear way. This is a consequence of the stochastic nature of the \textit{N}-body interactions. Giant impacts that strip the H/He envelope are an additional effect that is important for the radii that cannot occur in the single embryo population. Two consequences of giant impacts are obvious: first, they populate the evaporation valley with cores that would otherwise be too massive for the envelope to be lost only via atmospheric escape. The fact that the valley appears rather too blurred in the 100 embryo population compared to observations \citep{2017AJFulton,2018AJPetigura} could be an observational hint that impact stripping might be overestimated in the model and should be improved in further model generations, for example along the lines of \citet{2020MNRASDenman}. Second, in the 100 embryo population, in the group of planets at around $a=\SI{1}{au}$ and with radii between about \num{1.5} and \SI{2}{\rearth} (which is above the evaporation valley), there is a region of mixed planets with some possessing H/He, and others without it. In the single embryo population, all planets above the evaporation valley possess in contrast H/He envelopes. The black points in the 100-embryos population are thus the results of giant impact envelope stripping.

The quantitative comparison of the populations with transit surveys is as mentioned beyond the scope of this overview paper, but we note that many similar features are also found in the observed population.
 This reflects that the Generation III Bern is in contrast to older model generations able to simulate the formation and evolution also of close-in planets that are observationally particularly important.

\subsection{Distance-magnitude plot, mass-magnitude relation and giant planets at large orbital distances}

While transit surveys probe the planetary population at close-in orbital distance, direct imaging surveys like GPIS \citep{2019AJNielsenA}, NACO LP \citep{2017AAVigan}, or SPHERE SHINE \citep{2021AAVigan} probe young giant planets at large orbital distances.

In the bottom left panel of Figure~\ref{fig:araH}, the population with 100 embryos (NG76) is shown in the plane of apoastron distance versus absolute magnitude in the H band. The magnitude was calculated using the AMES COND atmospheric grid \citep{2012PTRSAAllard} assuming a solar-composition atmosphere. Magnitudes are a strong function of planet mass, therefore they accentuated massive giant planets.

Note that similarly to the radii, also here, the magnitudes were not obtained from some pre-computed mass-time-luminosity (or magnitude) relation or fit, but from solving the planetary internal structure of all planets during their entire ``life'', i.e., from a planet's birth as a lunar-mass embryo to present day, possibly as a massive deuterium-burning planet. To the best of our knowledge, the Generation III Bern model is currently the only global model predicting self-consistently besides the orbital elements and masses also the radii, luminosities, and magnitudes.

The plot shows that the synthesis only predicts few giant planets outside of about 5 AU. The main cause for this absence is rapid inward migration, explaining why in the single embryo population, there are no giant planets at all outside of about \SI{3}{\au} (Fig.~\ref{fig:1emb}).  As mentioned above (see the formation tracks of group I in Fig.~\ref{fig:am_tracks}), in the multiple embryo populations, giant planets at larger orbital distances are the result of violent scattering events among several concurrently forming giant planets \citep{2019AAMarleau}. Such events take preferentially place in very massive and metal-rich discs, explaining why the distant giant planets are massive (about 3 to several ten Jovian masses), in particular more massive than the ``normal'' giants in the pile-up at about \SI{1}{\au}. This is reflected in the apoastron-magnitude plot by an absence of distant planets with higher magnitudes (i.e., fainter planets). Compared to the mass-distance plot, the clustering is amplified by another effect \citep{2017A&AMordasini}: we see that there is a pile up of planets that have similar magnitudes of 11 to 9. To understand this pile-up, we need to consider the bottom right panel showing the mass - H magnitude relation at \num{20}, \num{50}, and \SI{100}{\mega\year}. This plot is equivalent to the mass-radius plot shown above in connecting fundamental observable characteristic (here mass and luminosity).

Besides the expected general decrease of the magnitude with mass, we also see that there is a bump in the relation. It is caused by deuterium burning  which is modelled as described in \citet{2012A&AMolliere}. At \SI{20}{\mega\year}, the bump is centered at around \SI{20}{\mj}, and at around 13-\SI{15}{\mj} at later times. Deuterium burning delays the cooling, and causes planets of a relatively large mass range (at \SI{20}{\mega\year} about \num{15} to \SI{35}{\mj}) to fall in the same aforementioned magnitude range. This leads to the pile-up seen on the left.

In terms of the statistical properties and frequency of distant giants in the 100 embryo population, we find (see Table~\ref{tab:props}) that giant planets (\SI{>300}{\mearth}) are found outside of 5, 10, and \SI{20}{\au} for only 3.5, 1.6, and \SI{0.8}{\percent} of all stars (compared to \SI{18}{\percent} for all orbital distances). For comparison, in the SPHERE SHINE survey, the observed fraction of stars with at least one planet with $\num{1}-\num{75}{\mj}$ and $a=\num{5}-\SI{300}{\au}$ is $5.9^{+4.5}_{-2.7}$ \% \citep{2021AAVigan}. In this paper, a statistical analysis of the NG76 population in the context of the SPHERE SHINE survey can be found. The distant synthetic planets are also eccentric (mean eccentricity of about 0.4-0.6), found around high [Fe/H] stars (mean: 0.2-0.3), and their multiplicity is unity, i.e., there is only one distant giant planet. They do, however, often have another massive companion closer in. For example, of the 8 giant planets with $a>\SI{20}{\au}$ in the population, 5 have a giant companion inside of \SI{5}{\au}. These properties are all signposts of the violent formation pathway of these planets.

In the future, comparisons with direct imaging surveys should include besides the planet frequency such architectural aspects and also  that due to different formation histories, at a given moment in time, there is no unique mass - magnitude conversion \citep{2017A&AMordasini} as it is traditionally often assumed. This is again visible in the bottom right plot, where the magnitudes as a function of mass obtained in the synthesis are compared to the well known \citet{2015AABaraffe} models. They start form arbitrary hot initial conditions. The general trend is as expected similar in the two cases, and the magnitudes are very similar at lower masses \SI{<5}{\mj} at \num{50} and \SI{100}{\mega\year}, but above there are differences of almost 1 mag. The peak caused by D-burning is clearly sharper in our simulations. This is partially due to the coarse sampling in the \citet{2015AABaraffe} tables, but not only. This could affect that analyses of direct imaging surveys. We also see the intrinsic spread in the self-consistent model population model especially at young ages which comes from the different formation histories. The spread includes now in particular also the effects of giant impacts. The spread means that there is no 1-to-1 conversion from magnitude to mass, even if all other complexities (like cold vs. hot start, atmospheric composition, clouds etc.) would be solved. At \SI{20}{\mega\year}, the spread induces a fundamental uncertainty in the mass-magnitude relation of maximum \SI{1}{\mj} at lower masses without D-burning. In the mass range where D burning occurs, the impact is much larger, inducing an uncertainty of up to \SI{10}{\mj}.

\section{The planetary mass function (PMF)}
\label{sec:mass}

\begin{figure}
	\includegraphics{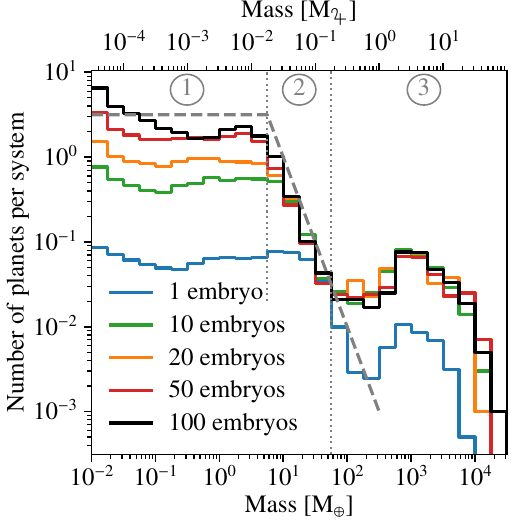}
	\includegraphics{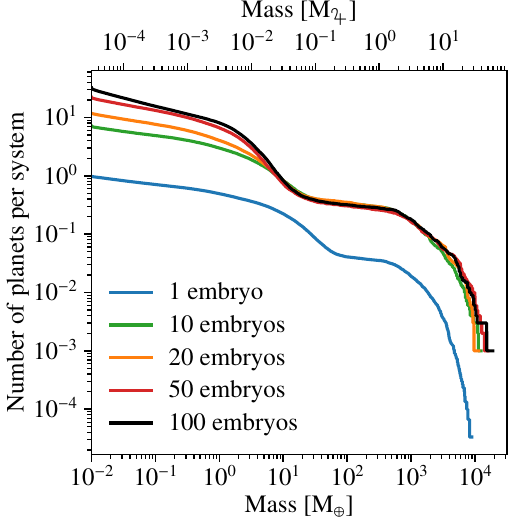}
	\caption{Histogram (\textit{top}) and reverse cumulative distribution (\textit{bottom}) of the planet masses for the four populations presented in this study. The values are normalised by the number of systems in each population. Only planets that reached the end of the formation stage are counted; the maximum number of planets per system (the top left ending of the cumulative curves) can then be lower than the initial number of embryos.}
	\label{fig:mass}
\end{figure}

The prediction of the planetary mass function \citep[PMF;][]{2004ApJIda1,2009A&AMordasinib} is a fundamental outcome of any population synthesis. The PMF is a key quantity because of its observability and because it bears the imprint of the formation mechanism. We show the PMF and its reverse cumulative distribution of the different populations in Fig.~\ref{fig:mass}. Both give the average number of number planets per systems, i.e. the total number of planets divided by the number of systems in each population. The intersection of the curves with the left axis gives the average number of planets per system in each population. In the single-embryo population, the number is close to one, as all planet reach the end of the formation stage and can only be lost by tidal migration during the evolution phase. In the multi-embryos populations, giant impacts lead to the loss of embryos, especially which is especially important in the populations with the largest initial number of embryos. For instance, in the 100-embryos population, on average, only 32 embryos per system reach \SI{5}{\giga\year}. To improve clarity, we will stick to the same colour code throughout the remainder of this article when comparing the populations: black curve denotes the 100-embryos population, red the 50-embryos population, orange the 20-embryos population, green the 10-embryos population, and blue the single-embryo one.

When comparing the overall results, we may divide the relative behaviour in three different regions, as shown with the grey dotted lines on Fig.~\ref{fig:mass}. Region 1, a relatively flat region in the histogram, with its upper boundary depending on the population: about \num{5} to \SI{10}{\mearth} for the multi-embryos populations and \SI{30}{\mearth} for the single-embryo population. Region 2 shows a drop in the occurrence rates, up to the \SI{50}{\mearth} where the cumulative distribution indicates that we have an increased percentage of planets with the population with 50 embryos compared to the one with 20. Finally, region 3 of the giant planets, where there is first a minimum of occurence rate at about \SI{200}{\mearth} followed by a local maximum located in the \num{1000}--\SI{2000}{\mearth} range.

In the first region, the increase of the number of embryos results in the corresponding increase of planets, there is thus virtually no other effect occurring. The only different is the end of this region, which gradually tends toward lower masses as the number of embryos increases. For the population with a single embryo, we observe a steeper drop of the cumulative curve in the 20--\SI{80}{\mearth} range. Planets that are contributing to this feature are actually located at the inner disc edge; these are planets that migrated inward without accreting substantial material during their migration.  Furthermore, we note that the first bin in the histogram has a greater value that the other ones; this is due to the far out embryos that do not grow, or only very little during the formation process.

\subsection{Independence on the number of embryos for the giant planets}

For the planets above \SI{10}{\mearth}, the mass function shows limited variations in all the populations with multiple embryos. The highest mass achieved in each populations shows a trend with the number of embryos. Except for that, the results we obtain are robust. This includes the common slope in the histogram for masses below \SI{100}{\mearth} and the ``planetary desert'' \citep{2004ApJIda1} for planets around \SI{200}{\mearth}.

Thus, to obtain a mass function for planets above $\sim\SI{50}{\mearth}$, the number of embryos is unimportant. The single-embryo population show an overall lower number of planets, but this is due to missed opportunities to form giant planets because it is unlikely that the embryo will start at a location which is needed to form these planets. Applying a correction factor on the outcome of that population is also a possibility to retrieve the mass function obtained from multi-embryos populations while limiting the computational needs (because the \textit{N}-body is the most resource-intensive part of the model). We will use this to study the effects of the model parameters in the subsequent papers of this series.

\subsection{Location of the giant planets}
\label{sec:res-giant-loc}

\begin{figure}
	\includegraphics{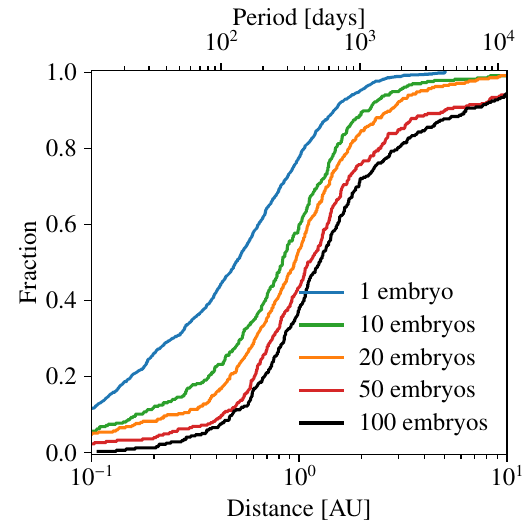}
	\caption{Cumulative distribution of the distance of the giant planets (mass greater than \SI{300}{\mearth}) for the five populations presented in this study. The higher the number of embryos, the more distant the giant planets.}
	\label{fig:dist-giant}
\end{figure}

\begin{figure}
	\includegraphics{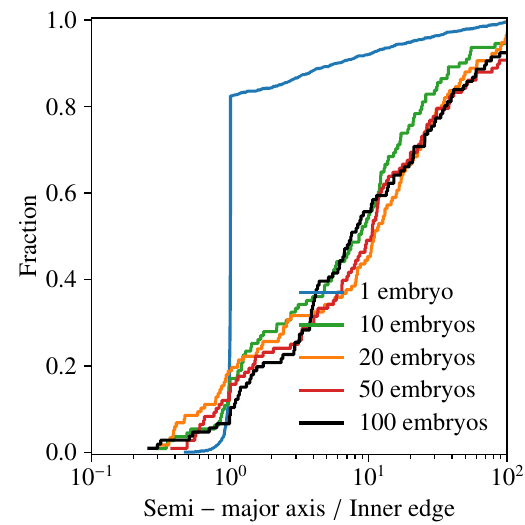}
	\caption{Cumulative distribution of the location of the planets between \num{30} and \SI{300}{\mearth} with respect to the inner edge of the disc for the five populations presented in this study. If only one embryo per disc is present, more than \SI{80}{\percent} of all planets in this mass range end up at the inner edge of the gas disc.}
	\label{fig:dist-subgi}
\end{figure}

However, while the mass function of the giant planets is similar between the populations, the location of the giant planets is not. We find that there is a steady increase in the distance as the number of embryos grows. To illustrate this effect, we provide cumulative distributions of the giant planet's distances for the different populations in Fig.~\ref{fig:dist-giant}. Also, both the 50- and 100-embryos population have \SI{5}{\percent} of the giant planets beyond \SI{10}{\au}.

Nevertheless, all the populations show a similar pattern in the distribution of these planets. We observe a pileup of planets around \SI{1}{\au}, which is consistent with results that suggest a maximum occurrence rate close to the ice line \citep{2019ApJFernandes}. In our populations, the median location of the ice line is at \SI{2.81}{\au}, while the median location of the giant is \num{0.49}, \num{0.83}, \num{0.95}, \num{1.10} and \SI{1.26}{\au} in the 1, 10, 20, 50 and 100-embryos populations respectively. The giant planets are further in than the ice line, which is caused by the gas-driven migration.

We note that there are two causes for this change. First is the reduction of the importance of migration. We have already discussed in Sects.~\ref{sec:res-form-mid} and~\ref{sec:res-form-giant} that in the 100-embryos population the final location of the planets is closer to the starting location of the embryos than in the single-embryo case. The second cause is the increase of the close-encounters that put planets on wide orbits. This effect is responsible for the increase of the distant planets.

All the populations have a similar percentage of planets in the region between \num{0.7} and \SI{2}{\au}. The differences remain in the inner or outer locations, where the populations with a higher number of embryos have more planets beyond \SI{2}{\au}, while the populations with less embryos have more planets inside \SI{0.7}{\au}. Thus, the number of planets in the middle region, between \num{0.7} and \SI{2}{\au} is independent of the initial number of embryos. This feature will be useful, as it allows to study this region using populations with a limited number embryos, which are less computationally expensive.

It was discussed in Sect.~\ref{sec:am-mig-acc} that the single-embryo population exhibit a different accretion pattern that the multi-embryos populations. In the former case, only very massive cores ($\gtrsim\SI{50}{\mearth}$) can undergo runaway gas accretion because the luminosity due to the accretion of solids does not drop during the inward migration. This means many planets will end up at the inner edge of the disc. To illustrate the effect, we show in Fig.~\ref{fig:dist-subgi} the location of the planets in the \num{30} to \SI{300}{\mearth} range, normalised to the inner edge of the gas disc. It can be observed that for the single-embryo population more than \SI{80}{\percent} of the planets are located within or at the inner edge of the gas disc, while we see no special pile-up of planets at the inner edge for the other populations. For the multi-embryos populations, unlike for the giant planets, we do not obtain any systematic shift between the populations. They are also closer-in, with the median distance being \num{0.6} to \SI{0.8}{\au}.

\subsection{A common slope for medium-mass planets}

For planets between \num{5} and \SI{50}{\mearth}, it can be seen that all population show the same behaviour in the histogram. To highlight this point, an additional dashed grey line with a slope $\partial\log{N}/\partial\log{M}=-2$ has been superimposed. Here $N \textrm{d}\log{M}$ is the number of planets whose masses are between $\log{M}$ and $\log{M}+\textrm{d}\log{M}$ (the bin sizes being constant in the logarithm of the planet mass). This corresponds to $N\propto M^{-2}$, as well as $P\propto M^{-2}+C$, where $P$ is the total number of planets whose masses are larger than $M$ (the cumulative distribution) and $C$ a constant of integration. In the case of the single-embryo population, the mass range where the number of planets is similar to the ones of the other populations is limited to masses above \SI{\sim30}{\mearth}, but on the other hand the distribution follows the line to larger masses, up to \SI{200}{\mearth}.

The bulk of the planets in this range are those that migrated close to the inner edge of the gas disc due to the fast inward migration in this mass range. There are two effects that are needed to obtain this peculiar slope: 1) gas accretion and 2) planetary migration.

Concerning the first point, these planets hold a significant amount of H/He, although they have not undergone runaway gas accretion. In the majority of these planets, the dominant component is the core. This slope cannot be achieved only with solids accretion, because without gas, the most massive planets in this range cannot be reproduced. Nevertheless, this is the range where the most massive planets would be found, were it not for gas accretion.

The second effect is planetary migration as the planets that cause the slope are almost all located inside \SI{0.2}{\au}. In the in-situ case (without any migration at all), the decrease in the occurrence rate begins before \SI{5}{\mearth}, because planets can accrete only up to their isolation mass. With migration however, planets can access a larger mass reservoir. Changing the planetary migration prescription in the models also affects the slope. It is however unclear to us the mechanism that causes this precise slope.

In the four multi-embryos populations, the end location where the common slope is encountered is similar, but not for the beginning location. The single-embryo population is different, first because it start to follow the slope at higher masses (about \SI{30}{\mearth}) and second because the end is also for larger masses, at about \SI{300}{\mearth}. As we mentioned before, the slope only occurs where planets are core-dominated. What is different with the single-embryo population with respect to the others is the different behaviour of accretion and migration, as we saw in Sect.~\ref{sec:am-mig-acc}. The resulting planets are mostly located at the inner edge of the disc (Fig.~\ref{fig:dist-subgi}). This being the case, the maximum gas accretion of those planets remains low, as the gas surface density is low (and as we use the Bondi rate to compute the maximum gas accretion rate, \paperone). This explain the shift to larger masses for the change in behaviour of the single-embryo population compared to the multi-embryos one. This interaction, which can also be seen as a competition for solids, can shift the location of where this common slope is found, but will not change it fundamentally.

\subsection{Convergence for the lower masses}

For small masses, the histograms flatten, which is the expected behaviour with our setup. To highlight this, let us remember that the initial surface density of solids is follows $\sigmasol\propto r^{-\betas}$ and define $b$ the half-width of the feeding zone given in terms of the Hill radius (\paperone). Then, let us assume that all bodies grow to their isolation mass,
\begin{equation}
\miso = \frac{\left(4\pi b r^2 \sigmasol\right)^\frac{3}{2}}{(3\mstar)^\frac{1}{2}} \propto r^{\frac{3}{2}\left(2-\betas\right)}
\end{equation}
\citep{1987IcarusLissauer}. As we place the embryos with a uniform probability in the logarithm of the distance $r$, we have $\mathrm{d} P\propto \mathrm{d}\log{r}$. Substituting for the isolation mass, we have $\mathrm{d} P\propto (3/2)(2-\betas)\mathrm{d}\log{\miso}$. So as long as $\betas\neq 2$, we have $\mathrm{d} P\propto \mathrm{d}\log{\miso}$, i.e. $\mathrm{d} P\propto \left(1/\miso\right) \mathrm{d}\miso$. This relationship results in flat histogram when the bin sizes are uniform in the logarithm of the mass, as it is the case in Fig.~\ref{fig:mass}.

There are other mechanisms affecting the mass distribution. For instance, not all planets will grow to their isolation mass, especially the ones at large separation. This results in the number of planets decreasing with the mass, as distant planets, the ones with the largest isolation mass, will need more time. Close-in planets, whose isolation mass is low, will have short accretion times compared to that of the protoplanetary disc and will not suffer from time constraints. However, this effect alone is not able to explain the shape of the distribution for the small-mass planets.

Another mechanism that will affect the distribution is planetary migration. The consequence is that embryos will have access to a larger reservoir of solids that they can accrete. As migration efficiency increases with the mass in the range under consideration here \citep[e.g.][]{2002ApJTanaka,2014PPVIBaruteau}. This will lead to planets that would attain a mass of $\sim\SI{1}{\mearth}$ to migrate and have access to new planetesimals. This pushes the mass distribution toward larger values, which should tend to flatten the curve. The reduced occurrence rate of planets that occur at about \num{e-1} (for the populations with 1, 10, 20, and 50 embryos) and \SI{1}{\mearth} (for the 100 embryos population) are due to planetary migration.

\section{Planet types and system-level analysis}
\label{sec:types}

So far, we have only performed population-level analysis, disregarding the properties planets of planetary systems. Here, we will define different planet types (or categories). This allows to separate the diverse planets from our population and analyse them separately. In addition, this will help to quantitatively compare certain regimes of our populations with the known exoplanets.

The results presented in this section assume that all stars have had a protoplanetary disc during their formation. This assumption can lead to an overall overestimation of the frequency of planets; however observation results shown in Fig.~\ref{fig:lifetimes} show that roughly \SI{80}{\percent} of stars have such a disc until about \SI{2}{\mega\year}. Thus, while our models misses short lived discs, they represent a small fraction of the total and will not significantly affects the analysis presented here.

\subsection{Definitions of planet categories}
\label{sec:types-def}

\begin{table}
    \centering
    \caption{Constraints for the different planet categories}
    \label{tab:type-defs}
    \begin{tabular}{lllll}
        & Min. & Max. & Min. & Max. \\
        & mass & mass & dist. & dist. \\
        Type & [\si{\mearth}] & [\si{\mearth}] & [\si{\au}] & [\si{\au}] \\
        \hline
        Mass \SI{>1}{\mearth} & 1 & \ldots & \ldots & \ldots \\
        \hline
        Earth-like & 0.5 & 2 & \ldots & \ldots \\
        super Earth & 2 & 10 & \ldots & \ldots \\
        Neptunian & 10 & 30 & \ldots & \ldots \\
        Sub-giant & 30 & 300 & \ldots & \ldots \\
        Giant & 300 & \ldots & \ldots & \ldots \\
        D-burning & 4322 & \ldots & \ldots & \ldots \\
        \hline
        Earth-like \SI{<1}{\au} & 0.5 & 2 & \ldots & 1 \\
        super Earth \SI{<1}{\au} & 2 & 10 & \ldots & 1 \\
        Neptunian \SI{<1}{\au} & 10 & 30 & \ldots & 1 \\
        Sub-giant \SI{<1}{\au} & 30 & 300 & \ldots & 1 \\
        Giant \SI{<1}{\au} & 300 & \ldots & \ldots & 1 \\
        \hline
        Habitable zone & 0.3 & 5 & 0.95 & 1.37 \\
        Kepler \citep{2018AJPetigura} & \multicolumn{2}{c}{see Eq.~(\ref{eq:def-kepler-petigura})} & \ldots & 0.88 \\
        Kepler \citep{2018ApJZhu} & \multicolumn{2}{c}{see Eq.~(\ref{eq:def-kepler-zhu})} & \ldots & 1.06 \\
        \hline
        Hot Jupiter & 100 & \ldots & \ldots & 0.15 \\
        Jupiter analogues & 105.3 & 953.4 & 3 & 7 \\
        Giant \SI{>5}{\au} & 300 & \ldots & 5 & \ldots \\
        Giant \SI{>10}{\au} & 300 & \ldots & 10 & \ldots \\
        Giant \SI{>20}{\au} & 300 & \ldots & 20 & \ldots \\
        \hline
    \end{tabular}
\end{table}

The planet categories were selected as follows: we first have a series that are constrained only by the planet masses: Earth-like plants are between \num{0.5} and \SI{2}{\mearth}, then super Earth up to \SI{10}{\mearth}, Neptunian to \SI{30}{\mearth}, Sub-giant to \SI{300}{\mearth} and giant above. In addition we also provide Deuterium-burning planets for masses larger than \SI{13.6}{\mj}, which overlaps with the giant planets category. The mass range of the sub-giants was chosen so that the category is located where the planetary desert discussed in Sect.~\ref{sec:mass} is found in the multi-embryo population. We also set categories for the same masses, but for planets inside \SI{1}{\au}. The presence of the second series of categories is to avoid counting embryos that did not finish growth during the formation stage of our model (see the discussion in Sect.~\ref{sec:res-form-low}).

We defined planets in the habitable zone as planets between 0.3 and \SI{5}{\mearth} in mass and located between 0.95 and \SI{1.37}{\au} \citep{2014PNASKasting}. We also include two different category that relate to the Kepler's observatory biases. The first follows \citet{2018AJPetigura}, which contains planets with a period $P<\SI{300}{\day}$ (\SI{0.88}{\au}) that also satisfy
\begin{equation}
\frac{\rtot}{\si{\rearth}}>1.37\left(\frac{P}{\SI{100}{\day}}\right)^{0.19},
\label{eq:def-kepler-petigura}
\end{equation}
with $\rtot$ the planet's radius. The second follows \citet{2018ApJZhu}, with planets that have a period $P<\SI{400}{\day}$ (\SI{1.06}{\au}) and satisfy
\begin{equation}
\frac{\rtot}{\si{\rearth}}>2\left(\frac{\aplanet}{\SI{0.7}{\au}}\right)^{0.31}.
\label{eq:def-kepler-zhu}
\end{equation}

Finally we have several categories related to giant planets: hot Jupiters have more than \SI{100}{\mearth} and are located within \SI{0.1}{\au}, Jupiter analogues have masses between $1/3$ and \SI{3}{\mj} and semi-major axis between 3 and \SI{7}{\au}, and three categories for giant planets (mass above \SI{300}{\mearth}) further out than 5, 10 and \SI{20}{\au}. These categories were chosen to identify planets that lie outside of the bulk of giants. Such planets are prime targets for direct imaging surveys. For instance, there are no giant planets outside \SI{5}{\au} in the single-embryo population (see Fig.~\ref{fig:1emb}), so they ended up there because of planet-planet interactions in the multi-embryos populations. All these definitions are summarised in Table~\ref{tab:type-defs}.

\subsection{Occurrence rates and multiplicity as function of the number of embryos}
\label{sec:types-conv}

\begin{table*}
	\caption{Percentage of systems/stars with specific planetary types ($f_\mathrm{s}$) and their mean multiplicity ($\mu_\mathrm{p}$) for the different populations. The multiplicity is the mean number of planets of a given type per star for those stars which have at least one planet of this type. The multiplicity of the single-embryo population is not given because it is 1 per definition.}
	\label{tab:types}
	\begin{center}
		\begin{tabular}{lrrrrrrrrr}
			\hline
			& \multicolumn{9}{c}{Initial umber of embryos} \\
			& \multicolumn{1}{c}{1} & \multicolumn{2}{c}{10} & \multicolumn{2}{c}{20} & \multicolumn{2}{c}{50} & \multicolumn{2}{c}{100} \\
			Type & $f_\mathrm{s}$ & $f_\mathrm{s}$ & $\mu_\mathrm{p}$ & $f_\mathrm{s}$ & $\mu_\mathrm{p}$ & $f_\mathrm{s}$ & $\mu_\mathrm{p}$ & $f_\mathrm{s}$ & $\mu_\mathrm{p}$ \\
			\hline
			\hline
			Mass \SI{>1}{\mearth} & \SI{49.3}{\percent} & \SI{91.8}{\percent} & 3.3 & \SI{93.2}{\percent} & 4.3 & \SI{94.6}{\percent} & 7.0 & \SI{96.0}{\percent} & 8.4 \\
            \hline
            Earth-like & \SI{15.5}{\percent} & \SI{63.7}{\percent} & 2.1 & \SI{69.0}{\percent} & 3.3 & \SI{83.7}{\percent} & 4.9 & \SI{90.1}{\percent} & 5.2 \\
            Super Earth & \SI{19.1}{\percent} & \SI{73.6}{\percent} & 2.1 & \SI{75.5}{\percent} & 2.8 & \SI{79.3}{\percent} & 4.8 & \SI{82.1}{\percent} & 5.6 \\
            Neptunian & \SI{13.3}{\percent} & \SI{34.4}{\percent} & 1.2 & \SI{33.0}{\percent} & 1.3 & \SI{27.2}{\percent} & 1.3 & \SI{30.3}{\percent} & 1.4 \\
            Sub-giant & \SI{5.4}{\percent} & \SI{10.0}{\percent} & 1.1 & \SI{10.1}{\percent} & 1.2 & \SI{8.9}{\percent} & 1.2 & \SI{8.5}{\percent} & 1.2 \\
            Giant & \SI{3.6}{\percent} & \SI{19.8}{\percent} & 1.5 & \SI{19.1}{\percent} & 1.5 & \SI{17.4}{\percent} & 1.5 & \SI{18.1}{\percent} & 1.6 \\
            D-burning & \SI{0.1}{\percent} & \SI{3.0}{\percent} & 1.0 & \SI{3.5}{\percent} & 1.0 & \SI{4.5}{\percent} & 1.0 & \SI{4.5}{\percent} & 1.0 \\
            \hline
            Earth-like \SI{<1}{\au} & \SI{12.9}{\percent} & \SI{52.8}{\percent} & 1.8 & \SI{53.3}{\percent} & 2.9 & \SI{58.7}{\percent} & 3.7 & \SI{57.2}{\percent} & 2.8 \\
            Super Earth \SI{<1}{\au} & \SI{17.8}{\percent} & \SI{66.9}{\percent} & 1.9 & \SI{63.1}{\percent} & 2.5 & \SI{62.0}{\percent} & 3.7 & \SI{66.1}{\percent} & 3.7 \\
            Neptunian \SI{<1}{\au} & \SI{13.1}{\percent} & \SI{32.8}{\percent} & 1.2 & \SI{30.9}{\percent} & 1.2 & \SI{24.8}{\percent} & 1.2 & \SI{26.2}{\percent} & 1.4 \\
            Sub-giant \SI{<1}{\au} & \SI{5.1}{\percent} & \SI{7.7}{\percent} & 1.0 & \SI{6.9}{\percent} & 1.1 & \SI{6.7}{\percent} & 1.1 & \SI{6.5}{\percent} & 1.1 \\
            Giant \SI{<1}{\au} & \SI{2.8}{\percent} & \SI{14.0}{\percent} & 1.2 & \SI{12.6}{\percent} & 1.2 & \SI{9.3}{\percent} & 1.1 & \SI{9.2}{\percent} & 1.1 \\
            \hline
            Habitable zone & \SI{1.0}{\percent} & \SI{14.4}{\percent} & 1.2 & \SI{29.9}{\percent} & 1.3 & \SI{49.4}{\percent} & 1.5 & \SI{43.7}{\percent} & 1.3 \\
            Kepler \citep{2018AJPetigura} & \SI{40.8}{\percent} & \SI{78.2}{\percent} & 3.0 & \SI{76.8}{\percent} & 3.5 & \SI{72.2}{\percent} & 4.6 & \SI{76.7}{\percent} & 4.4 \\
            Kepler \citep{2018ApJZhu} & \SI{37.2}{\percent} & \SI{78.1}{\percent} & 2.8 & \SI{77.1}{\percent} & 3.3 & \SI{71.6}{\percent} & 4.3 & \SI{65.7}{\percent} & 4.5 \\
            \hline
            Hot Jupiter & \SI{0.8}{\percent} & \SI{2.8}{\percent} & 1.0 & \SI{2.2}{\percent} & 1.1 & \SI{0.9}{\percent} & 1.1 & \SI{0.3}{\percent} & 1.0 \\
            Jupiter analogue & \SI{0.0}{\percent} & \SI{0.5}{\percent} & 1.0 & \SI{0.4}{\percent} & 1.0 & \SI{0.2}{\percent} & 1.0 & \SI{0.8}{\percent} & 1.0 \\
            Giant \SI{>5}{\au} & \SI{0.0}{\percent} & \SI{0.7}{\percent} & 1.0 & \SI{1.2}{\percent} & 1.0 & \SI{2.6}{\percent} & 1.0 & \SI{3.5}{\percent} & 1.0 \\
            Giant \SI{>10}{\au} & \SI{0.0}{\percent} & \SI{0.3}{\percent} & 1.0 & \SI{0.3}{\percent} & 1.0 & \SI{1.4}{\percent} & 1.0 & \SI{1.6}{\percent} & 1.0 \\
            Giant \SI{>20}{\au} & \SI{0.0}{\percent} & \SI{0.1}{\percent} & 1.0 & \SI{0.1}{\percent} & 1.0 & \SI{0.9}{\percent} & 1.0 & \SI{0.8}{\percent} & 1.0 \\
            \hline
		\end{tabular}
	\end{center}
\end{table*}

\begin{figure*}
	\includegraphics{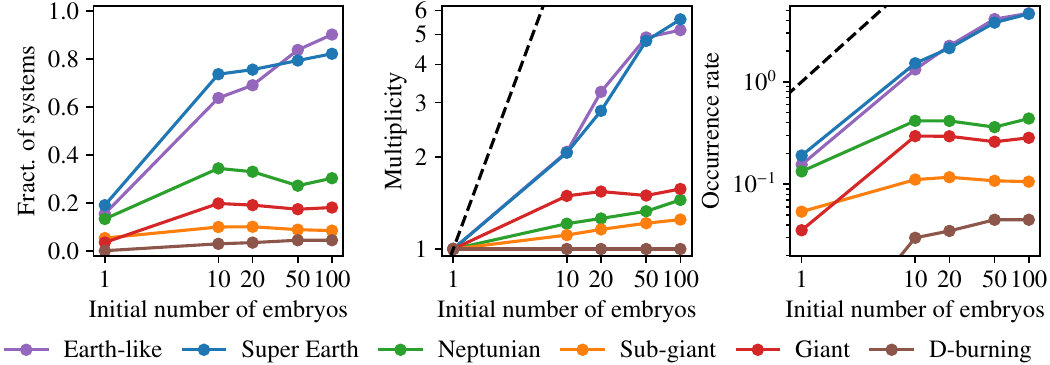}
	\caption{Graphical representation of the fraction of systems (stars) containing at least one planet of this category ($f_\mathrm{s}$), multiplicity ($\mu_\mathrm{p}$, mean number of planets of this category per star including only these stars with at least 1 planet of this category), and occurrence rate ($o_\mathrm{p}=f_\mathrm{s}\mu_\mathrm{p}$) as function of the number of embryos for six planet categories that depend on the masses. The underlying data is provided in Table~\ref{tab:types}. The dashed black lines in the two last panels show the identity function.}
	\label{fig:convergence-mass}
\end{figure*}

\begin{figure*}
	\includegraphics{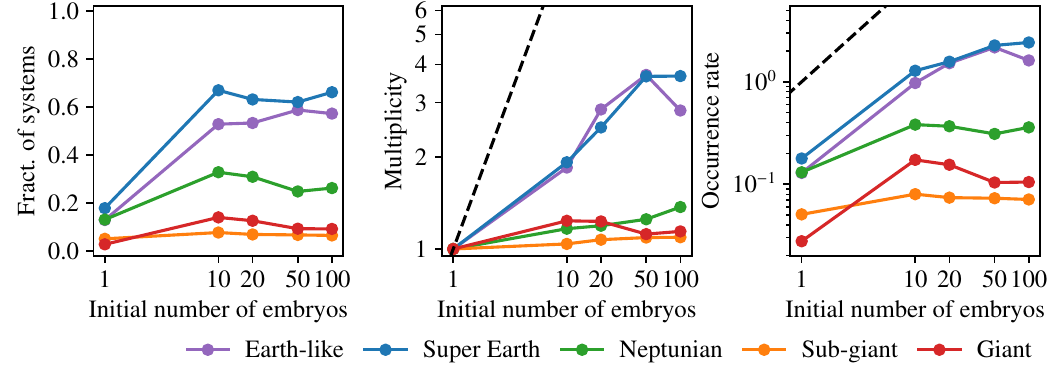}
	\caption{Graphical representation of the fraction of systems (stars) containing at least one planet of this category ($f_\mathrm{s}$), multiplicity ($\mu_\mathrm{p}$, mean number of planets of this category per star including only these stars with at least 1 planet of this category), and occurrence rate ($o_\mathrm{p}=f_\mathrm{s}\mu_\mathrm{p}$) as function of the number of embryos for five planet categories that depend on the masses, but only accounting for the inner planets, i.e. inside \SI{1}{\au}. The underlying data is provided in Table~\ref{tab:types}. The dashed black lines in the two last panels show the identity function.}
	\label{fig:convergence-close}
\end{figure*}

One of the goal of this work is to determine the convergence of our formation model with respect to the initial number of embryos. For this, we provide the occurrence rates and the multiplicity of these categories of planets in Table~\ref{tab:types}. These quantities are computed as follows. The total number of systems in each population is $\nsystot$, whose value is \num{1000} in the multi-embryos population and \num{30000} in the single-embryo population. The number of planet in each category is $\npla$ and the number of systems where at least one such planet is present is $\nsys$. From these, we define the occurrence rate $o_\mathrm{p}=\npla/\nsystot$, the fraction of systems harbouring such planets $f_\mathrm{s}=\nsys/\nsystot$ and the mean multiplicity of such planets $\mu_\mathrm{p}=\npla/\nsys$. It follows that $o_\mathrm{p}=f_\mathrm{s}\mu_\mathrm{p}$.

A graphical representation of the values for the categories of planets as function of their masses for any location is provided in Fig.~\ref{fig:convergence-mass}, while the same information for planets inside \SI{1}{\au} is provided in Fig.~\ref{fig:convergence-close}. In the latter case, the category of Deuterium-burning planets has been left out has it is always empty. In addition to the overall occurrence of these kinds of planets (as we discussed in Sect.~\ref{sec:mass} for the mass function), this gives additional insights on the distribution of planet types in the different systems.

Overall, the results confirm what we discussed in the previous section: convergence is achieved with a smaller number of embryos for the most massive planets than for the lower-mass ones. In the low-mass range (habitable zone, Earth-like and Super-Earth planets) the trend is an increasing number of planets along with the number of embryos. As we already discussed in \paperone{} and Sect.~\ref{sec:res-form-low}, the growth of the planetary bodies is not finished for larger separation by the time our model switched from the formation stage to evolution. Thus, the bodies that are further out may not reflect the end state of planetary systems. For this reason, we also provide categories accounting bodies that are inside \SI{1}{\au}, where growth should be mostly finished at the end of the formation stage of our model (\SI{20}{\mega\year}), and whose result we show in Fig.~\ref{fig:convergence-close}. In that plot, we may note that the multiplicity of the Earth-like planets drops in the 100-embryos population compared to the 50-embryos one. This effect is related to how the growth of small-mass planets is followed up to a giant-impact phase only in the 100-embryos population. With less embryos, the planets do not disturb their orbits to the same extent, and the final phase of planetary growth via giant impacts is missing. This is corroborated by the median mass of the planets in this category: in the 20 and 50-embryos populations, these value is \num{0.95} and \SI{0.96}{\mearth} while in the 100-embryos population it increases to \SI{1.14}{\mearth}.

For the most massive planets (Neptunian, sub-giants, giants and Deuterium-burning) however, we obtain similar numbers in the populations that have at least 10 embryos. Nevertheless, we still see some trends. The first three categories (Neptunian, sub-giants and giants) have slight reductions in their fraction of systems as the initial number of embryos increase while the multiplicity slightly increases so that the overall number of such planets remain quite constant. On the other hand, for the last category (deuterium-burning) we observe first an increase of the occurrence rate along with the number of embryos. Then it becomes constant at \SI{4.5}{\percent} for both the 50 and 100-embryos populations.

For the location of the giant planets, the different categories based on the separation show results that are consistent with what is shown on Fig.~\ref{fig:dist-giant}. The fraction of systems with Hot-Jupiters peaks for the 10-embryos population at \SI{2.8}{\percent} of systems, down to \SI{0.3}{\percent} for the 100-embryo population. For comparison, the observed occurrence rate of these planets is 0.5-\SI{1}{\percent} \citep[e.g.][]{2012ApJSHoward}. Thus, only the 50-embryo with \SI{0.9}{\percent} shows a value that is consistent with the observations. As we have discussed previously, the overall separation of the giant planets increase along with the number of embryos (see Sect.~\ref{sec:am-nemb}), so that 100-embryos population has very few inner giant planets. The decrease of the number of hot-Jupiters is consistent with the decrease of the efficiency of migration with increasing embryo number that we observed in Sect.~\ref{sec:res-form-giant}, as the embryos forming hot-Jupiters come mostly from beyond the ice-line.

Conversely, for the most distant giants, their number increase along with the initial amount of embryos. The fraction Jupiter analogues increases, with an occurrence rate of up to \SI{0.8}{\percent} in the 100-embryos population. Observational estimates for this class of planets are: \SI{3.3\pm1.4}{\percent} \citep{2011ApJWittenmyer}, \SI{2.7\pm0.8}{\percent} \citep{2008PASPCumming} and \SI{\approx3}{\percent} \citep{2016ApJRowan}. We thus obtain values lower than the observational results for this class, even for the 100 embryos population. The same increase with the initial number of embryos applies for the distant giant planets (beyond \SI{5}{\au}). It should be noted that there is a value that is the same for all categories and populations: there is never more that a single distant giant planet in any system.

\subsection{Multiplicity of the different types of planets}
\label{sec:multi-types}

\begin{figure*}
	\centering
	\includegraphics{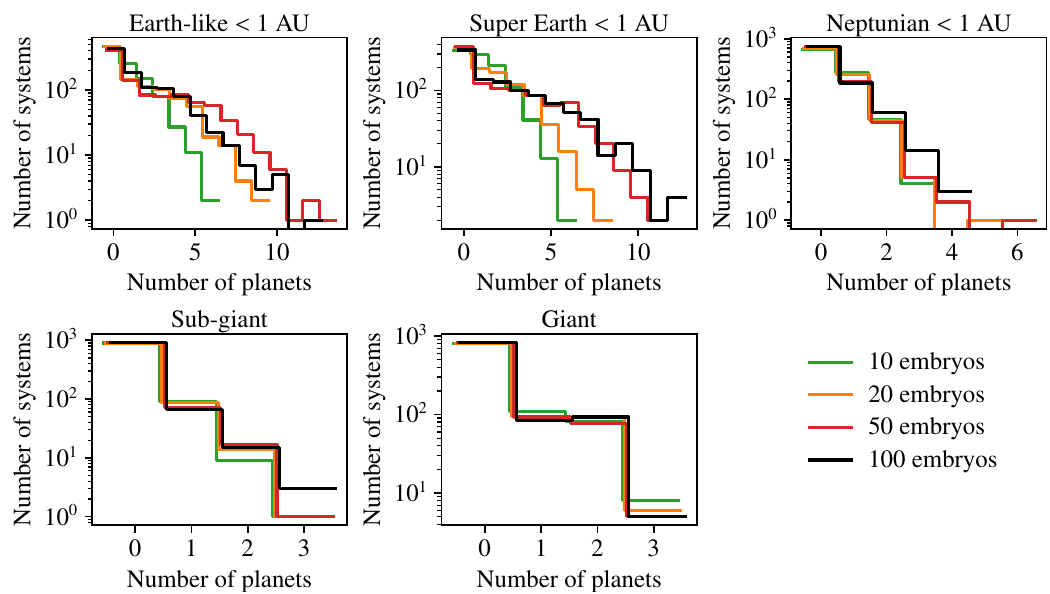}
	\caption{Histogram of the multiplicity of different types of planets, for the four multi-embryo populations presented in this study. In each panel, the curves are slightly shifted horizontally from another to make them more visible. We see for example that for the giant planets, out of the 1000 systems about 800 do not contain any giant planets. About an equal number (100 each) have one or two giants. Less than ten out of the 1000 systems contain 3 giants, which is the highest number per system that occurs.}
	\label{fig:hist}
\end{figure*}

To investigate the distributions of multiplicities in a more detailed fashion than just the mean values shown in Figs.~\ref{fig:convergence-mass} and~\ref{fig:convergence-close}, and Table~\ref{tab:types}, we provide in Fig.~\ref{fig:hist} histograms of these for five types of planets. These categories are the five ones defined in Sect.~\ref{sec:types-def} that have a mass criterion. For the first three (Earth-like, Super-Earth and Neptunian) we use the categories that are limited to planets inside \SI{1}{\au} while for the last two (Sub-giant and Giant) we selected the categories without restriction on the planet' distance. This choice is consistent with our discussion about the lack of convergence for the smaller-mass planets at larger distances.

The results here are very similar to our previous discussion. For the giant and sub-giant categories, all the multi-embryos populations show a similar distribution. Although we do not show it, this is valid for both set of categories (all distances or only within \SI{1}{\au}). Thus, for the most massive planets, the number of embryos does play a role for the final multiplicity, as long as that number is at least around 10. This result is in line with \citetalias{2013A&AAlibert}. It can be noted that there are roughly the same number of systems with giant planets, that have a multiplicity of 1 and 2. This result is consistent with the results of \citet{2016ApJBryan} that half of the systems with a giant planet inside \SI{5}{\au} have a companion planet.

For Neptunian and super Earth planets inside \SI{1}{\au}, we also see that the distributions of multiplicity converge. The Neptunian category does not such much variation between the population, as for the sub-giant and giants planets. However, the convergence of the Super Earths is only achieved between the two populations with the most embryos per system. In the 10-embryos population a steady decrease of the number of systems for higher multiplicities, while in the populations with more embryos it is more likely to find systems with several such planets than lower counts. The Earth-like category shows a similar behaviour, except for the 100-embryos population. Here, the 100-embryos shows less systems with high multiplicity than the 50-embryos population. This is most likely related to the formation of the terrestrial planets that we discussed in \paperone{} and Sect.~\ref{sec:res-form-low}. Thus, for planets above \SI{0.5}{\mearth}, increasing the further the number of embryos would not increase the planet count further.

It should also be noted that unlike for the other categories or other populations, the Earth-like and super Earth categories in the 50- and 100-embryos populations show a plateau for the low-multiplicity counts. Here, the multiplicities between 1 and 3 have similar probabilities and they account for \SI{32}{\percent} of the systems with Earth-like planets and \SI{21}{\percent} of the systems with Super-Earths in the 100-embryos population.

In summary, we find that convergence for the overall multiplicity (that is, the total number of planets of a given type divided by the number of systems having such planets) is a good indicator for the convergence of the underlying distribution of multiplicities. The multiplicity of the sub-giant and giant planets at all locations are similar in all multi-embryos populations (though not their locations, see Sect.~\ref{sec:res-giant-loc}); the same applies for the Neptunian planets inside \SI{1}{\au}. For the inner Super-Earths, only the 50 and 100-embryos populations show similar results while for inner Earth-like planets, the 100-embryos population show a decrease of the multiplicity of the Earth-like planets. The 100-embryos population should be the only one used to analyse Earth-like planets.

\subsection{Statistical results on the 100 embryos population}
\label{sec:res-100emb}

\begin{table*}
	\caption{Properties of different planet kinds from the population with 100 embryos per system. See main text for the precise definition of the kinds.}
	\label{tab:props}
	\begin{center}
		\begin{tabular}{lrrrrrrr}
			\hline
			& Number of & Number of & Fraction   & Occurrence & Multi- &  Mean [Fe/H] & Mean ecc. \\
			Type & planets   & systems   & of systems & rate       & plicity & $\pm$ std. dev. & $\pm$ std. dev. \\
			\hline
			\hline
			All & 32030 & 1000 & \SI{100.0}{\percent} & \num{32.03} & \num{32.03} & \num{-0.03\pm0.20} & \num{0.05\pm0.13} \\
            Mass \SI{>1}{\mearth} & 8065 & 960 & \SI{96.0}{\percent} & \num{8.06} & \num{8.40} & \num{-0.02\pm0.20} & \num{0.05\pm0.10} \\
            \hline
            Earth-like & 4660 & 901 & \SI{90.1}{\percent} & \num{4.66} & \num{5.17} & \num{-0.04\pm0.20} & \num{0.07\pm0.12} \\
            Super Earth & 4603 & 821 & \SI{82.1}{\percent} & \num{4.60} & \num{5.61} & \num{-0.03\pm0.19} & \num{0.04\pm0.08} \\
            Neptunian & 438 & 303 & \SI{30.3}{\percent} & \num{0.44} & \num{1.45} & \num{0.05\pm0.17} & \num{0.05\pm0.09} \\
            Sub-giant & 106 & 85 & \SI{8.5}{\percent} & \num{0.11} & \num{1.25} & \num{0.12\pm0.16} & \num{0.06\pm0.16} \\
            Giant & 284 & 181 & \SI{18.1}{\percent} & \num{0.28} & \num{1.57} & \num{0.14\pm0.15} & \num{0.12\pm0.18} \\
            D-burning & 45 & 45 & \SI{4.5}{\percent} & \num{0.04} & \num{1.00} & \num{0.19\pm0.15} & \num{0.14\pm0.24} \\
            \hline
            Earth-like \SI{<1}{\au} & 1618 & 572 & \SI{57.2}{\percent} & \num{1.62} & \num{2.83} & \num{-0.09\pm0.18} & \num{0.07\pm0.11} \\
            Super Earth \SI{<1}{\au} & 2421 & 661 & \SI{66.1}{\percent} & \num{2.42} & \num{3.66} & \num{-0.01\pm0.18} & \num{0.03\pm0.08} \\
            Neptunian \SI{<1}{\au} & 359 & 262 & \SI{26.2}{\percent} & \num{0.36} & \num{1.37} & \num{0.05\pm0.17} & \num{0.05\pm0.10} \\
            Sub-giant \SI{<1}{\au} & 71 & 65 & \SI{6.5}{\percent} & \num{0.07} & \num{1.09} & \num{0.13\pm0.15} & \num{0.07\pm0.11} \\
            Giant \SI{<1}{\au} & 105 & 92 & \SI{9.2}{\percent} & \num{0.10} & \num{1.14} & \num{0.11\pm0.16} & \num{0.13\pm0.15} \\
            \hline
            Habitable zone & 560 & 437 & \SI{43.7}{\percent} & \num{0.56} & \num{1.28} & \num{-0.11\pm0.18} & \num{0.03\pm0.06} \\
            Kepler \citep{2018AJPetigura} & 3344 & 767 & \SI{76.7}{\percent} & \num{3.34} & \num{4.36} & \num{0.00\pm0.19} & \num{0.03\pm0.10} \\
            Kepler \citep{2018ApJZhu} & 2934 & 657 & \SI{65.7}{\percent} & \num{2.93} & \num{4.47} & \num{0.02\pm0.18} & \num{0.03\pm0.09} \\
            \hline
            Hot Jupiter & 3 & 3 & \SI{0.3}{\percent} & \num{0.00} & \num{1.00} & \num{0.19\pm0.05} & \num{0.07\pm0.04} \\
            Jupiter analogues & 8 & 8 & \SI{0.8}{\percent} & \num{0.01} & \num{1.00} & \num{0.19\pm0.11} & \num{0.14\pm0.20} \\
            Giant \SI{>5}{\au} & 35 & 35 & \SI{3.5}{\percent} & \num{0.04} & \num{1.00} & \num{0.22\pm0.12} & \num{0.35\pm0.26} \\
            Giant \SI{>10}{\au} & 16 & 16 & \SI{1.6}{\percent} & \num{0.02} & \num{1.00} & \num{0.24\pm0.12} & \num{0.58\pm0.24} \\
            Giant \SI{>20}{\au} & 8 & 8 & \SI{0.8}{\percent} & \num{0.01} & \num{1.00} & \num{0.25\pm0.14} & \num{0.55\pm0.28} \\
			\hline
		\end{tabular}
	\end{center}
\end{table*}

\begin{figure*}
	\centering
	\includegraphics{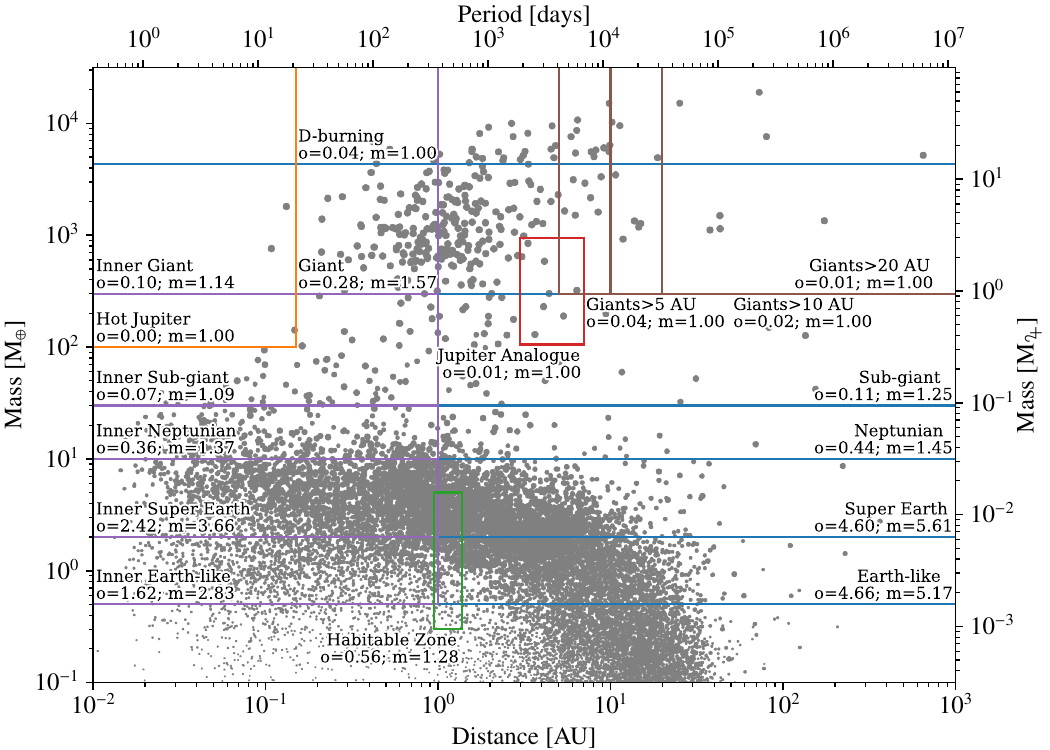}
	\caption{Mass-distance diagram of the population with 100 embryos per system overlaid with the different planet categories. The point size is related to the logarithm of the planet's physical size. The different categories provided in Table~\ref{tab:types} have their boundaries marked, and the principal characteristics of the planets falling within each category are repeated: $o$ denotes the occurrence rate and $m$ the multiplicity.}
	\label{fig:am_legend}
\end{figure*}

\begin{figure}
	\centering
	\includegraphics{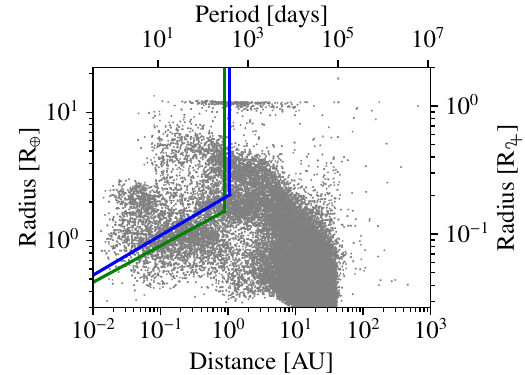}
	\caption{Radius-distance diagram of the population with 100 embryos per system overlaid with two Kepler-related categories. The green line shows the criterion following \citet{2018AJPetigura}, while the blue line shows the criterion following \citet{2018ApJZhu}.}
	\label{fig:ar_legend}
\end{figure}

For the 100-embryos population, we provide key statistical characteristics of the different kinds of planets in Table~\ref{tab:props}, which constitutes the overall demographic predictions of our formation model. The column fraction of systems is the same as in Table~\ref{tab:types}. The mean [Fe/H] column denotes the mean host star metallicity of systems where the relevant kinds of planets are found. We provide an annotated graphical view in Fig.~\ref{fig:am_legend}. This figures shows the same as the bottom right panel of Fig.~\ref{fig:ame}, but the colouring has been removed and dot sizes go with the logarithm of the planets' physical radii. Following the discussion Sects.~\ref{sec:res-form-low} and~\ref{sec:res-time}, the Earth-like, super Earth, and Neptunian at all distances should be taken cautiously. For the two Kepler-related criterion that use the radius rather the mass, we provide their graphical representation in Fig.~\ref{fig:ar_legend}. We remind the reader that these (absolute) results depend on the model parameters, likely in a stronger fashion than general trends and relative correlations, as discussed at the beginning of Sect. \ref{sect:popsynt} and in Sect. \ref{subsect:resultspops}.

The values of the occurrence rate column for the ``All'', ``Mass \SI{>1}{\mearth}'' and ``Giant'' categories give of the cumulative distribution shown in the bottom panel of Fig.~\ref{fig:mass} at \num{0.01}, \num{1} and \SI{300}{\mearth} respectively. Out of the initial \num{100000} embryos (\num{1000} systems with 100 embryos each), only \num{32030} remain at \SI{5}{\giga\year}. Most of the embryos (\num{63124}) were lost due to giant impacts, \num{2675} were ejected, \num{1869} ended in the central star following close-encounters during formation stage, \num{292} ended in the central star due to tidal migration in the evolution stage, and \num{10} were fully evaporated during the evolution stage. Thus, on average, 32 embryos per disc remain. But these are mainly embryos that did not grow, in outer parts of disc where accretion is very slow. Of the average of 32 embryos per disc that remain, only 8.4 have a mass larger than \SI{1}{\mearth}, as indicated by the ``Mass \SI{>1}{\mearth}'' category. For comparison, the solar system has five planets matching the same criterion (plus Venus that has a mass of \SI{0.8}{\mearth}). The values are hence not different. The multiplicity is larger for systems with only terrestrial planets, as giants will usually lead to the removal of terrestrial planets \citepalias{NGPPS1}.

Most of the sub-giants are also found to be in the inner part of the disc, with \SI{69}{\percent} of them being within \SI{1}{\au}. These planets either form late or had their envelope ejected just before the dispersal of the gas disc to have a limited time in the runaway gas accretion. They spent then more time when their masses where in the \SI{10}{\mearth} range, which means they experienced more migration than giant planets that had to form quicker or terrestrial planets that are largely unconstrained by the life time of the protoplanetary gas disc. It can also be seen that the multiplicity of sub-giant planets is the lowest, as it is unlikely for two planets to be in the same situation in the same system.

The multiplicity of the distant giant planets is always unity. This means that in no systems we find two (or more) giant planets beyond \SI{5}{\au}. As we discussed Sect.~\ref{sec:res-form-giant}, these planets mostly originate from seeds that were initially positioned within \SI{10}{\au}. They are then moved to their final location by one or more close encounters with other massive planets. Out of those systems, nearly the half only have one giant planet remaining, that is, the one beyond \SI{5}{\au} while the other have one (or in one case, two) other giant planets further in. Nevertheless all these systems had two giant planets at some point, some of which were subsequently lost, mostly by ejections. The study of systems with giant planets will be the subject of further work \citep[\citetalias{NGPPS11}]{NGPPS11}.

The comparison between the inner categories and the others allows to recover some information about the location of these planets. Only few systems have multiple Sub-giant and giant planets inside \SI{1}{\au}, as we can see in Table~\ref{tab:props}. What we can learn in addition here is that these do not occur for systems with the highest metallicity, but rather for moderate values.

\subsection{Effect of metallicity}
\label{sec:feh}

\begin{figure*}
	\centering
	\includegraphics{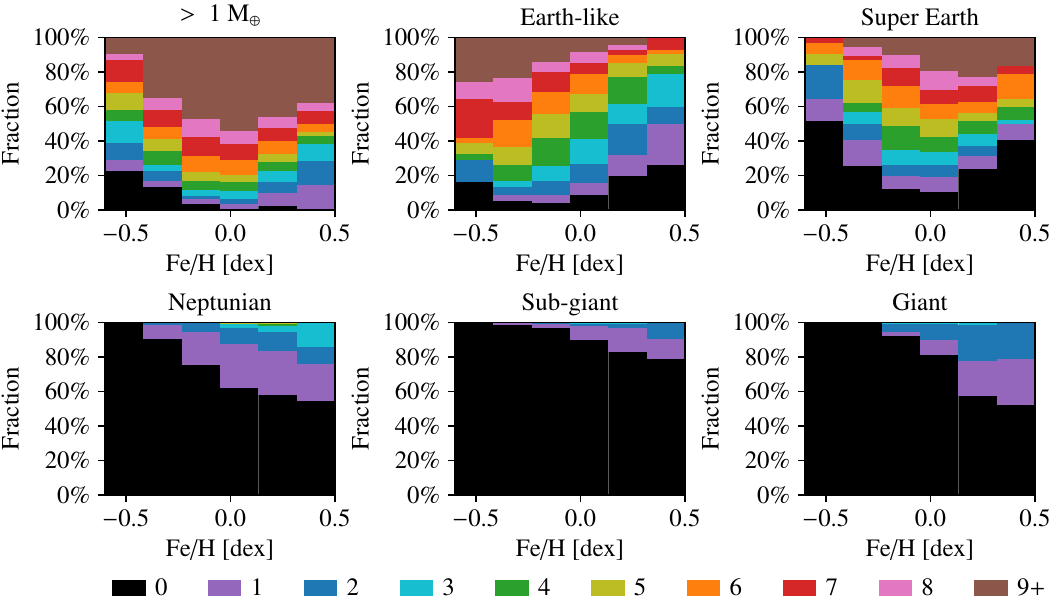}
	\caption{Histogram of the multiplicities of different planet categories versus the stellar metallicity for the 100-embryos population. All these categories do not have constraints on the locations of the planets. The top-left panel shows the histogram of all planets larger than \SI{1}{\mearth} while the others are the categories discussed previously. They were defined in Sect.~\ref{sec:types-def}.}
	\label{fig:feh-all}
\end{figure*}

\begin{figure*}
	\centering
	\includegraphics{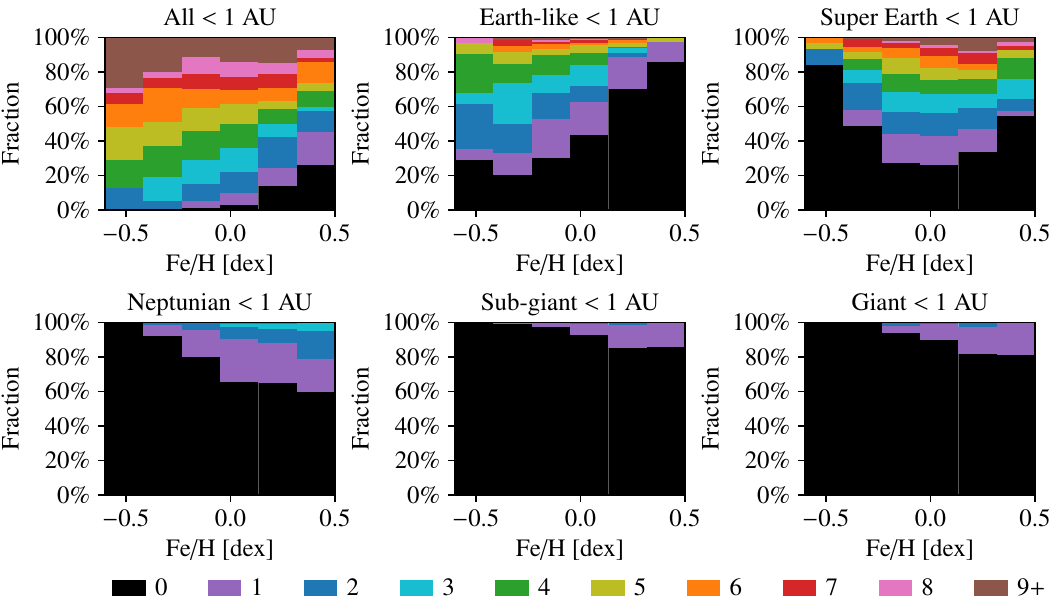}
	\caption{Histogram of the fraction of systems hosting planets of five categories versus the stellar metallicity for the populations presented in this study. The categories have the same boundaries as in Fig.~\ref{fig:feh-all} (with the exception of the top left panel that show planets of all masses), but they only account for the inner planets, i.e. inside \SI{1}{\au}.}
	\label{fig:feh-close}
\end{figure*}

The occurrence rate of giant planets is known to be correlated with the host star's metallicity \citep{1997MNRASGonzalez,2004A&ASantos,2005ApJFischer}. Lower-mass planets on orbits of less than \SI{10}{\day} are also preferentially found around metal-rich stars, but the correlation is weaker for other planets \citep{2016AJMulders,2018AJPetigura}. This finding, particularly in the case of the giant planets, has been an argument to promote the core accretion paradigm, as the formation of a sufficiently massive core takes less time when more solids are present, leaving more time for gas accretion \citep{2004ApJIda2,2009A&AMordasinib,2018ApJWang}.

The mean stellar metallicity of the systems harbouring the different kind of planets is provided in Table~\ref{tab:props}. For both sets of categories that depend of the planet masses (all distances and inside \SI{1}{\au} only), the mean metallicity increases with the masses. The means for Earth-like (\num{-0.04}) and Super-Earths (\num{-0.03}) planets are close to the one of the overall population (\num{-0.03}), so it is for the inner Super-Earths. This means that there is almost no metallcity effect for these planet kinds. However, systems with Earth-like planets inside \SI{1}{\au} and habitable zone are more metal-poor (\num{-0.09 \pm 0.18} and \num{-0.11 \pm 0.18}); these are the only two categories whose mean is lower than the one of the overall population. The mean of the systems with Neptunian and Sub-giant increase, but they are similar each for all distance and inner planets. This suggests that there are no dependency on the stellar metallicity for the location of these planets. Giant planets behave similarly, although the mean of the systems with giants inside \SI{1}{\au} is slightly lower that the one for all distances. On the other hand, the 3 hot-Jupiters have again a higher mean metallicity hosts than distant giant planets, as have the Jupiter analogues and those beyond \SI{5}{\au}. Our results are consistent with observational results for hot Jupiters \citep[e.g.][]{2018ApJBuchhave,2019arXivMoeKratter} and distant, eccentric giants \citep{2018ApJBuchhave}. However, we are unable to reproduce the observation of \citet{2018ApJBuchhave} for cold and circular giants forming around stars with near-Solar metallcities. This suggests that another formation channel exists for this planets, such as pebble accretion of gravitational instability.

The trend of increasing stellar metallicity with planet mass continues to the brown dwarfs (deuterium-burning). This is compatible with the results of \citet{2019GeoSciAdibekyan}, who found that the brown dwarfs can be explained by the core accretion paradigm, as we do in this work. They also found that it is possible for massive brown dwarfs to form around star with solar-like metallicity, but this is for more massive stars that we do not model in this work.

We also note that there is a trend of the metallicity for giant planets at intermediate and large orbital distances. The ones at larger separation are found still over more metal-rich stars than the general population: \num{0.22\pm0.12} and \num{0.25\pm0.14} for the ones beyond \SI{5}{\au} and \SI{20}{\au} versus \num{0.14\pm0.15} otherwise. We remember that all these systems formed more than one giant planet, some of which were subsequently lost (see Sect.~\ref{sec:res-100emb}). Most of the distant giants were brought to their distant orbits after one or more close encounter with other massive planets. These encounters happen after the planets have undergone runaway gas accretion, though the planets may continue to accrete after being sent on wide orbits. Hence, it is necessary for multiple giant planets to form in a single systems for close encounters to strong enough to alter the orbits from \SI{\approx 1}{\au} to more than \SI{20}{\au}.

\subsection{Correlation between multiplicity and metallicity}
\label{sec:res-feh-mult}

Another way to check for a metallicity effect is to look at the correlation between the numbers of certain types of planets as a function of the stellar metallicity. The results of this analysis of the 100-embryos population are provided in Fig.~\ref{fig:feh-all} for the categories encompassing all distances and Fig.~\ref{fig:feh-close} for the ones restricted to planets inside \SI{1}{\au}. The systems are divided in six equally spaced metallicity bins spanning metallicities from \num{-0.6} to \SI{0.5}{dex}.

The results for the most massive planets exhibit the expected behaviour: the fraction of systems with massive planets (Neptunian, sub-giants, and giants) increases monotonically with stellar metallicity. The lowest-metallicity bin does not have any system with Neptunian planets or above. The second bins has some systems with Neptunian planets, very few systems with Sub-giants and none with giants. The next bins show a gradual increase of the fraction of systems with these kind of planets, with roughly the half of the systems in the highest metallicity bin. Additionally, we can see the dependency of the multiplicity on the metallicity. For the sub-giants, we observe that as the metallicity increase, there are first systems with only one such planet, and the further on systems with two and for a few systems even three appear, starting roughly with a solar metallicity. For the giant planets however, the story is interestingly different. In this case, we have that, at the metallicities high enough to form giant planets, the percentage of systems with a single giant planet with respect to the population of systems with any number of giant planets increases. This comes to say that the mean multiplicity is anticorrelated to the metallicity. This is visible by the fact that systems with two giant planets are less frequent in the highest metallicity bin than in the one below. Similarly, the five systems with three giants are not in the highest metellicity bin.

Giant planet formation is then a self-limiting process. The more giant planets are formed, the more likely is that these systems will be unstable. When an instability occurs, it will lead to the loss of planets, by collisions between planets, ejections, or, in small fraction of the cases, accretion by the central star.

In \citetalias{NGPPS11}, dedicated to giant planets, we will quantify the number of giants lost in collisions with other planets and the star, and by the ejection out of the system where they become rogue planets.

A effect is happening for the systems with the highest metallicity: the number of inner planets decreases. All the low-metallicity systems have some inner planets, although they can be very low mass (as there quite less planets that are Earth-like or more). However, this does not mean that these systems do not form planets; it can be seen that all these systems have at least one Earth-mass planet at least (top left panel of Fig.~\ref{fig:feh-all}). What happens in these systems form several massive planets; due their number, the systems become dynamically unstable and the inner planets are lost. Much of these planets collide or are ejected, some fall in the star. In all but one of the resulting systems, a giant planet remains beyond \SI{1}{\au}. In the last case, a smaller planet remains, but its low mass is due to envelope ejection.

\section{Summary and conclusions}

In this work, we use the new Generation III Bern model of planetary formation and evolution presented in \paperone{} to compute synthetic planetary populations of solar-like stars. The model assumes that planets form according to the core accretion paradigm. During the formation stage (0 to \SI{20}{\mega\year}), the model self-consistently evolves a 1D radial constant-$\alpha$ gas disc with internal and external photoevaporation, and the dynamical state of planetesimals under viscous stirring and damping by gas drag. Accretion of solids by the protoplanets includes both planetesimals and giant impacts, while gas accretion is obtained by solving the 1D spherically-symmetric internal structure equations. The model also includes gas-driven planetary migration and gravitational interactions between the protoplanets by means of the \texttt{mercury} \textit{N}-body integrator. During the evolutionary phase (\SI{20}{\mega\year} to \SI{10}{\giga\year}) we follow the thermodynamic evolution (cooling and contraction) of the individual planets including the effects of atmospheric escape, bloating, and stellar tides.

To synthesise populations, we vary four disc initial conditions of the the model according to observed (or observationally motivated) distributions. These Monte Carlo variables are: the initial mass of the gas disc \citep{2018ApJSTychoniec}, the dust-to-gas ratio which is tied to the stellar [Fe/H] \citep{2005AASantos}, the external photoevaporation rate which is distributed such that the synthetic discs have a lifetime distribution compatible with the observed one (see Sect.~\ref{sec:mwind}), and the inner edge of the protoplanetary disc \citep{2017AAVenuti}. Lunar-mass (\SI{e-2}{\mearth}) planetary seeds are put with a uniform probability in the logarithm of the distance into the disc. We compute five populations, each with a different the initial number of seeds per system (or disc).

One aim of this study is to determine the convergence of the model with respect to this free parameter. Our results for this part are:
\begin{itemize}
	\item There is a strong difference between the single and multi-embryos populations. We find that migration in the single-embryo is more effective than in the multi-embryos population.
	\item The properties of the giant planets are only weakly affected by the number of embryos, as long as the latter is at least about 10, consistent with previous work \citepalias{2013A&AAlibert}. For example, the fraction of stars with giant planets and their multiplicity is \SI{19.8}{\percent} and 1.5 in the 10 embryo case, and \SI{18.1}{\percent} and 1.6 in the 100 embryo case.
	\item For the lower-mass planets, a higher number of embryos is necessary. Only the 100-embryos population is able to track the formation of the lower-mass planets up to giant impacts stage (large embryo-embryo collisions).
\end{itemize}
There are two main reasons for these changes. The first is the dynamical interactions between the embryos, as we discussed in \paperone. A tighter spacing between the embryos increases their mutual gravitational interactions, which gives them access to more planetesimals to accrete. This helps small-mass systems to accrete a large percentage of the planetesimals at small separation during the time of our formation models (\SI{20}{\mega\year}).
For the larger-mass planets however, the increased number of embryos results in more competition for solids. When the embryos grow to several Earth masses, they undergo gas-driven migration, which result in access to a larger mass reservoir. However, other embryos will have accreted planetesimals at different places of the disc, resulting in migrating embryos experiencing a sudden drop in their growth rate. The more embryos there are, the less migration embryos must have performed before experiencing this effect. This in turn can trigger runaway gas accretion (see discussion in Sect.~\ref{sec:am-mig-acc}).
The last effect is presence of multiple large embryos. With many embryos, it is more likely to form multiple giant planets. This means that the protoplanets can experience giant impacts. They can lead to envelope stripping of some giant planets. Thus, we find a small proportion of massive cores with a tiny envelope compared to the usual scenario provided by the core accretion paradigm. Systems with many embryos offer a greater diversity of envelopes mass fractions. The increase of dynamical interactions with the number of embryos has repercussions on the formation tracks, with planets being scattered to wide and eccentric orbits.

One of the reason for this study is to determine if results of the population with many embryos per systems can be recovered by populations with a lower initial embryo count. The more embryos are put in each systems, the larger the computational requirements are (mainly due to the \textit{N}-body). For future work where we want to study the effects of model parameters, it is then more efficient to run the simulations with a lower number of embryos. From this study, we find that planets whose masses are roughly \SI{10}{\mearth} or more are insensitive to this parameter provided there are at least 10 embryos per system. There some effects of including more embryos, such as an overall increased distance for the giant planets (see Sect.~\ref{sec:res-giant-loc} and Fig.~\ref{fig:dist-giant}), but this is small compared to other changes in the outcomes of the model (for instance, between the single and multi-embryos populations), so it should not constitute a major problem. The single-embryo population is different from the others and most of its properties are not recovered in multi-embryos populations. Nevertheless, some outcomes, such as the mass function for planets above roughly \SI{10}{\mearth} can be retrieved. This means that the study of gas accretion in the detached phase or the overall fraction of giant planets (provided a correction factor is taken into account) can be done with these simple populations that require very limited computational resources.

Based on our population with the highest number of embryos per system (100), we computed properties of different planet kinds that are provided in Table~\ref{tab:props} and graphically in Fig.~\ref{fig:am_legend}. These values represent the key demographic predictions of our formation model. The main points are:
\begin{itemize}
    \item Overall, planetary systems contain on average 8 planets larger than \SI{1}{\mearth}. The fraction of systems with giants planets at all orbital distances is \SI{18}{\percent}, but only \SI{1.6}{\percent} have one further than \SI{10}{\au}. System with giants contain on average 1.6 giants. This value is consistent with observations \citep{2016ApJBryan}.
    \item Inside of \SI{1}{\au}, the planet type with the highest occurrence rate and multiplicity are super Earth (2.4 and 3.7), followed by Earth-like planets (1.6 and 2.8). They are followed by Neptunian planets, but with an already clearly reduced occurrence rate and multiplicity (0.4 and 1.4).
    \item The planet mass function varies as $M^{-2}$ between \num{5} and \SI{50}{\mearth}. Both at low and high masses, it follows approximately $M^{-1}$.
    \item The frequency of terrestrial and super Earth planets peaks at a stellar metallicity of -0.2 and 0.0 respectively. At lower metallicities, they are limited by a lack of building blocks and at higher metallicities by detrimental growth of more massive, potentially dynamically active planets, which results in accretion or ejection of terrestrial planets and super Earths. The frequency of more massive planet types (Neptunian, giants) increases in contrast monotonically with [Fe/H].
\end{itemize}
These results support observations about the metallicity effect for  giant planets (see Figs.~\ref{fig:feh-all} and~\ref{fig:feh-close}). It should be noted that the quantitative demographic results presented here like the (absolute) occurrence rates and multiplicities are functions of the chosen parameters and underlying model assumptions. We think that relative results and trends are  more robust against parameter variations than such absolute results. To assess the impact of at least two model parameters and assumptions, we also studied two non-nominal populations starting with 100 embryos per disc (Appendix \ref{sec:modparams}): in-situ formation without gas-driven orbital migration, and a population with a steeper slope of the initial planetesimal surface density, inspired by recent planetesimal formation models \citep{2020AAVoelkel}.

In future work, we will compare these populations with observational data, in a similar fashion that was already done for radial-velocity surveys \citep{2009A&AMordasinib} and transit \citep{2019ApJMulders}. This will determine how our populations statistically compare to the known exoplanet population. This should allow us to make  steps towards the development of a standard model of planetary system formation and evolution. Observationally, the syntheses represent a large data set that can be searched for comparison synthetic planetary systems that show how observed systems may have come into existence. The systems, including their full formation and evolution tracks are available online. Knowing the underlying population will also help to understand the pathways certain categories of system follow to reach their final stage and the initial conditions they require. It would also permit to make predictions on the yet-unobserved regions of the parameter space, which is important for the development of future exoplanet discovery and characterisation missions.

\begin{acknowledgements}
The authors thank Ilaria Pascucci, Rachel~B. Fernandes, and Barbara Ercolano for fruitful discussions. We also thank the anonymous reviewer, whose comments helped improve this manuscript.
A.E. and E.A. acknowledge the support from The University of Arizona.
A.E. and C.M. acknowledge the support from the Swiss National Science Foundation under grant BSSGI0\_155816 ``PlanetsInTime''. R.B. and Y.A. acknowledge the financial support from the SNSF under grant 200020\_172746. Parts of this work have been carried out within the frame of the National Center for Competence in Research PlanetS supported by the SNSF.
Calculations were performed on the Horus cluster at the University of Bern.
The plots shown in this work were generated using \textit{matplotlib} \citep{2007CSEHunter}.
\end{acknowledgements}

\bibliographystyle{aa}
\bibliography{manu,add}

\appendix

\section{Influence of model parameters}
\label{sec:modparams}

\begin{figure*}
	\centering
	\includegraphics{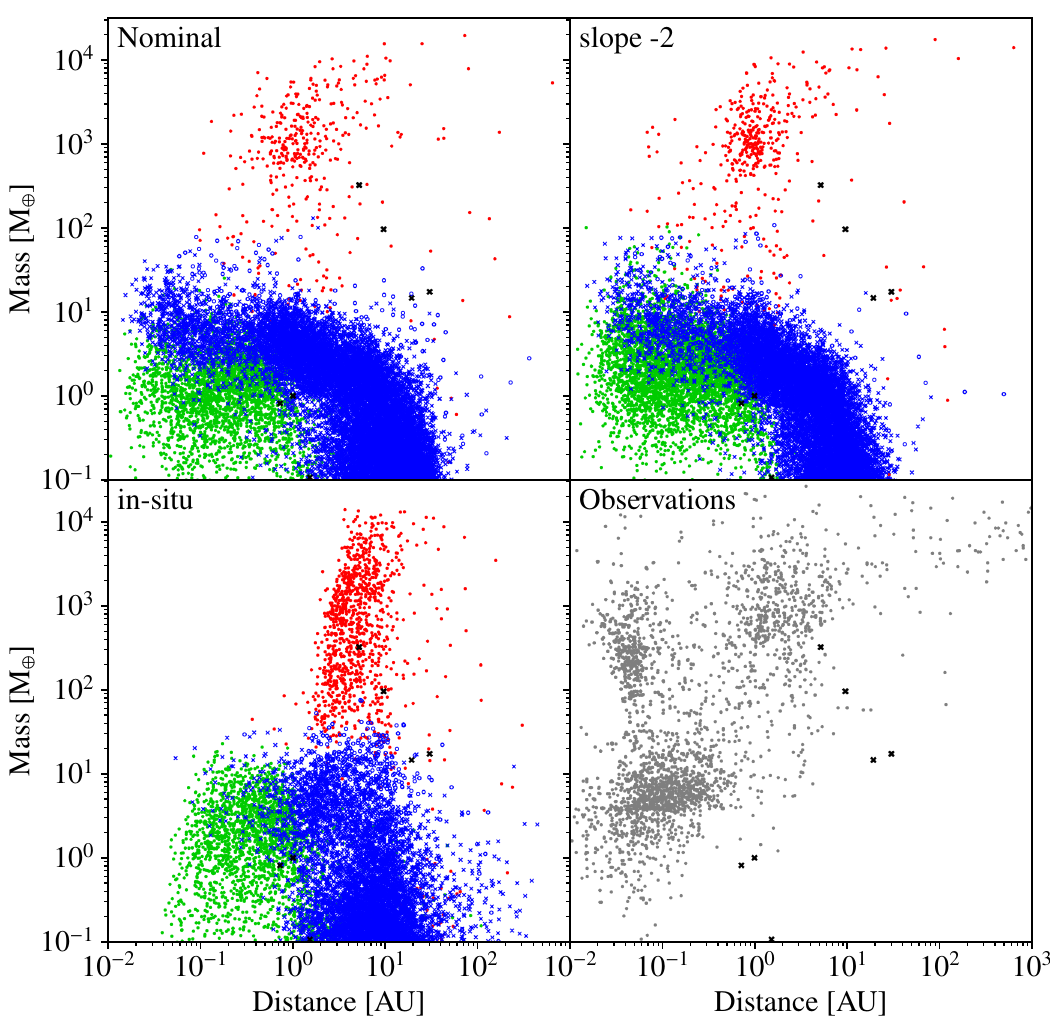}
	\caption{Mass-distance diagram for the comparison of 100~embryos populations generated with different model parameters. The upper left panel shows the nominal population, the upper right panel shows one where the index of the power law for the initial distribution of solids was changed to $\betas=2$ (so that the isolation mass is constant with distance), while the lower left panel shows a population where no gas-driven migration is included. The bottom right panel shows the known exoplanets as of 18~June~2021. It should be noted that this does not account for detection biases, which favour the discovery of hot-Jupiters. This gives the incorrect impression that the model severely fails to reproduce those planets.}
	\label{fig:comp-100emb}
\end{figure*}

\begin{table}
	\caption{Percentage of systems/stars with specific planetary types ($f_\mathrm{s}$) and their mean multiplicity ($\mu_\mathrm{p}$) for the different populations. Similar to Table~\ref{tab:types}, but comparing the non-nominal populations.}
	\label{tab:comp-100emb}
	\begin{center}
		\begin{tabular}{lrrrrrr}
			\hline
			& \multicolumn{2}{c}{Nominal} & \multicolumn{2}{c}{slope -2} & \multicolumn{2}{c}{In-situ} \\
			Type & $f_\mathrm{s}$ & $\mu_\mathrm{p}$ & $f_\mathrm{s}$ & $\mu_\mathrm{p}$ & $f_\mathrm{s}$ & $\mu_\mathrm{p}$ \\
			& [\si{\percent}] & & [\si{\percent}] & & [\si{\percent}] & \\
			\hline
			\hline
			Mass \SI{>1}{\mearth} & \num{96.0} & 8.4 & \num{96.2} & 8.7 & \num{96.2} & 4.7 \\
            \hline
            Earth-like & \num{90.1} & 5.2 & \num{89.9} & 5.3 & \num{54.9} & 3.8 \\
            Super Earth & \num{82.1} & 5.6 & \num{77.6} & 5.8 & \num{78.9} & 2.7 \\
            Neptunian  & \num{30.3} & 1.4 & \num{32.3} & 1.7 & \num{33.2} & 1.2 \\
            Sub-giant & \num{8.5} & 1.2 & \num{12.8} & 1.2 & \num{28.8} & 1.3 \\
            Giant & \num{18.1} & 1.6 & \num{21.3} & 1.5 & \num{45.4} & 1.4 \\
            D-burning & \num{4.5} & 1.0 & \num{3.8} & 1.0 & \num{9.3} & 1.0 \\
            \hline
            Earth-like \SI{<1}{\au} & \num{57.2} & 2.8 & \num{49.6} & 3.8 & \num{37.4} & 1.8 \\
            Super Earth \SI{<1}{\au} & \num{66.1} & 3.7 & \num{69.5} & 4.3 & \num{63.3} & 1.4 \\
            Neptunian \SI{<1}{\au} & \num{26.2} & 1.4 & \num{30.3} & 1.6 & \num{8.0} & 1.1 \\
            Sub-giant \SI{<1}{\au} & \num{6.5} & 1.1 & \num{10.7} & 1.2 & \num{0.6} & 1.0 \\
            Giant \SI{<1}{\au} & \num{9.2} & 1.1 & \num{14.0} & 1.1 & \num{0.0} & \ldots \\
            \hline
            Habitable zone & \num{43.7} & 1.3 & \num{49.4} & 1.3 & \num{26.5} & 1.1 \\
            Kepler P18 & \num{76.7} & 4.4 & \num{90.0} & 4.7 & \num{61.9} & 1.4 \\
            Kepler Z18 & \num{65.7} & 4.5 & \num{83.2} & 4.4 & \num{37.8} & 1.2 \\
            \hline
            Hot Jupiter & \num{0.3} & 1.0 & \num{1.4} & 1.0 & \num{0.0} & \ldots \\
            Jupiter analogue & \num{0.8} & 1.0 & \num{0.3} & 1.0 & \num{22.0} & 1.0 \\
            Giant \SI{>5}{\au} & \num{3.5} & 1.0 & \num{2.2} & 1.0 & \num{30.0} & 1.0 \\
            Giant \SI{>10}{\au} & \num{1.6} & 1.0 & \num{1.3} & 1.0 & \num{6.9} & 1.0 \\
            Giant \SI{>20}{\au} & \num{0.8} & 1.0 & \num{0.7} & 1.0 & \num{2.1} & 1.0 \\
            \hline
		\end{tabular}
	\end{center}
\end{table}

To study the effect of some model parameters, we performed additional populations with initially 100 embryos per system. Two additional populations were generated with 1000 stars/systems each; the first one has the power-law index of the radial slope of initial planetesimals distribution was set to $\betas=2$ (instead of the nominal 1.5) so that the planetesimal isolation mass remains constant with distance \citep{1987IcarusLissauer} except for the jumps at the ice lines, and in line with the results of \citet{2019ApJLenz} regarding the slope of the planetesimal disc as predicted by their planetesimal formation model. In the second population, gas-driven migration is not included, though \textit{N}-body interactions remain; we refer to this population as ``in-situ''. These two populations, along with the nominal one are shown in Fig.~\ref{fig:comp-100emb}. For comparison, we also plot the confirmed exoplanets as of 18 June 2021, without accounting for observational biases. Table \ref{tab:comp-100emb} lists the fundamental demographic results.

Comparing first the the nominal population and that with the modified power-law index, we note little differences for the giant planets. The number of such planets is \num{315} is the population with the modified slope compared to \num{284} in the nominal one, or a \SI{11}{\percent} increase. These planets are still piled up around \SI{1}{\au} in the synthetic populations, or slightly closer-in than the observed population. Some differences remain, like a larger number of hot Jupiters, and of giant planets inside of \SI{1}{\au} in the populations with the modified slope. The number of distant giants is in contrast reduced, meaning that the more centrally condensed distribution of solids also leads to more compact planetary systems, as one might naively expects. In terms of the $f_{\rm s}$ and $\mu_{\rm p}$ listed in Table \ref{tab:comp-100emb}, these differences are certainly visible, but still do not correspond to a really fundamental change of the demographic predictions. Thus, we conclude the power-law index has a rather limited effect on the giant planets, affecting mostly the  occurrence of such planets as a function of orbital distance in the form of an inward shift. Similar, rather limited changes are seen in the $f_{\rm s}$ and $\mu_{\rm p}$ of several other planet types. An interesting difference between the two populations is seen in the hot and warm-Neptune-mass planets (about \SI{10}{\mearth} and inside \SI{0.5}{\au}). In the nominal population, most of these planets come outside the ice line, as there is limited mass reservoir in the inner region of the disc. With a power-law index of $\betas=2$ however, even planets of several Earth masses are able to grow locally. Migration still brings planets from regions beyond the ice line, but it is no longer the only mechanism able to form such planets. This difference is visible in Figure Fig.~\ref{fig:comp-100emb} by the high number of green points in the aforementioned part of the $a-M$ diagram compared to the nominal case, where this part is predominantly populated with blue points.  The percentage of stars with planets in the region probed by Kepler is increased in the slope -2 population, whereas their multiplicity is not significantly changed.

Conversely, the in-situ population shows large differences to the others in many aspects, which is visible both in the $a-M$ diagram and the demographic data of Table \ref{tab:comp-100emb}. For a start, the number of giant planets is increased by a factor of more than two to \num{621}. The fraction of stars with giants is about 45\%, compared to about 20\% in the other populations. The increase of the number of giant planets is due to the large efficiency of the in-situ model at forming these planets; for reference only 8 giants were accreted by the star in the nominal population, while the in-situ population has 3 such occurrences. Secondly, the planetary desert desert between \num{30} and \SI{300}{\mearth} is no longer observed. There are still slightly less planets in this range than giants, but the different is much smaller. The reason for these differences (number of planets and the depth of the desert) with the mechanism of interplay between accretion and migration discussed in Sect.~\ref{sec:am-mig-acc}. In the in-situ population, the mechanism is not effective at all, as there is no migration. This means that the formation pattern observed by \citet{1996IcarusPollack}, with core growth first followed by a long intermediate stage with little accretion before the final gas runaway, is possible in this case. In contrast, when migration is included the planets cannot remain at intermediate masses (\num{\sim10}--\SI{50}{\mearth}) for too long or they will end up taken to the inner edge of the gas disc where gas accretion is impossible. Thirdly, without migration, giant planets are found mostly beyond the ice line, and thus further away than the detected sub-population of warm giants. The in-situ population does not contain a single giant inside of \SI{1}{\au}, in clear contrast to observations. This is because only the high-eccentricity migration channel remains \citep{2018ARAADawson} and the we do not account for it in our populations. Further, the in-situ population also fails to reproduce the warm Neptune and hot-Jupiter sub-populations. This is due to the low mass of solid building blocks available in the inner region of the disc. This means that gas-driven migration is necessary to reproduce the observed population (in particular the location of the giant planets, the presence of close-in massive planets, and the super Earth and sub-Neptunes inside of about \SI{0.1}{\au}). Our nominal prescription results in contrast in too large migration. A certain reduction of the gas-driven migration would lead to a better match with the observations. This could be explained by the following: 1) a weaker migration than what is usually assumed \citet{2018ApJIda}, 2) by lower viscosity, which would reduce migration speed, or 3) a combination of strong and weak migration. The last point stems out from the fact that the observed population shows characteristics that are reproduced partly in either cases. We do not account for the last possibility in our populations, as the viscosity parameter is taken to be constant across all discs. This effect could be reproduced however by the parameter being varied, such as if were treated as one of the Monte Carlo variables of our populations.

Finally, it is worth mentioning that the in-situ simulation populates the positions and masses of Jupiter and Saturn much more than the populations with orbital migration. The latter populations do contain systems with an architecture that is qualitatively the same as the Solar System with low-mass planets inside, followed by two giants, and then ice giant planets outside; however, in such synthetic systems the giants are shifted inward relative to the Solar System, with the inner giant planet residing at about 2-3 au and the outer at 5-8 au. The in-situ population provides a better match, but fails, as we have just seen, in reproducing several fundamental constraints coming from the extrasolar planets. An obvious candidate for a mechanism that would help to conciliate Solar System and exoplanets is the Masset-Snellgrove mechanism \citep{2001MNRASmassetsnellgrove} which leads to outward migration of two giant planets in a mean motion resonance. This is not possible in the Type II migration model we currently use. Including outward migration of two giants in the 1D disc/migration framework is thus another important line of future work.

\end{document}